\documentclass[smallcondensed]{svjour3}

\usepackage{geometry}
\usepackage[left]{lineno} 
\usepackage{hyperref}

\usepackage{graphicx} 			
\usepackage{latexsym}			
\usepackage{scrtime}  			
\usepackage[dont-mess-around]{fnpct} 	
\usepackage[utf8]{inputenc}
\usepackage{color}
\usepackage{etoolbox} 			
\usepackage{hyperref}

\usepackage{natbib}

\usepackage{amsmath}
\usepackage{amssymb}
\usepackage{mathtools}
\usepackage{nccmath}
\usepackage{epstopdf}
\usepackage{gensymb}
\usepackage{lipsum}

\usepackage{array}
\usepackage{hhline}
\usepackage{xcolor}
\usepackage{colortbl}
\usepackage{booktabs}
\usepackage{multirow}
\usepackage{threeparttable}
\usepackage{float}
\usepackage{pifont}
\usepackage{multicol}
\usepackage{esint} 
\usepackage{setspace}

\usepackage{textcomp}

\usepackage[titletoc,title]{appendix}

\usepackage[super]{nth} 

\usepackage[T1]{fontenc}

\usepackage{algorithm, algorithmic}

\newcommand*{\affaddr}[1]{#1} 
\newcommand*{\affmark}[1][*]{\textsuperscript{#1}}

\makeatletter
\newcommand{\mathleft}{\@fleqntrue\@mathmargin0pt}
\newcommand{\mathcenter}{\@fleqnfalse}
\makeatother

\makeatletter
\newcommand*\bigcdot{\mathpalette\bigcdot@{0.7}} 
\newcommand*\bigcdot@[2]{\mathbin{\vcenter{\hbox{\scalebox{#2}{$\m@th#1\bullet$}}}}}
\makeatother

\newcommand\GammaN{\mathrm{\Gamma}}

\newcommand\PhiN{\mathrm{\Phi}}

\makeatletter
\newcommand\arraybslash{\let\\\@arraycr}
\makeatother

\makeatletter
\renewcommand*\env@matrix[1][\arraystretch]{%
  \edef\arraystretch{#1}%
  \hskip -\arraycolsep
  \let\@ifnextchar\new@ifnextchar
  \array{*\c@MaxMatrixCols c}}
\makeatother

\definecolor{lightgray}{RGB}{238,238,238}

\DeclareTextCommandDefault{\registered}{\textsuperscript{\tiny\textregistered\text{ }}}

\begin{document}


\title{Topology optimization for \textcolor{black}{stationary} fluid-structure interaction problems with turbulent flow
}
\titlerunning{Topology optimization for fluid-structure interaction problems with turbulent flow}

\author{
	Renato Picelli \protect\affmark[1] \and
	Shahin Ranjbarzadeh  \protect\affmark[2] \and
	Raghavendra Sivapuram \protect\affmark[3]
	Rafael dos Santos Gioria \protect\affmark[4] \and
	Emílio Carlos Nelli Silva  \protect\affmark[2] 
} 

\authorrunning{
	Renato Picelli  \and
	Shahin Ranjbarzadeh  \and
	Raghavendra Sivapuram \and
	Rafael dos Santos Gioria \and
	Emílio Carlos Nelli Silva
}

\institute{
	Renato Picelli \at \email{rpicelli@usp.br} \\ ORCID: https://orcid.org/0000-0003-4456-0213
	\and Shahin Ranjbarzadeh \at \email{ranjbarzadeh@usp.br} \\ ORCID: https://orcid.org/0000-0003-3269-6850]
	\and Raghavendra Sivapuram \at \email{rsivapur@eng.ucsd.edu} \\ ORCID: https://orcid.org/0000-0002-1639-9349
	\and Rafael dos Santos Gioria \at \email{rafaelgioria@usp.br} \\ ORCID: http://orcid.org/0000-0001-8715-5125
	\and Emílio Carlos Nelli Silva \at \email{ecnsilva@usp.br} \\ ORCID: http://orcid.org/0000-0003-1715-1713
	\\\\
	\affaddr{\affmark[1] Department of Naval Architecture and Ocean Engineering, Polytechnic School of the University of São Paulo, SP, Brazil \\
	\affmark[2] Department of Mechatronics and Mechanical Systems Engineering, Polytechnic School of the University of São Paulo, SP, Brazil\\
	\affmark[3] Structural Engineering Department, University of California, San Diego, CA, USA \\
	\affmark[4] Department of Mining and Petroleum Engineering, Polytechnic School of the University of São Paulo, SP, Brazil \\
	}
}

\date{Received: date / Accepted: date}

\maketitle


\begin{abstract}
Topology optimization methods face serious challenges when applied to structural design with fluid-structure interaction (FSI) loads, specially for high Reynolds fluid flow. This paper devises an \textcolor{black}{explicit boundary} method that \textcolor{black}{employs separate} analysis and optimization grids in FSI systems. A geometry file is created after extracting a smooth contour from a set of binary design variables that defines the structural design. The FSI problem can then be modeled with accurate physics and explicitly defined regions. The Finite Element Method is used to solve the fluid and structural domains. This is the first work to consider a turbulent flow in the fluid-structure topology optimization framework. The fluid flow is solved considering the $k-\varepsilon$ turbulence model including standard wall functions at the fluid and fluid-structure boundaries. The structure is considered to be linearly elastic. \textcolor{black}{Semi-automatic} differentiation is employed to compute sensitivities and an optimization problem using binary design variables is solved via sequential integer linear programming. The fluid loading is linearly interpolated in order to provide the \textcolor{black}{sensitivities} of the fluid flow on the fluid-structure interfaces. Results show that the proposed methodology is able to provide structural designs with smooth boundaries considering loads from low and high Reynolds flow.

\end{abstract}



\section{Introduction}
\label{sec:intro}

The design of elastic structures in engineering projects highly benefits from computational methods. Most of the times, simplification of the loads acting on the structure are enough to yield efficient and useful designs \citep{Keshavarzzadeh19}. These simplifications can include, e.g., the analytical approximation of loads arising from different physics. In some other applications, the physics surrounding the structure represents complex and design-dependent loads that cannot be easily simplified or should not be ignored \citep{Zhang17}. This can be the case of fluid-structure interaction (FSI). One may recall the Tacoma Narrows bridge case \citep{Billah91}, in which the dynamic loads arising from the fluid flow \textcolor{black}{lead to erratic motion and eventually catastrophic failure}. In that case, the proposed structural layout changed the fluid path and could not prevent the structure to fail under the loads from the new fluid flow. With the increase \textcolor{black}{in} complexity of the modern engineering systems and tighter economic and environmental requirements, the design of such structural layouts or topologies goes beyond intuition when considering FSI and new computational methods need to be devised \citep{Fourestey04}.

Fluid-structure interaction is an extremely important engineering problem present in a wide range of applications, e.g., acoustic systems, aerospace structures, dams, pumps, turbines and others. Usually, the structure is governed by the linear or nonlinear elasticity equation \citep{Bungartz06}. Depending on the assumptions for the fluid, the problem can be categorized in different ways. Hydrostatics can consider static fluid pressure on submerged structures in certain applications, e.g., deep sea systems in cases where the fluid movement is negligible \citep{Karamanos04}. If the fluid is also at rest but governed by the wave equation, one has acoustic-structure interaction problems \citep{Axisa}. The term fluid-structure interaction is usually employed to identify the case of a fluid flowing over a structural surface and the effects both physics have on each other \citep{Bathe09}. In a more broad way, all the aforementioned problems have FSI and present the same coupling conditions: stress equilibrium and kinematic compatibility. The imposition of these conditions must be carried out with attention when optimizing fluid-structure systems. The challenge regards the location of the coupling fluid-structure interface that can possibly change during optimization. More general than parametric and shape optimization, this work focuses on structural topology optimization, where the idea is to distribute solid material inside a design domain \citep{Bendsoe03}. The method presented here is employed to design stiff (linearly elastic) structures under FSI loads when a viscous fluid is in motion and governed by the incompressible Navier-Stokes (NS) or Reynolds-averaged Navier-Stokes (RANS) equations.

In topology optimization of fluid loaded structures, the case where the fluid-structure \textcolor{black}{interface} is allowed to change \textcolor{black}{is the most challenging one}. In such case, the fluid loads change as the structural surfaces change during material distribution. The challenges include to keep track of the fluid-structure interfaces location and to correctly model the loads generated by the fluid physics \citep{Jenkins16,Picelli20}. For instance, the surrounding fluid pressures and velocities depend on the structural boundary position (\textcolor{black}{a feature} so-called design-dependency). \textcolor{black}{In this case, the fluid and structural domains change during optimization}. This is a much more complex case to be solved if compared to \textcolor{black}{the case} where the fluid-structure interface remains the same at all steps \citep{Maute04} \textcolor{black}{and the fluid and structural optimization domains remain unchanged (still allowing large deformations as in \citet{Jenkins16})}. Structural topology design under viscous fluid flow loads has been considered first by \cite{Yoon10,Yoon14} in the case \textcolor{black}{with interfaces change}. The author developed a SIMP (Solid Isotropic Material with Penalization) model coupled to a monolithic fluid-structure approach. \cite{Lundgaard18} revisited the method with extensive and clear studies and comparisons, \textcolor{black}{also highlighting the importance of coupled fluid-structure sensitivities}. \textcolor{black}{The methodology by \cite{Lundgaard18}} is based on the distribution of interpolated structural and fluid properties (also called densities in topology optimization). In this way, the elasticity and NS equations are solved in mixed and overlapping domains. \textcolor{black}{On the other hand, in order to favour the numerical analysis, it is reasonable to devise methods that solve the FSI problem with separate domains.} The methods based on level sets \citep{Jenkins16,Feppon20,Li22} and binary design variables \citep{Picelli20} are suitable for that. In comparison, standard binary methods are based on local material distribution, usually a much easier methodology to implement. On the other hand, they present jagged boundaries. Standard level set methods are based on shape sensitivities and move the smooth and implicit boundaries/surfaces in a level set propagation. Both approaches can be quite attractive when dealing with \textcolor{black}{interfaces changes in} fluid-structure optimization. This work devises a new \textcolor{black}{explicit boundary} method for FSI optimization that keeps the (``digital'') material distribution feature from the binary methods but produces designs with smooth boundaries.

The present work uses the Topology Optimization of Binary Structures (TOBS) method by \cite{Sivapuram18} to solve the $\{$0,1$\}$ design problem. TOBS is a gradient-based topology optimization method that employs sequential integer linear programming (SILP). For years, integer programming in topology optimization was deemed intractable due to large computational expenses and inability to effectively handle constraint nonlinearities. However, this can now be reevaluated. TOBS uses off-the-shelf efficient \textit{branch-and-bound} solvers (such as CPLEX from IBM) without significant added cost. Another option is to use the Canonical relaxation algorithm \textcolor{black}{tailored} by \cite{Liang22} to solve the SILP problem. These methods offer clear distinction between solid and void phases, similarly to the Bi-directional Evolutionary Structural Optimization (BESO, \cite{Huang07}) method. On the other hand, TOBS' formal mathematical programming approximates the method to the density-based approach, being able to address multiple \textcolor{black}{(non-linear) constraints \citep{Sivapuram18,Picelli20edu,Mendes22}}, \textcolor{black}{although these constraints were not explored yet in the context of fluid-structure interaction}. For pedagogical purposes, a 101-line MATLAB code is available in \cite{Picelli20edu}. Recently, \cite{Picelli20} associated the ILP solver from TOBS with a geometry trimming procedure in FSI design with low Reynolds flow. The authors also classified the different types of FSI design in topology optimization. For instance, \cite{Jenkins16} applied an immersed boundary approach and \cite{Feppon20} and \cite{Li22} used remeshing, all of these works \textcolor{black}{are based} on the level set framework. The TOBS-based method developed by \cite{Picelli20} applied binary (also called discrete) variables but allowed the finite element package to remesh the design during optimization.

Although the aforementioned methods showed to effectively produce fluid-structure designs, several challenges remain so they can be applied to practical problems. One of them is the consideration of high Reynolds (Re) flow, i.e., turbulent fluid flow. Up to date, the available topology optimization methods were able to consider a fluid flow up to Re = 120 in fluid-structure design problems \citep{Jenkins16,Lundgaard18,Feppon20,Picelli20,Li22}. In order to consider turbulent flow, the fluid walls, including the fluid-structure interfaces, must be explicitly defined so the turbulence equations can be solved with wall functions under a reasonable computational cost. Besides, jagged walls from the standard binary methods are highly detrimental to turbulence simulation. \textcolor{black}{Topology optimization of FSI problems with turbulent flow is a challenging problem that has not been addressed before}. This work builds upon the geometry trimming idea from \cite{Picelli20} and \textcolor{black}{further} employs a boundary smoothing technique and remeshing to include high Reynolds (turbulent) flow in the optimization of FSI systems. With that, turbulence wall functions can be directly included. Herein, the $k-\epsilon$ turbulence model is applied. COMSOL Multiphysics is used as a Finite Element Analysis (FEA) package to solve the governing equations and to provide \textcolor{black}{semi-automatic symbolic} differentiated sensitivities. \textcolor{black}{The combination of the TOBS method with the geometry trimming (GT) idea with a boundary smoothing procedure} was first developed and called TOBS-GT method by \cite{Picelli22} for minimizing fluid flow energy dissipation \textcolor{black}{in a single physics system. In \cite{Picelli22}, the Brinkman term was included to aid the derivation of the objective function and to find optimized fluid channels. In the present multiphysics work, the Brinkman term is not required as the focus is on the structural side of the fluid-structure system and the material models are used to interpolate the stiffness and design-dependent loading, differently from \cite{Picelli22}.} In summary, the TOBS-GT solves the problem for a structured grid of optimization points, obtaining a binary $\{0,1\}$ set of design variables. Then, a geometry file is produced by reading the variables and trimming out the void regions represented by $\{0\}$. The wall contours are smoothed via a Savitzky-Golay filter \citep{Savitzky64}, in 2D, and a CAD wrapping tool, in 3D. Any new geometry is freely meshed with COMSOL Multiphysics based on physics requirements, e.g., modelling of boundary layers or adaptive refinement. This can be of some benefit when solving complex FSI problems that need local mesh quality \citep{Zienkiewicz05}. The previous FSI binary topology optimization methods \citep{Picelli17} were criticized for not considering the sensitivities of the fluid flow loading on the structural surface. This problem is solved in this work by linearly interpolating the fluid loads via the stress equilibrium coupling condition. To the best of the authors' knowledge, this is the first work to carry out fluid-structure topology optimization with turbulent fluid flow and the first binary method that includes coupled sensitivities. The minimum compliance problem with single or multiple volume constraints is solved by considering the case of linearly elastic structures under turbulent fluid flow loading. The contributions of the proposed method can be outlined as it follows:

\begin{itemize}
    \item Turbulence models with wall functions are directly included in the fluid-structure topology optimization framework without the need of developing new interpolation models.
    \item The TOBS-GT is the first \textcolor{black}{explicit boundary} method that employs a binary optimization solver and produces smooth boundary designs in FSI systems.
    \item The \textcolor{black}{use of separate} optimization and FEA grids allows the \textcolor{black}{method} to have \textcolor{black}{more than one optimization point per finite element, generating} convergent and \textcolor{black}{computationally viable finite element} meshes, an important feature when employing turbulence models.
    \item Coupled sensitivities are used for the first time in the binary topology optimization framework by interpolating the fluid flow loading.
    \item High Reynolds flow is solved, herein up to a case with Re = 50,000.
\end{itemize}

The remainder of the paper is as follows. In Sec. \ref{sec:FEA}, it is described the basic FEA-based formulation to simulate the turbulent fluid-structure problem. In Sec. \ref{sec:topopt}, the topology optimization framework and the computational procedure are detailed. In Sec. \ref{sec:results} numerical results are presented and discussed while in Sec. \ref{sec:con} some conclusions are inferred.

\section{Fluid-structure interaction model}
\label{sec:FEA}

\subsection{FE-based turbulent fluid flow}

\begin{figure}[ht]
\centering
\includegraphics[scale=1]{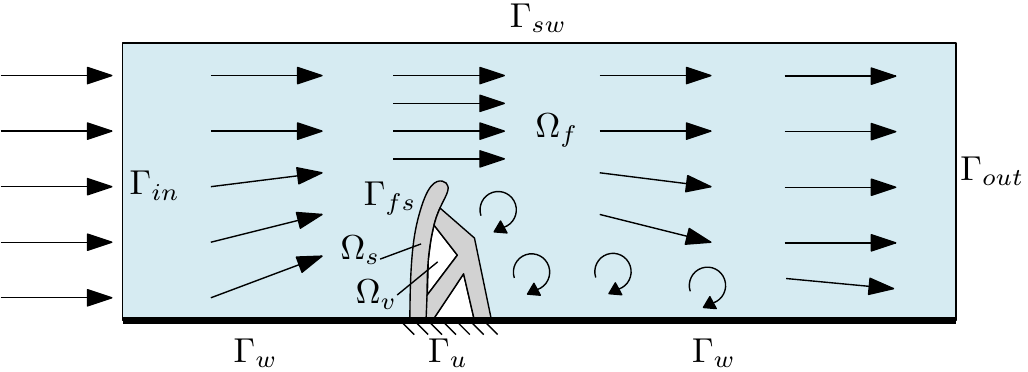}
\caption{Illustration of the fluid-structure interaction problem. A fluid domain $\Omega_f$ with inlet $\Gamma_{in}$ and outlet $\Gamma_{out}$ boundaries interacts with a structure $\Omega_s$ via the fluid-structure interface $\Gamma_{fs}$. Fixed displacements in the structure are prescribed at $\Gamma_u$ and slip and no-slip conditions are imposed at $\Gamma_{sw}$ and $\Gamma_{w}$ boundaries of the fluid domain, respectively. A void domain $\Omega_v$ can be created when topology optimizing the structure.}
\label{fig:fsi_model}
\end{figure}

The turbulent fluid flow domain $\Omega_f$ (see Fig. \ref{fig:fsi_model}) can be described by the RANS equations \citep{Wilcox98}. Assuming a homogeneous, steady-state, isothermal and incompressible fluid flow with constant viscosity and density and no body forces, these equations can be given in the tensor form by
\begin{equation}\label{sec2_fluid_mom}
{\rho_f}\left( {{\mathbf{v}}} \cdot \nabla {\mathbf{v}} \right)   = \nabla \cdot \left[ -p \mathbf{I} + \boldsymbol{\tau}_f \right] \qquad \mbox{in $\Omega_{f}$,} 
\end{equation}
\begin{equation}\label{sec2_fluid_con}
\rho_f  \nabla  \cdot \left({{{\mathbf{v}}}} \right) = 0 \qquad \mbox{in $\Omega_{f}$,}
\end{equation}
where $\rho_f$ is the fluid density, $\mathbf{v}$ is the averaged velocity field, $p$ is the fluid pressure, $\mathbf{I}$ is the identity matrix, and $\boldsymbol{\tau}_f$ is related to viscosity, given by
\begin{equation}\label{sec2_stress_tensor}
\begin{aligned}
\boldsymbol{\tau}_f = \left(\mu + \mu_t \right) \boldsymbol{\epsilon} \left(\mathbf{v} \right) \mbox{,} \\
\boldsymbol{\epsilon} \left(\mathbf{v}\right)  = \left({\nabla}{{\mathbf{v}}} + \left({\nabla}{{\mathbf{v}}} \right)^T \right) \mbox{,}
\end{aligned}
\end{equation}
where $\mu$ is the fluid dynamic viscosity and $\mu_t$ is the isotropic eddy viscosity calculated with a turbulence model.


In high Reynolds flow, the isotropic eddy viscosity $\mu_T$ considers the normal and shear stresses on the fluid caused by turbulent eddies. The calculation of $\mu_T$ depends on the engineering application, presenting particular conveniences or numerical limitations. The ways of computing $\mu_T$ define a turbulence model. The $k-\varepsilon$ is a common choice and is used in this paper for illustration.

\subsubsection{$k-\epsilon$ turbulence model}

The inclusion of the turbulent effects via the $k-\varepsilon$ model is based on the transport of two additional turbulent quantities, namely the turbulent kinetic energy $k$ and its dissipation rate $\varepsilon$ \citep{Wilcox98}. In steady-state, these two additional balance equations can be written as
\begin{equation}\label{sec2_turb_k}
\rho_f {{{\mathbf{v}}} \cdot \nabla} k = \nabla \cdot \left( \left( \mu + \frac{\mu_T}{C_k} \right) \nabla k \right) + P_k - \rho_f \varepsilon \mbox{,}
\end{equation}
\begin{equation}\label{sec2_turb_e}
\rho_f {{{\mathbf{v}}} \cdot \nabla} \varepsilon = \nabla \cdot \left( \left( \mu + \frac{\mu_T}{C_\varepsilon} \right) \nabla \varepsilon \right) + C_{\varepsilon 1} \frac{\varepsilon}{k} P_k - C_{\varepsilon 2} \rho_f \frac{\varepsilon^2}{k} \mbox{,}
\end{equation}
where $\mu$ and $\mu_T$ account for the molecular and turbulent effects, respectively, and the latter is defined in terms of the two added turbulent fields
\begin{equation}\label{sec2_nu_t}
    \mu_T = \rho_f C_{\mu} \frac{k^2}{\varepsilon} \mbox{.}
\end{equation}
The source terms in Eq. \ref{sec2_turb_k} and \ref{sec2_turb_e} are written as a function of $P_k$, defined as
\begin{equation}
    P_k = \mu_T \left( \nabla \mathbf{v}:\left( \nabla \mathbf{v} + \left( \nabla \mathbf{v} \right)^T \right) - \frac{2}{3} \left( \nabla \cdot \mathbf{v} \right)^2 \right) - \frac{2}{3} \rho_f k \nabla \cdot \mathbf{v} \mbox{.}
\end{equation}
The $k-\varepsilon$ model has five constants, $C_{\mu}$ = 0.09, $C_{\varepsilon 1}$ = 1.44, $C_{\varepsilon 2}$ = 1.92, $C_k$ = 1.0 and $C_{\varepsilon}$ = 1.3.

The wall functions in COMSOL Multiphysics are such that the computational domain is assumed to be located a distance $\delta_w$ from the wall \citep{COMSOL}. The distance $\delta_w$ is automatically computed so that
\begin{equation}
\delta_w^{+} = \dfrac{\rho_f v_{\tau} \delta_w }{\mu} \mbox{,}
\end{equation}
where \textcolor{black}{$v_{\tau} = C_{\mu}^{1/4} \sqrt{k} = 11.06$} is the friction velocity. The boundary conditions for the velocity is a no-penetration condition $\mathbf{v} \cdot \mathbf{n} = 0$ and a shear stress condition
\begin{equation}
    \boldsymbol{\tau}_f \cdot \mathbf{n} - \left( \mathbf{n} \cdot \boldsymbol{\tau}_f \cdot \mathbf{n} \right) \mathbf{n} = - \rho_f v_{\tau} \frac{\mathbf{v}}{v^{+}} \mbox{,}
\end{equation}
where
\textcolor{black}{\begin{equation}
    v^{+} = \frac{\mathbf{v}}{v_{\tau}} \mbox{,}
\end{equation}
and}
\begin{equation}
    v_{\tau} = \text{max} \left( \frac{|\mathbf{v}|}{\dfrac{1}{\kappa_{V}} \ln{\delta_w^{+}} + B  } , C_{\mu}^{1/4} \sqrt{k} \right) \mbox{,}
\end{equation}
being $\kappa_V$ the von Kárman constant and $B$ a constant that by default is set to 5.2 \citep{COMSOL}.

\subsubsection{Boundary conditions on the fluid}

In order to solve the RANS equations, the following boundary conditions are applied:
\begin{equation}\label{sec2_NS_bc1}
\mathbf{v} = \mathbf{v}_0 \left( \Gamma_{xyz}, v_{in} \right)  \qquad \mbox{on $\Gamma_{in}$,}
\end{equation}
\begin{equation}\label{sec2_NS_bc3}
\boldsymbol{n}^T \left[ -p \boldsymbol{I} + \boldsymbol{\tau}_f  \right] \boldsymbol{n} = -\hat{p}_0  \qquad \mbox{on $\GammaN_{out}$,}
\end{equation}
\begin{equation}\label{sec2_NS_bc4}
    \hat{p}_0 \leq p_{out} \mbox{.}
\end{equation}
representing the velocity profile given at the inlet boundary $\Gamma_{in}$ (Eq. \ref{sec2_NS_bc1}) depending on the coordinates $\Gamma_{xyz}$ and the maximum velocity $v_{in}$ and the stress free condition at the outlet boundary $\Gamma_{out}$ (Eq. \ref{sec2_NS_bc3} and \ref{sec2_NS_bc4}), where $p_{out}$ is the outlet pressure. The velocity profiles used are given in each example. The laminar sublayers at the fluid walls are not resolved when wall functions are imposed. Thus, fluid velocity at walls are non zero. When prescribing $\Gamma_{sw}$, a slip condition is imposed.

The default turbulent flow settings from COMSOL Multiphysics are used \citep{COMSOL}. For the inlet $\Gamma_{in}$, they are the turbulent intensity $I_T$ and turbulence length scale $L_T$, which are related to the turbulence variables via the following equations:
\begin{equation}
    k = \dfrac{3}{2} \left(|\mathbf{v}| I_T \right)^2 \qquad \mbox{and} \qquad \varepsilon = C_{\mu}^{3/4} \dfrac{k^{3/2}}{L_T} \mbox{,} \qquad \mbox{on $\Gamma_{in}$,}
\end{equation}
when solving for the $k-\varepsilon$ model. The default values used are $I_T$ = 0.05 and $L_T$ = 0.01 m. For the outlet $\Gamma_{out}$, the boundary conditions for the turbulence quantities are $\mathbf{n} \cdot \nabla k = 0$ and $\mathbf{n} \cdot \nabla \varepsilon = 0$.

At the walls $\Gamma_{f}$ and $\Gamma_{fs}$, the turbulent kinetic energy is subject to a homogeneous Neumann condition $\mathbf{n} \cdot \nabla k = 0$ \citep{COMSOL}. When solving the $k-\varepsilon$ equations, the boundary condition for $\varepsilon$ is given by
\textcolor{black}{\begin{equation}
\varepsilon = \dfrac{C_{\mu}^{3/4} k^{3/2}}{\kappa_V \delta_w} \qquad \mbox{on $\Gamma_{f}$ and $\Gamma_{fs}$.}
\end{equation}}

\subsection{Structural analysis and coupling}

The solid domain $\Omega_s$ illustrated in Fig. \ref{fig:fsi_model} is assumed to be linearly elastic and under viscous fluid flow loads \citep{Zienkiewicz05}. Without body forces and considering a steady-state analysis, the structure is governed by
\begin{equation}\label{chap02_FSI_eq_solid}
\nabla  \cdot {\boldsymbol\sigma_{s}} \left( \mathbf{u} \right) + \mathbf{f}^{fsi} = 0 \mbox{,}
\end{equation}
where $\nabla  \cdot \boldsymbol\sigma_{s} \left( \mathbf{u} \right)$ is the divergence of the Cauchy stress tensor, $\mathbf{u}$ is the displacement field and $\mathbf{f}^{fsi}$ is the loading vector at the fluid-structure interface $\Gamma_{fs}$. A void domain $\Omega_{v}$ is also present (see Fig. \ref{fig:fsi_model}). Dirichlet boundary conditions at $\Gamma_u$ are applied as:
\begin{equation}
    \mathbf{u} = 0 \qquad \mbox{on $\Gamma_{u}$.}
\end{equation}

Herein, the displacements and deformation of the structural boundaries are considered to be small enough to not change the fluid flow path. Besides, the FSI coupling condition \citep{Bazilevs} is given as:
\begin{equation}\label{chap02_FSI_eq_coup2}
\boldsymbol\sigma_s \cdot \mathbf{n}_s = - \boldsymbol\sigma_f \cdot \mathbf{n}_f \qquad \mbox{on $\Gamma_{fs}$.}
\end{equation}
where $\boldsymbol\sigma_s$ and $\boldsymbol\sigma_f$ are the solid and fluid stress tensors, respectively, with ${\boldsymbol\sigma_f} = \nabla \cdot \left[ -p \mathbf{I} + \boldsymbol{\tau}_f \right]$ and $\mathbf{n}_s$, and $\mathbf{n}_f$ are the normal vectors outwards the solid and fluid domain, respectively. The focus of this work is to develop a method that includes turbulence in the FSI application via topology optimization. By considering small structural displacements, similarly as \cite{Yoon10,Lundgaard18,Picelli20} in topology optimization, the analysis is one-way coupled. Further analyses such as large deformations are target of future research.

\section{Topology optimization framework}
\label{sec:topopt}

\subsection{Optimization Problem}

In this work, structural compliance is minimized subject to volume constraints. This optimization formulation can be expressed as
\begin{equation}\label{eq:topoptproblem}
	\begin{split}
		\underset{\mathbf{x}}{\text{Minimize}}& ~C(\mathbf{x})\\
		\text{Subject to}& ~V_i(\mathbf{x})\le\overline{V}_i, ~i\in[1, N_g]\\
		&~x_j\in\{0, 1\}, ~j\in[1, N_d]
	\end{split}
\end{equation}
where $\mathbf{x}$ is the vector including the design variables $x_j$, $C(\mathbf{x})$ is the structural compliance or total deformation energy, $V_i$ is the volume fraction of the structure with respect to the initial design domain, $\overline{V}_i$ is the constrained volume fraction \textcolor{black}{and $N_g$ and $N_d$ are the total number of constraints and design variables, respectively}.

\subsection{Sensitivity analysis}\label{sec:sens}

The TOBS method is a gradient-based algorithm, therefore, the derivatives (sensitivities) of the objective and contraints functions are required. A general way of computing sensitivities of a function $L$ is via the adjoint method \citep{HaftkaBook}. The generic adjoint equation is expressed as
\textcolor{black}{\begin{equation}\label{eq:adjoint}
\left( \frac{\partial \bf{R}}{\partial \boldsymbol{\PhiN}} \right)^T \boldsymbol{\lambda}  = - \left( \frac{\partial f}{\partial \boldsymbol{\PhiN}} \right)^T \mbox{,}
\end{equation}
where $\boldsymbol{\PhiN}$ and $\boldsymbol{\lambda}$ are the vectors of state and adjoint variables, respectively,} $f$ is the vector of objective function and $\boldsymbol{R}$ is the residual. Sensitivities can then be computed as
\begin{equation}\label{eq:adjoint2}
\left( \frac{d L}{d \mathbf{x}} \right) = \left( \frac{\partial f}{\partial \mathbf{x}} \right)^T + \mathbf{\boldsymbol{\lambda}}^T \frac{\partial \bf{R}}{\partial \mathbf{x}} \mbox{.}
\end{equation}
This is a general formulation to compute the sensitivities of any function $f$, herein used to compute the sensitivities of the structural mean compliance $C$. In order to find analytical expressions, the stiffness of the structure is first interpolated. The SIMP material model can be used, expressed as
\begin{equation}\label{eq:material_model}
    E = x_j^\gamma E_0 \qquad \mbox{on $\Omega_{s}$,}
\end{equation}
where $E_0$ is the Young's modulus of the solid material, $\gamma$ is a penalty factor and $E$ is the interpolated property. To further consider the sensitivities of the fluid flow loading with the change of the fluid-structure surfaces, the stress equilibrium condition from Eq. \ref{chap02_FSI_eq_coup2} is replaced by
\begin{equation}\label{eq:material_model2}
\boldsymbol\sigma_s \cdot \mathbf{n}_s = - x_j \boldsymbol\sigma_f \cdot \mathbf{n}_f \qquad \mbox{on $\Gamma_{fs}$.}
\end{equation}
As the design variables $x_j$ are restricted to 0 or 1 and the TOBS-GT removes the regions with 0 variable, the finite element analysis falls back into the classic governing equations, e.g., when $x_j = 1$, Eq. \ref{eq:material_model2} is equivalent to Eq. \ref{chap02_FSI_eq_coup2}. The same is valid for the stiffness interpolation in Eq. \ref{eq:material_model}. Therefore, the physical analysis is not influenced by the material model. The penalization, however, is used to aid the \textcolor{black}{semi-automatic symbolic} differentiation module built-in the commercial software. Hence, we advocate that any sensitivity analysis method can be used as long as only the derived values for the $\{$0,1$\}$ bound variables are used. The sensitivities of the volume fraction function are also required to be computed. As the volume fraction of the structure is defined as
\begin{equation}\label{eq:volsens}
    V_i = \sum_{j = 1}^{N_d} x_j V_0  \mbox{,} 
\end{equation}
the sensitivity $\dfrac{\partial V_i}{\partial x_j} = V_0$, where $V_0$ is the volume fraction regarding the design variable $j$.

\subsection{TOBS method}\label{sec:tobs}

The TOBS method sequentially creates approximate integer linear optimization subproblems and solves the integer linear programs. Using Taylor's series expansion and truncating it in the linear part, the approximate compliance objective and volume constraint functions can be expressed as,
\begin{equation}\label{eq:linearization2}
    \begin{split}
        C(\mathbf{x}) &\approx C(\mathbf{x}^{k}) + \dfrac{ \partial C(\mathbf{x}^k)}{\partial\mathbf{x}}\cdot\Delta\mathbf{x}^k + O(\big|\big|\Delta\mathbf{x}^k\big|\big|_2^2) \mbox{,} \\
        V_{i}(\mathbf{x}) &\approx V_{i}(\mathbf{x}^{k}) + \dfrac{\partial V_{i}(\mathbf{x}^{k})}{\partial\mathbf{x}}\cdot\Delta\mathbf{x}^k + O(\big|\big|\Delta\mathbf{x}^k\big|\big|_2^2) \mbox{,}
    \end{split}
\end{equation}
with the truncation error being $O(\big|\big|\Delta\mathbf{x}^k\big|\big|_2^2)$. The vector $\Delta\mathbf{x}^k$ indicates the changes in design variables. These changes must be restricted in order to keep the design variables integer (and binary). For instance, in structural topology optimization $x_j = 1$ represents a solid element. In this case, one can choose $\Delta x_j \in\{-1, 0\}$ to prescribe that the element $j$ either turns void ($x_j = 0$) or remains solid. This is analogous for void elements. The bound constraints for $\Delta x_j$ can then be expressed as,
\begin{equation}\label{eq:intconstraint_brackets} 
	\begin{cases}
	0\le\Delta x_j^k \le 1 &\text{ if $x_j^k = 0$} \mbox{,}\\
	-1\le\Delta x_j^k \le 0 &\text{ if $x_j^k = 1$} \mbox{,}
	\end{cases} 
\end{equation}
or,
\begin{equation}\label{eq:intconstraint}
    \Delta x_j^k\in\{-x_j^k, 1-x_j^k\} \mbox{,}
\end{equation}
where $\Delta x_j^k \in \mathbb{Z}$. In order to maintain the linear approximation valid, the truncation error $O(\big|\big|\Delta\mathbf{x}^k\big|\big|_2^2)$ must be small enough. The truncation error is controlled by adding an extra constraint that restricts the number of flips $\Delta\mathbf{x}^k$ from 1 to 0 and vice-versa. The truncation error constraint can be written as
\begin{equation}\label{eq:flipconstraint}
    \big|\big|\Delta\mathbf{x}^k\big|\big|_1\le\beta N_d \mbox{.}
\end{equation}
\textcolor{black}{In this case, the total number of elements flipping between solid and void are constrained to a $\beta$ fraction of $N_d$, the amount of design variables. Small $\beta$'s mean that the number of flips is low as well as the truncation error.}

By using the sequential linear approximations from Eq. \ref{eq:linearization2} and the extra constraints from Eqs. \ref{eq:intconstraint} and \ref{eq:flipconstraint}, the approximate integer linear subproblem is written as
\begin{equation}\label{eq:linopt}
	\begin{split}
		\underset{\Delta\mathbf{x}^k}{\text{Minimize}}& ~\dfrac{\partial C(\mathbf{x}^k)}{\partial\mathbf{x}}\cdot\Delta\mathbf{x}^k \mbox{,}\\
		\text{Subject to}& ~\dfrac{\partial V_i(\mathbf{x}^k)}{\partial\mathbf{x}}\cdot\Delta\mathbf{x}^k\le\overline{V}_i-V_{i}\big(\mathbf{x}^k\big):=\Delta V_i^k, ~i\in[1, N_g] \mbox{,}\\
		&~\big|\big|\Delta\mathbf{x}^k\big|\big|_1\le\beta N_d \mbox{,}\\
		&~\Delta x_j^k\in\{-x_j^k, 1-x_j^k\}, ~j\in[1, N_d] \mbox{.}
	\end{split}
\end{equation}

\autoref{eq:linopt} expresses the sequential optimization subproblems in the standard TOBS formulation. The truncation error constraint (Eq. \ref{eq:flipconstraint}) restrains the topology from undergoing great changes. \textcolor{black}{This might lead to the infeasibility of some of the constraints in the current iteration $k$ when the bound $\Delta V_i^k=\overline{V}_i-V_{i}\big(\mathbf{x}^k\big)$ is used. To ensure the existence of feasible solutions in the optimization subproblems, the upper bounds of the constraints are relaxed. In this case,} the constraint bounds are modified by using
\begin{equation}\label{eq:constbounds}
	\Delta V_i^k = 
	\begin{cases}
	-\epsilon_i V_i\big(\mathbf{x}^k\big)&:\overline{V}_i<(1-\epsilon_i)V_i\big(\mathbf{x}^k\big) \mbox{,}\\
	\overline{V}_i-V_i \big(\mathbf{x}^k\big)&:\overline{V}_i\in[(1-\epsilon_i) V_i \big(\mathbf{x}^k\big), (1+\epsilon_i)V_i \big(\mathbf{x}^k\big)] \mbox{,}\\
	\epsilon_i V_i\big(\mathbf{x}^k\big)&:\overline{V}_i>(1+\epsilon_i)V_i\big(\mathbf{x}^k\big) \mbox{,}
	\end{cases}
\end{equation}
where $\epsilon_i$ is the relaxation parameter corresponding to each constraint $V_i$. Although the TOBS formulation is herein described for volume constraints, any other differentiable function can be used as constraint.

\textcolor{black}{Integer Linear Programming (ILP) can be used to solve the optimization subproblems described by Eq. \ref{eq:linopt}. The integer programming approach should be a natural choice for topology optimization as one can restrict the solutions to be $\{0,1\}$. Herein, the branch-and-bound method implemented in CPLEX is employed as the ILP solver \citep{Vanderbei14}.}

\section{Numerical implementation}

\subsection{Details of FEA software setup}

The fluid-structure governing equations are solved via a \textcolor{black}{segregated approach and} with separate domains, using the commercial FEA software COMSOL Multiphysics. The RANS equations including the $k-\varepsilon$ turbulence model are solved with the standard wall functions in the software. The Fluid-structure Interaction module in COMSOL Multiphysics automatically identifies the fluid-structure boundaries to impose the equilibrium conditions, e.g, wall functions. To prescribe the one-way coupled problem, the coupling condition is set to be \texttt{FluidLoadingOnly}. The sensitivity analysis required by the optimization is carried out with the \textcolor{black}{semi-automatic} differentiation module from COMSOL Multiphysics. The material models from Eqs. \ref{eq:material_model} and \ref{eq:material_model2} are prescribed by editing the material properties and stress coupling condition, respectively.

\subsection{TOBS with geometry trimming (GT)}

\begin{figure}[ht]
\centering
\includegraphics[scale = 0.35]{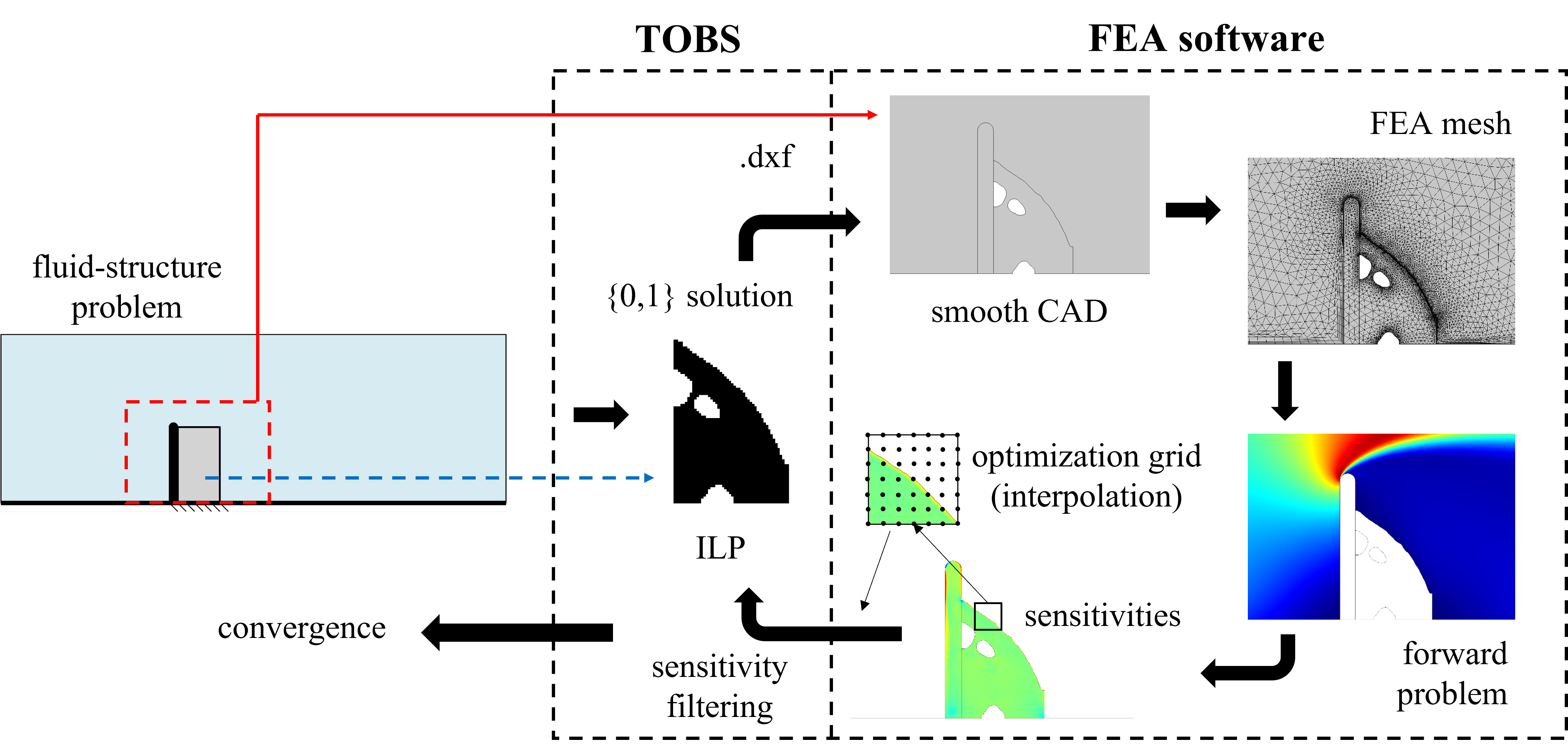}
\caption{Illustration of the optimization methodology. The key idea is the \textcolor{black}{separation} of optimization and analysis grids.}
\label{fig:tobs_fsi}
\end{figure}

The proposed approach is a material distribution method based on the \textcolor{black}{separation of} optimization grid and FEA mesh. The fluid-structure interaction and sensitivity analyses are carried out using COMSOL Multiphysics. An optimization grid is defined as design domain and a binary topology (usually fully solid in the beginning and solid-void during optimization) is assigned. The contour information of existent void regions (holes) is extracted from the set of binary design variables. Initially, this contour defines a design with jagged boundaries, as illustrated in Fig. \ref{fig:tobs_fsi}. Then, these boundaries are smoothed out by a Savitzky–Golay filter \citep{Savitzky64}, when in 2D, or by a shrink wrap tool, when in 3D, and saved as \texttt{.dxf} or \texttt{.stl}, respectively. In the FEA software, the void regions are trimmed out from the design domain to produce a smooth CAD (computer-aided design) geometry. The extracted contours also carry the information that indicates in which edges the holes are located or whether they are completely inside the design domain. Holes that are located at the initial fluid-structure interfaces must allow the fluid domain to exist at those regions. This process replaces the ``fluid flooding'' technique used in other available methodologies \citep{Chen01}. 

Contour extraction favours the combination of the material distribution optimization method with CAD/CAE softwares, specially when dealing with multiphysics problems. Geometry trimming interprets the topology under optimization and generates a CAD model that can be meshed accordingly. In this work, the geometry is freely meshed using the option \texttt{physics controlled} in COMSOL Multiphysics. This procedure should provide good quality approximation of the problem (also controled by FEA convergence within the solver options, if needed), which is advised when simulating fluid-structure interaction systems. Furthermore, different types of elements are also used, as quadrilateral elements are employed near fluid walls (boundary layers) and triangular elements are used in the remaining of the domain. This is the key point in \textcolor{black}{separating the} optimization variables and the analysis mesh. With that, the optimization grid can be refined in order to obtain \textcolor{black}{crisper} topologies while the FEA mesh can be maintained in a certain size with reasonable computational cost. Figure \ref{fig:tobs_fsi} illustrates the proposed procedure of reading the optimization grid, creating CAD geometries and carrying out the FEA.

\subsection{Sensitivity computation}

The forward problem is computed with a stationary study solver, which can include the sensitivity module. The mean compliance objective function is defined in COMSOL Multiphysics by calling \texttt{solid.Ws}\underline{ }\texttt{tot}. The sensitivities with respect to the material model can be exported via \texttt{fsens(dtopo1.theta}\underline{ }\texttt{c)/dvol}. In COMSOL Multiphysics, the variable \texttt{theta}\underline{ }\texttt{c} is used to represent the vector of interpolation variables as expressed in Eqs. \ref{eq:material_model} and \ref{eq:material_model2} by $x_j$. A set of grid points coincident with the optimization grid can be created in order to extract the computed sensitivities. Sensitivities for points in the void or fluid regions are set as zero. Back to the optimization module, standard spatial filtering should be used in the sensitivity field to avoid the well known checkerboard problem and to smooth out the problem, populating the void regions with sensitivities.

\subsection{Algorithm}

In summary, the algorithm of the proposed method is the following:
\begin{enumerate}
    \item Define the optimization parameters.
    \item Create the optimization grid and assign an initial $\{0,1\}$ topology.
    \item Extract the topology contour, smooth the boundaries and save holes as CAD files.
    \item Define the fluid-structure interaction problem in CAD and create the initial geometry.
    \item Trim the holes out of the geometry and create the smooth fluid-structure topology in CAD.
    \item Mesh the trimmed geometry.
    \item Solve fluid-structure interaction equations (\textcolor{black}{including the turbulence model}).
    \item Extract sensitivities in a grid coincident with the optimization grid.
    \item Apply spatial filtering on the sensitivity field considering the optimization grid position.
    \item Solve the linearized optimization subproblem from Eq. \ref{eq:linopt} with the branch-and-bound algorithm.
    \item Update design variables to build a new $\{0,1\}$ topology.
    \item If converged, stop. If not, iterate from step 3.
\end{enumerate}

In this work, steps from 4 to 8 are \textcolor{black}{carried} out in COMSOL Multiphysics, while the others are done in MATLAB using the TOBS implementation available at \url{www.github.com/renatopicelli/tobs}. 

\section{Numerical examples}
\label{sec:results}

This section presents numerical results for \textcolor{black}{mean compliance} minimization subject to volume constraints for applications on FSI problems with turbulence. The structure is considered to have Young's modulus $E$ = 2$\cdot$10$^5$ Pa and Poisson's ratio $\nu$ = 0.3. The fluid is chosen to be water ($\rho_f$ = 1000 kg/m$^3$, $\mu$ = 0.001 Pa$\cdot$s) or air ($\rho_f$ = 1.184 kg/m$^3$, $\mu$ = 1.85$\cdot10^{-5}$ Pa$\cdot$s). The turbulence model used is the $k-\varepsilon$ with standard wall functions. The optimization is considered to be converged after evaluating the change of the objective function over 6 consecutive iterations under a tolerance of $0.001$.

\subsection{The wall}
\label{sec:wall}

\begin{figure}[ht]
\centering
\includegraphics[scale=1]{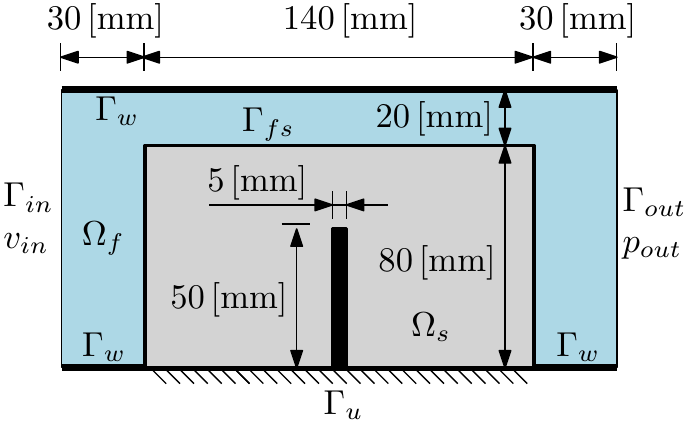}
\caption{The wall example. A $5\times50$ mm solid wall (non-design domain in black region) must be supported by a structure to be designed in the $140\times80$ mm area.}
\label{fig:the_wall_model}
\end{figure}

This first example considers a linearly elastic wall of $5\times50$ mm inside a $300\times100$ mm fluid flow channel, as illustrated in Fig. \ref{fig:the_wall_model}. The fluid is considered to be water. A region of $140\times80$ mm around the wall is considered as design domain where a structural support must comply with the fluid flow loads whilst it holds the solid wall. Low Reynolds and turbulent regimes are studied in this example. The fluid flow enters the channel \textcolor{black}{at the left boundary} with a parabolic velocity profile $\mathbf{v} = 6 v_{in} \frac{(H-Y) Y}{H^2}$, when in the low Reynolds regime, being $Y$ the vertical coordinates, or a constant velocity profile $\mathbf{v} = v_{in}$, when in the turbulent regime. The velocity $v_{in}$ is defined by choosing the Reynolds number $Re$ with respect to the inlet size $H$ and it is expressed as
\begin{equation}
    v_{in} = \frac{Re \cdot \mu}{\rho_f \cdot H} \mbox{.}
\end{equation}
Stress free condition (with $p_{out} = 0$) is imposed at the fluid right \textcolor{black}{boundary} outlet. The bottom \textcolor{black}{boundary} of the structure is clamped, i.e., $\mathbf{u}$ = 0.

\begin{figure}[ht]
\centering
\begin{tabular}{c}
\includegraphics[scale=0.27]{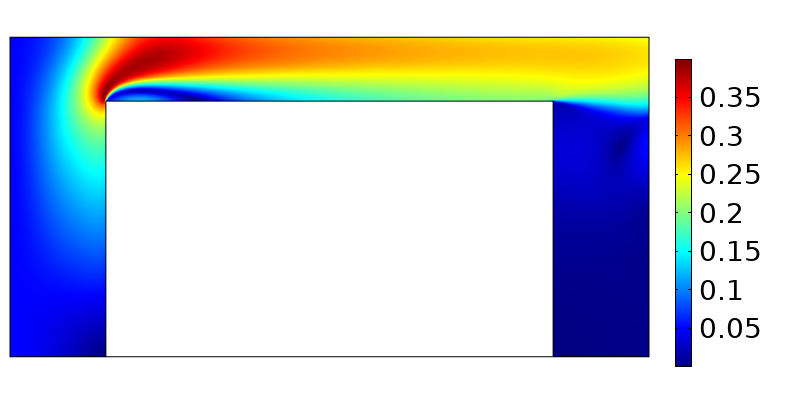} \\ (a) velocity (in m/s) \\ \includegraphics[scale=0.27]{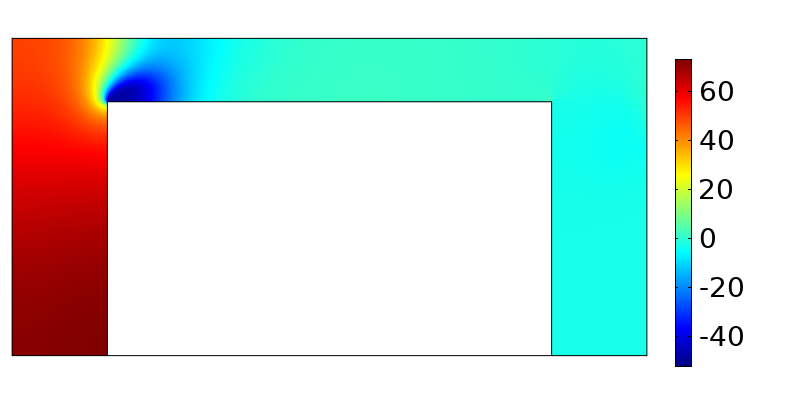} \\ \\ (b) pressure (in Pa) \\
\end{tabular}
\caption{\textcolor{black}{Velocity (a) and pressure (b) fields for the initial fluid flow domain for Re = 5,000.}}
\label{fig:the_wall_initial}
\end{figure}

\begin{figure}[ht]
\centering
\begin{tabular}{cc}
\includegraphics[scale = 0.27]{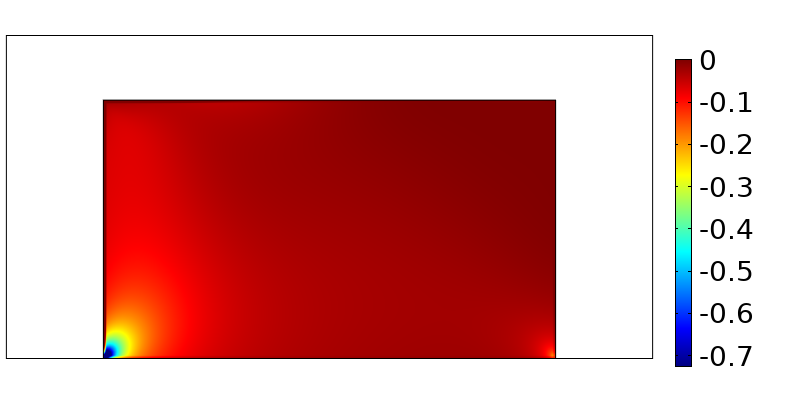} & \includegraphics[scale = 0.27]{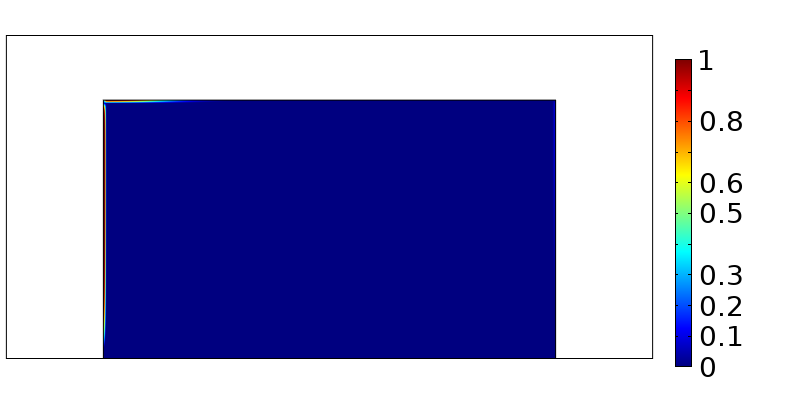} \\
(a) negative portion of the sensitivities (in 1/m$^2$) & (b) positive portion of the sensitivities (in 1/m$^2$) \\
\multicolumn{2}{c}{\includegraphics[scale = 0.27]{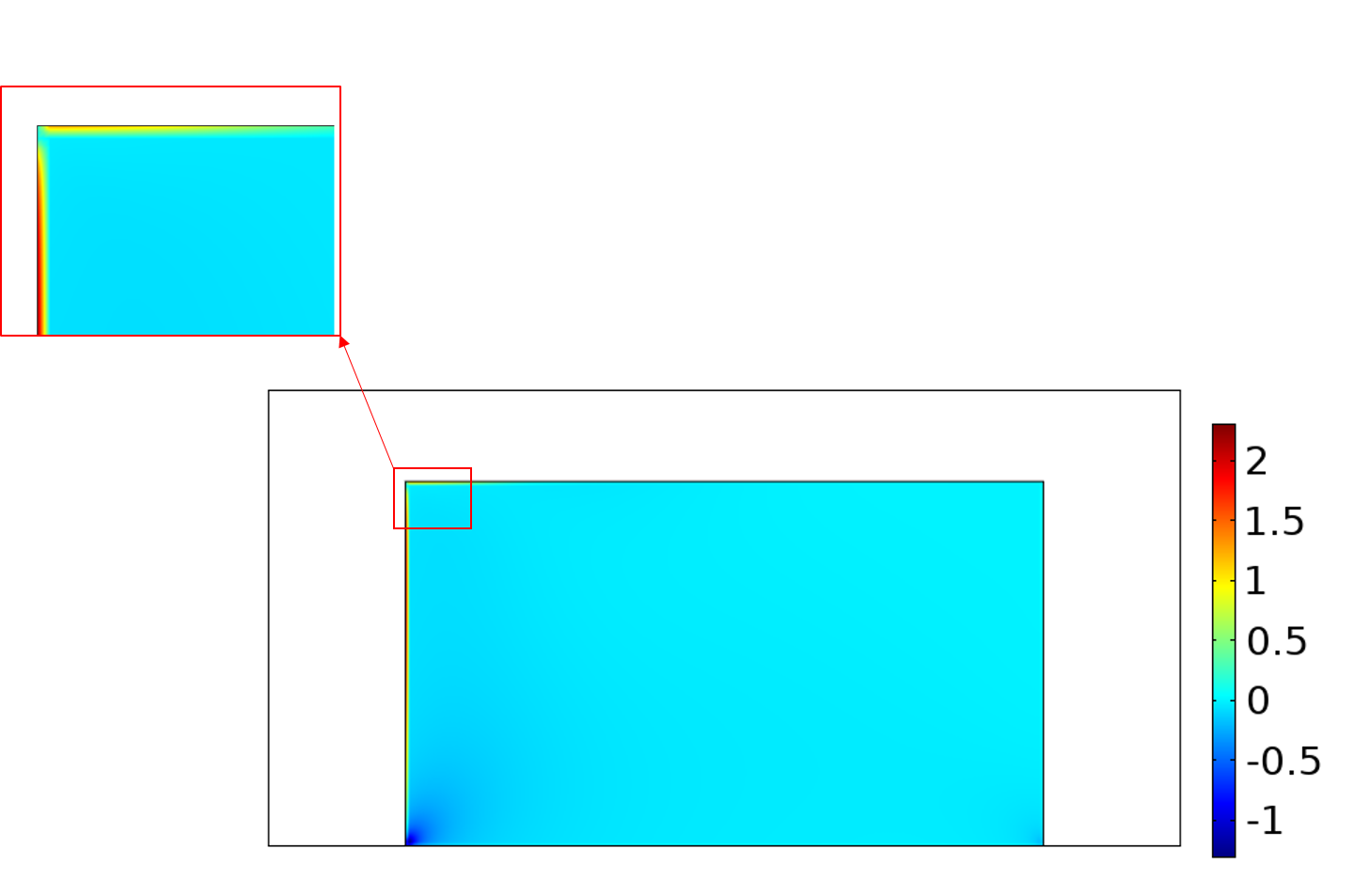}} \\
 \multicolumn{2}{c}{(c) complete sensitivity field with zoomed detail (in 1/m$^2$)} \\
\end{tabular}
\caption{\textcolor{black}{Sensitivity field for the wall example with an initial full solid design for Re = 5,000 and $p = 5$: (a) negative portion of the sensitivity field (in 1/m$^2$) clipped at an upper bound of 0, (b) positive portion of the sensitivity field (in 1/m$^2$) clipped at an upper bound of 1 and (c) complete sensitivity field (in 1/m$^2$) with a zoomed detail.}}
\label{fig:the_wall_initial_sens}
\end{figure}

\begin{figure}[ht]
\centering
\begin{tabular}{cc}
\includegraphics[scale=0.27]{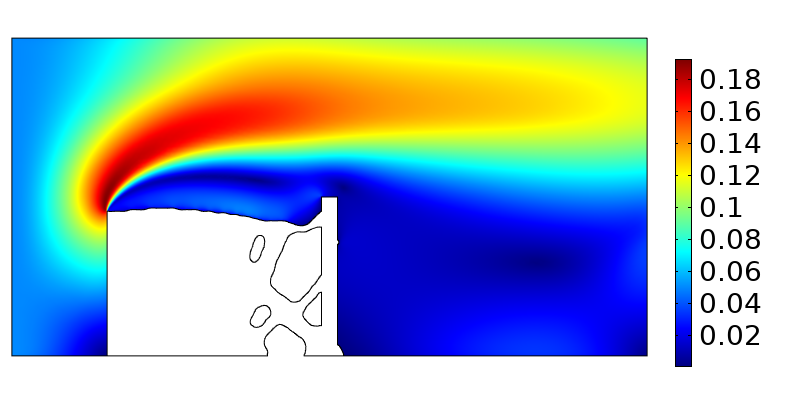} & \includegraphics[scale=0.27]{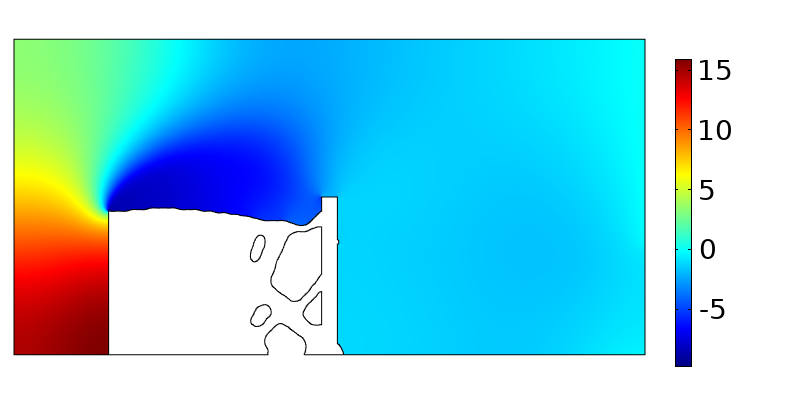} \\
\textcolor{black}{(a) velocity (in m/s)} & \textcolor{black}{(b) pressure (in Pa)} \\
\includegraphics[scale=0.27]{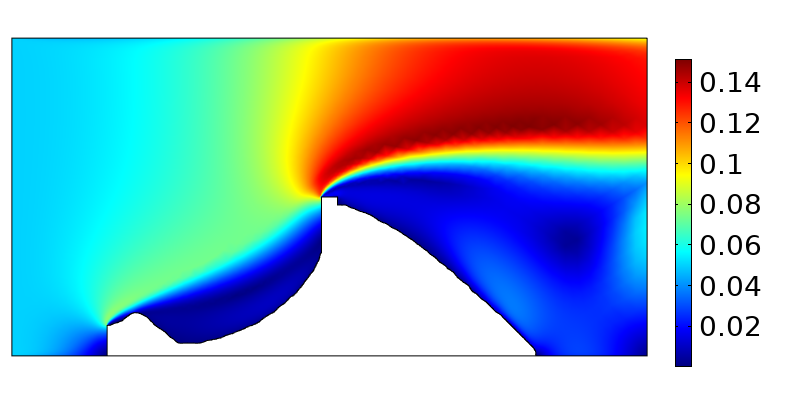} & \includegraphics[scale=0.27]{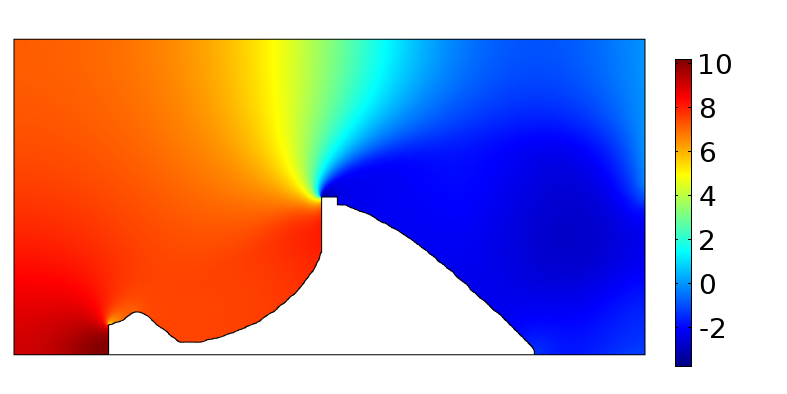} \\
\textcolor{black}{(c) velocity (in m/s)} & \textcolor{black}{(d) pressure (in Pa)} \\
\end{tabular}
\caption{\textcolor{black}{The solution for the wall example with $\bar{V}$ = 25$\%$, Re = 5,000 and $p = 5$ when (a-b) excluding only the material model from the FSI coupling condition given by Eq. \ref{eq:material_model2}, and (c-d) when using the complete sensitivity field (using both Eqs. \ref{eq:material_model} and \ref{eq:material_model2}).}}
\label{fig:the_wall_sens_no_sens}
\end{figure}

The wall example has been studied in different configurations, e.g., with the design domain around the solid wall \citep{Yoon10,Picelli17,Lundgaard18} or the design domain behind the wall \citep{Jenkins16,Picelli20,Li22}. The wall illustrated in Fig. \ref{fig:the_wall_model} was proposed by \cite{Lundgaard18} and not studied in any other works so far. \textcolor{black}{Figure \ref{fig:the_wall_initial} presents the velocity and pressure fields computed in the initial fluid flow domain for Re = 5,000. This example is the most challenging version of ``the wall'' problem as the boundaries (especially the left one) of the structural design domain are} under high pressure and shear loading. In this case, the \textcolor{black}{material model introduced} in Eq. \ref{eq:material_model2} is essential to indicate to the optimizer \textcolor{black}{which structural} region should be \textcolor{black}{modified} in order to decrease the loading on the structure. This information is \textcolor{black}{passed through the sensitivity field as shown in Fig. \ref{fig:the_wall_initial_sens}. The positive portion of the sensitivities at the structural boundaries (shown clipped in Fig. \ref{fig:the_wall_initial_sens}b)} indicates the regions to be removed in order to decrease the loading, and the \textcolor{black}{negative portion of the} sensitivities inside the structure (shown clipped in \textcolor{black}{Fig. \ref{fig:the_wall_initial_sens}a)} guides the optimizer to maximize stiffness. \textcolor{black}{This becomes more evident when solving the mean compliance minimization problem subject to a volume fraction constraint of $\bar{V}$ = 25$\%$. Figure \ref{fig:the_wall_sens_no_sens} presents the optimized topology solutions when not considering the material model in the coupling condition and when including it. For this problem, a 280$\times$160 optimization grid was used, a filter radius of 6 distances between optimization points was chosen and the optimization parameters were $\epsilon = 0.01$ and $\beta = 0.02$. The penalty $\gamma$ was set as 5. It can be observed that, without the material model in the coupling condition, the optimizer is not able to remove a large portion of solid material from the regions with high fluid loading (left side of the design domain), leading to a mean compliance value of $C$ = 1.6149$\times10^{-6}$ Nm. When including the material model in the coupling condition (as written in Eq. \ref{eq:material_model2}), i.e. evaluating the sensitivities of the fluid loading, more material is removed from high pressure zones and a more aerodynamic shape is obtained in a structure with $C$ = 5.5723$\times10^{-7}$ Nm, lower than in the other case. To further elaborate on the sensitivity field, Appendix A presents a finite differences check on the \textcolor{black}{semi-automatic} differentiated sensitivities and a study of the other common ``the wall'' problem, with the design domain behind the solid wall.}


\begin{figure}[ht]
\centering
\begin{tabular}{ccc}
  \includegraphics[scale = 0.25, trim={0cm 3.5cm 0cm 3.5cm}, clip]{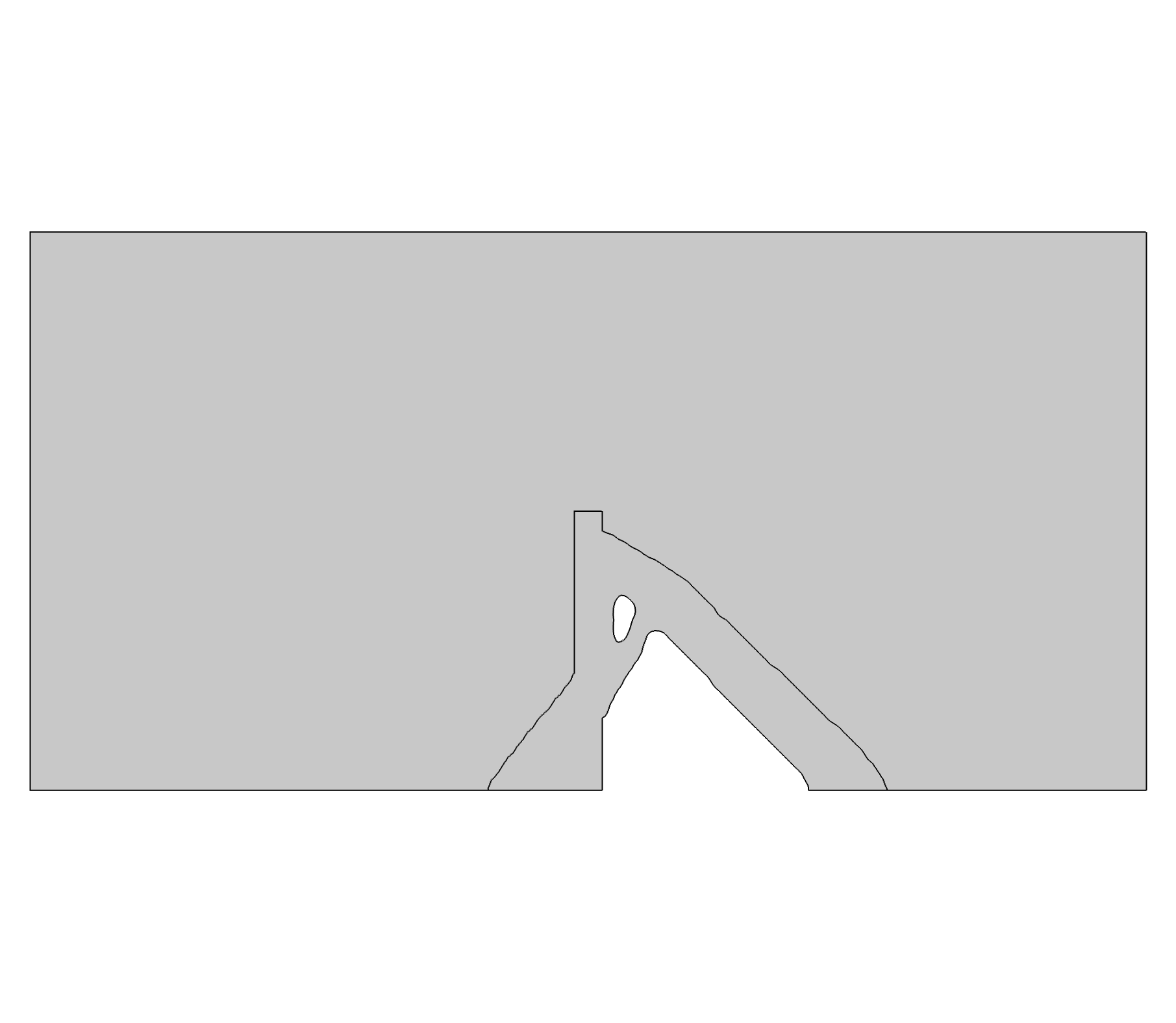} &
  \includegraphics[scale = 0.25, trim={0cm 3.5cm 0cm 3.5cm}, clip]{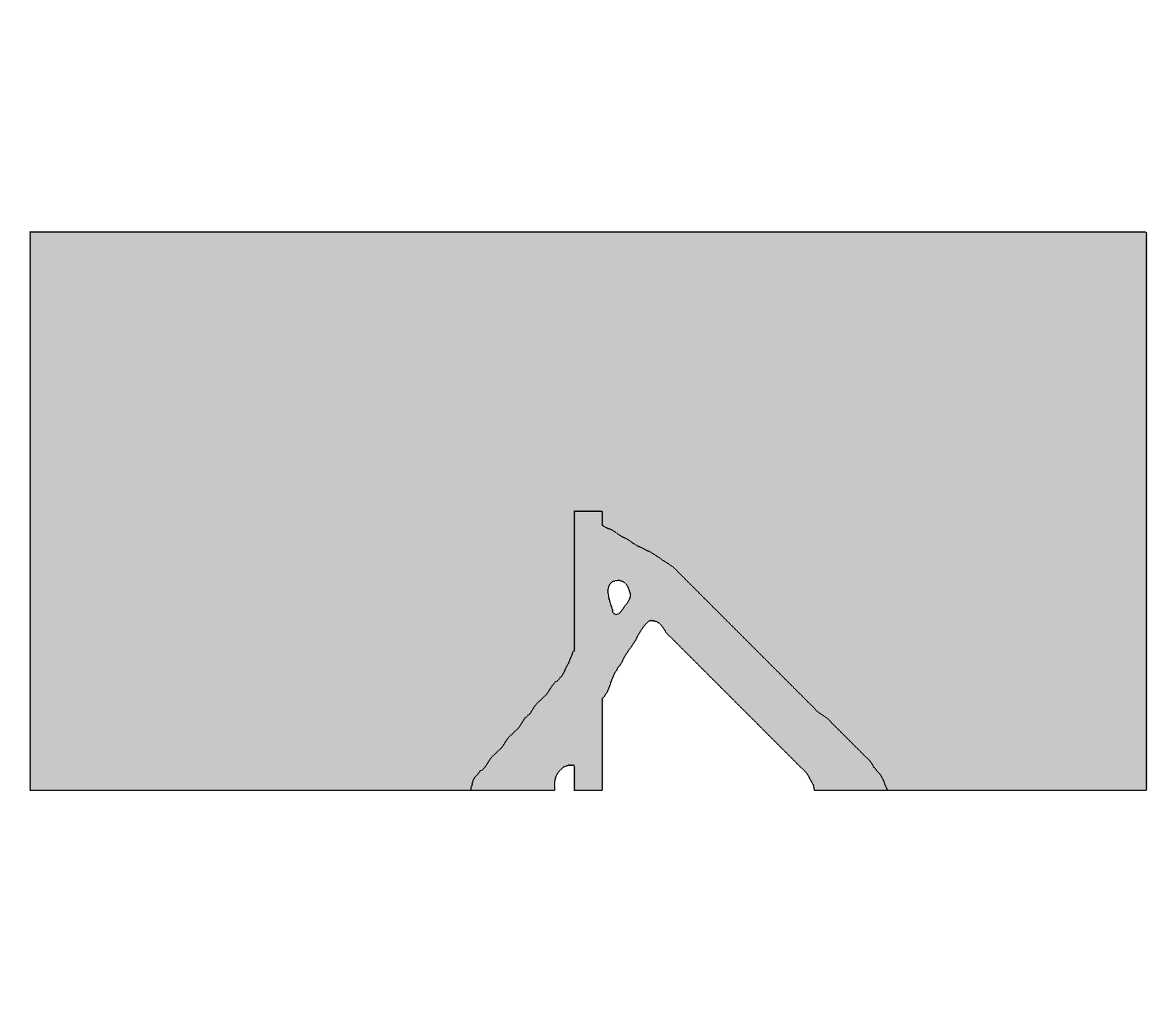} &
  \includegraphics[scale = 0.25, trim={0cm 3.5cm 0cm 3.5cm}, clip]{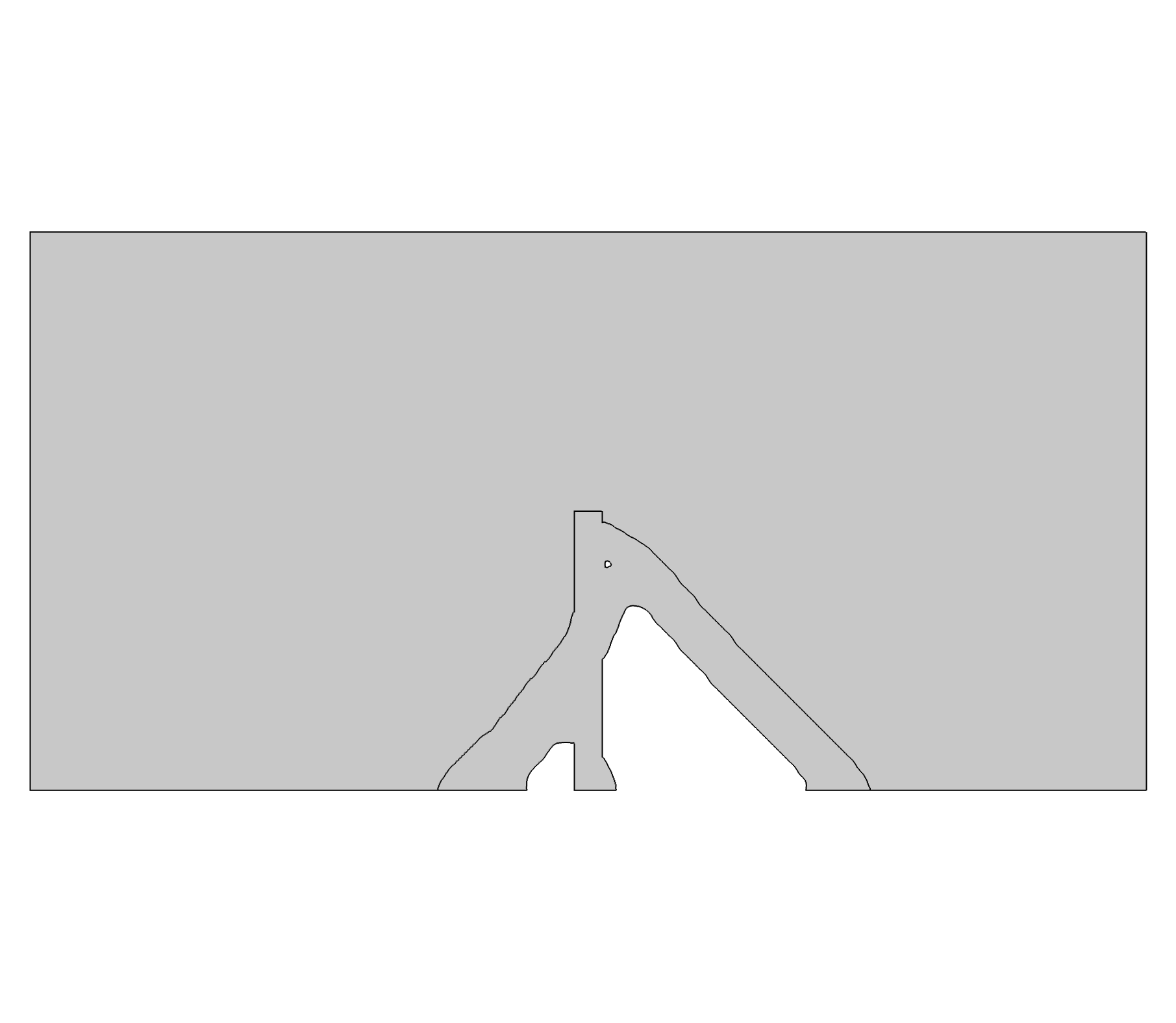} \\
  \textcolor{black}{(a)} Re = 10, $\gamma$ = 3 & \textcolor{black}{(b)} Re = 10, $\gamma$ = 5 & \textcolor{black}{(c)} Re = 10, $\gamma$ = 10 \\
  $C$ = 2.2958$\times10^{-16}$ Nm & $C$ = 2.1280$\times10^{-16}$ Nm & $C$ = 1.9115$\times10^{-16}$ Nm \\
  \includegraphics[scale = 0.25, trim={0cm 3.5cm 0cm 3.5cm}, clip]{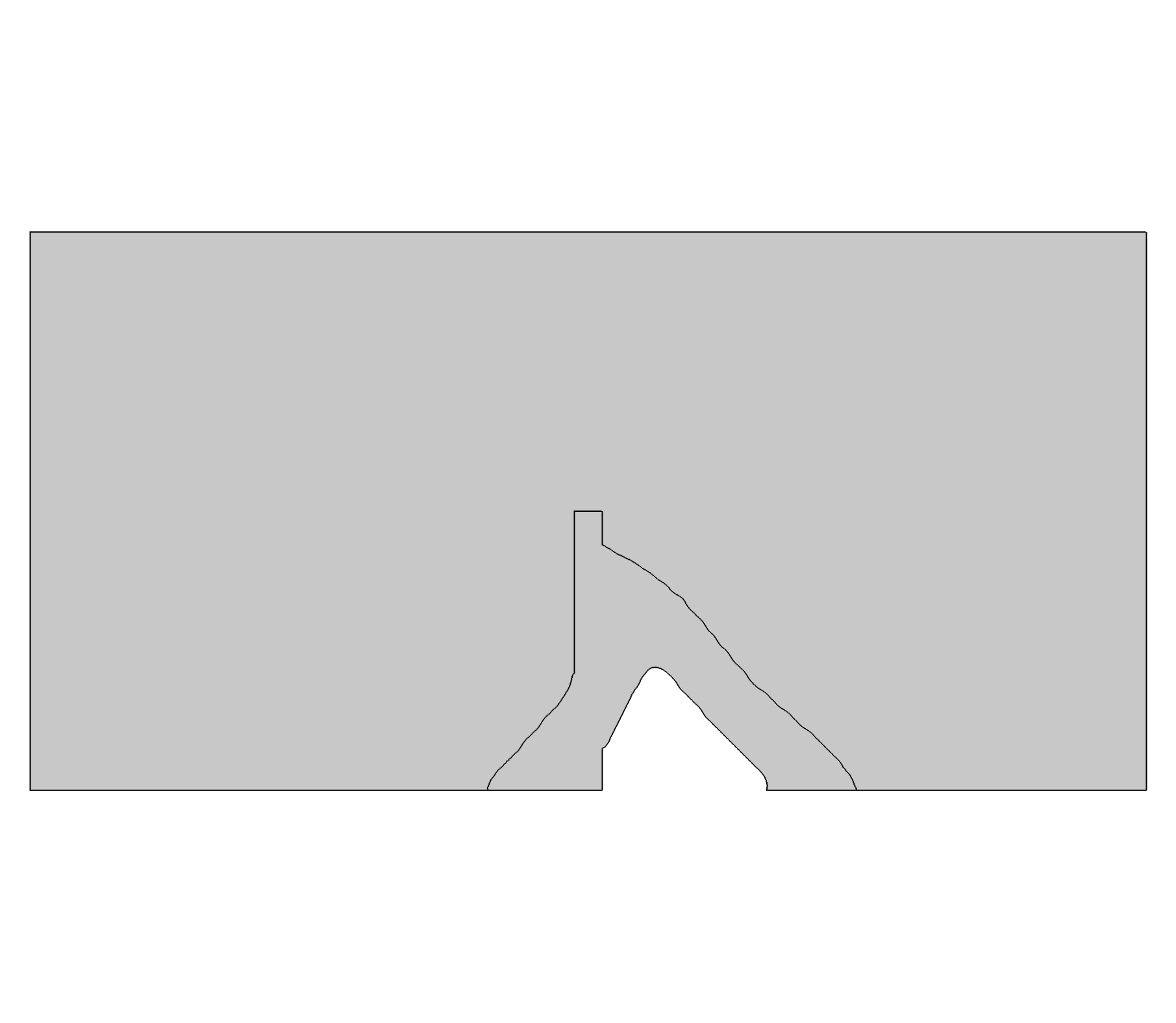} &
  \includegraphics[scale = 0.25, trim={0cm 3.5cm 0cm 3.5cm}, clip]{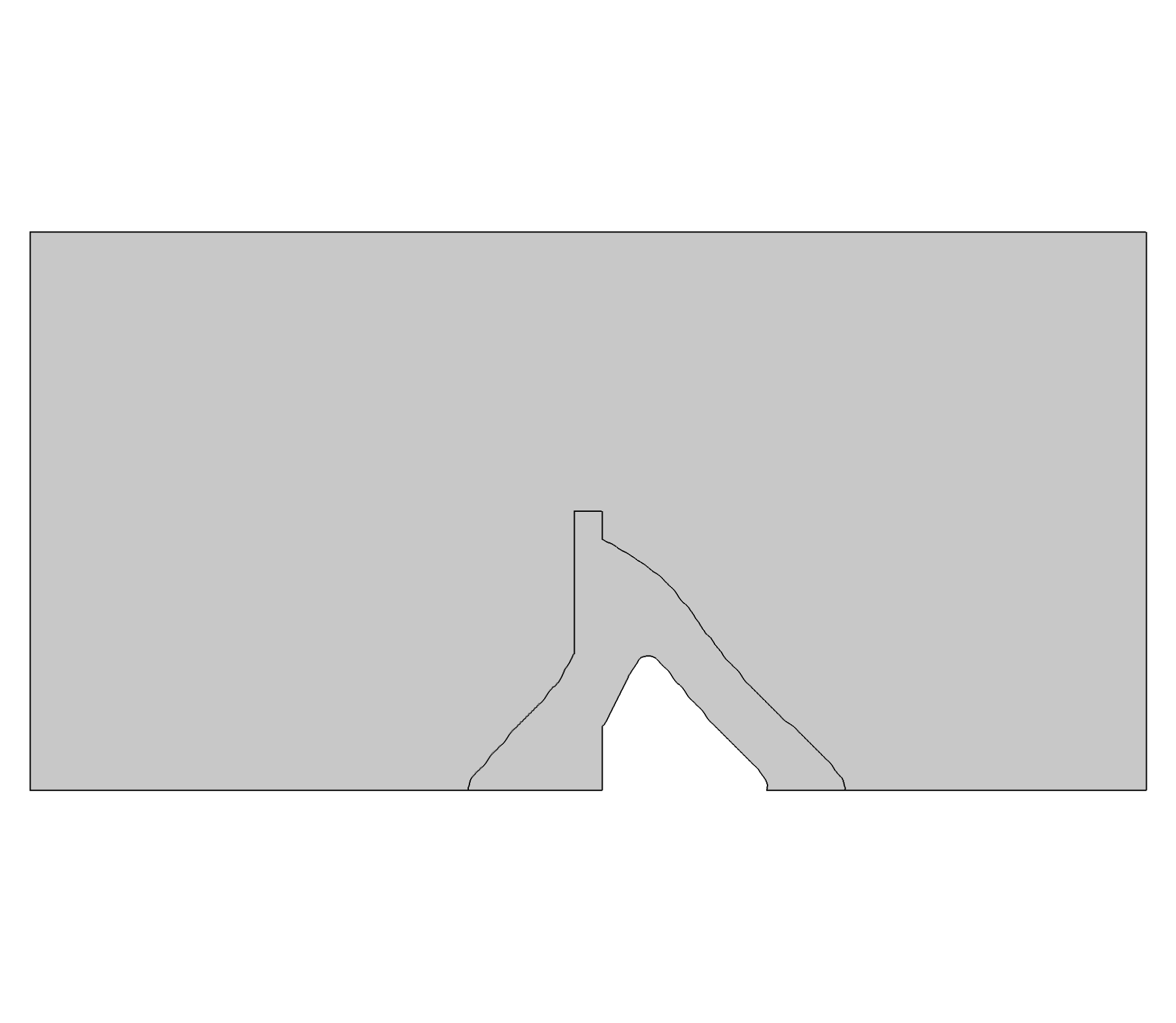} &
  \includegraphics[scale = 0.25, trim={0cm 3.5cm 0cm 3.5cm}, clip]{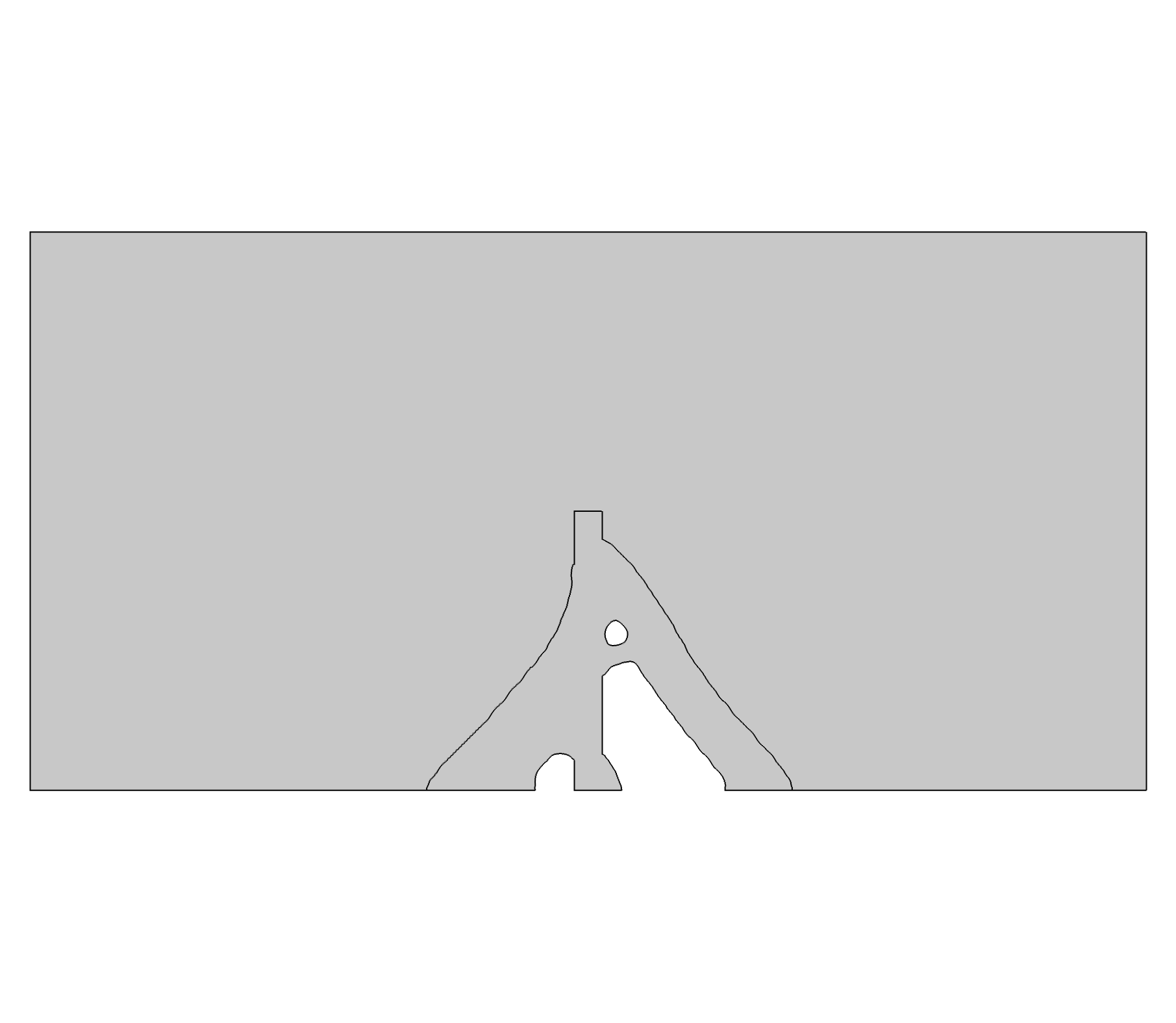} \\
  \textcolor{black}{(d)} Re = 100, $\gamma$ = 3 & \textcolor{black}{(e)} Re = 100, $\gamma$ = 5 & \textcolor{black}{(f)} Re = 100, $\gamma$ = 10 \\
  $C$ = 2.9572$\times10^{-13}$ Nm & $C$ = 2.6696$\times10^{-13}$ Nm & $C$ = 2.3899$\times10^{-13}$ Nm \\
  \includegraphics[scale = 0.25, trim={0cm 3.5cm 0cm 3.5cm}, clip]{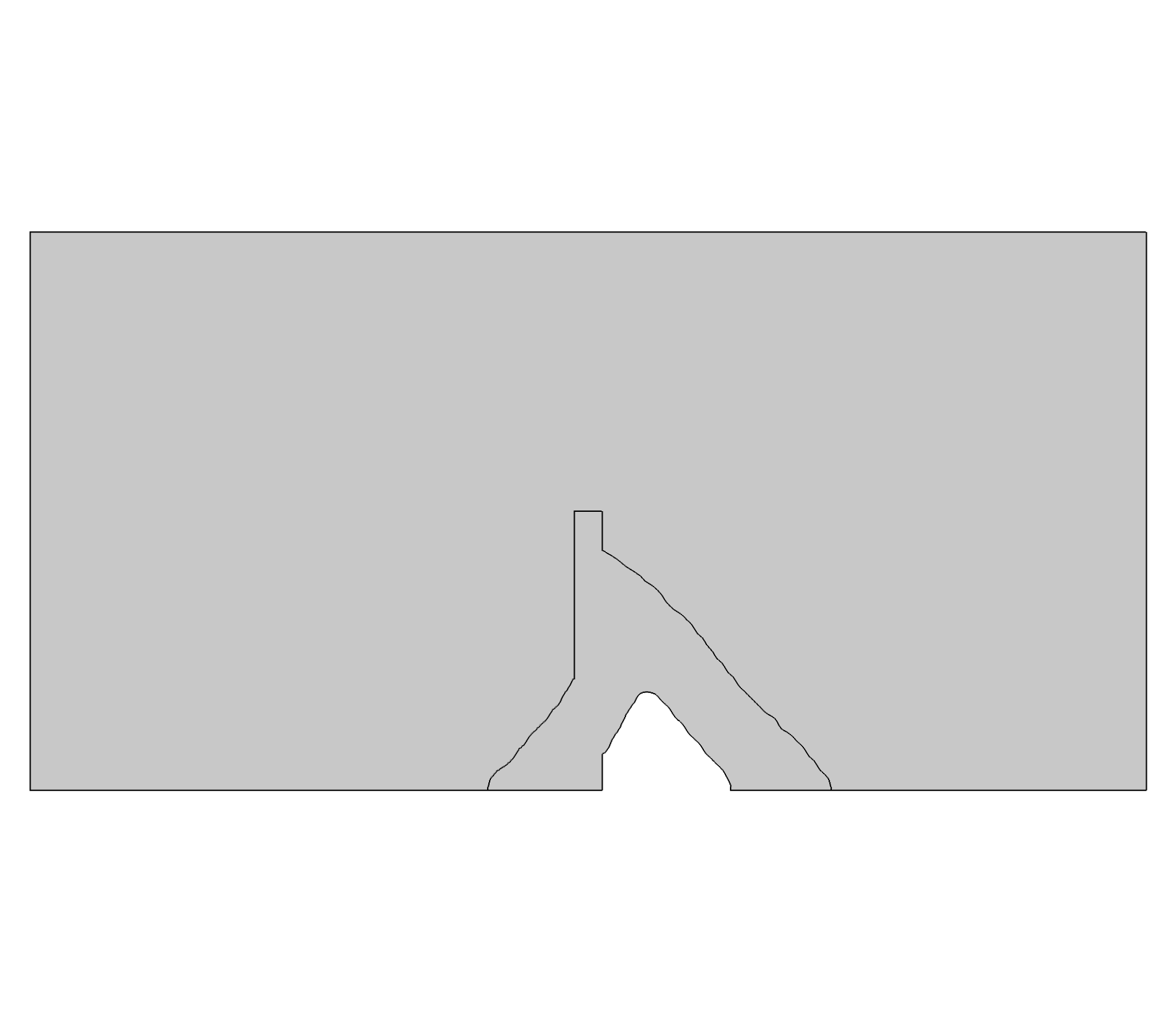} &
  \includegraphics[scale = 0.25, trim={0cm 3.5cm 0cm 3.5cm}, clip]{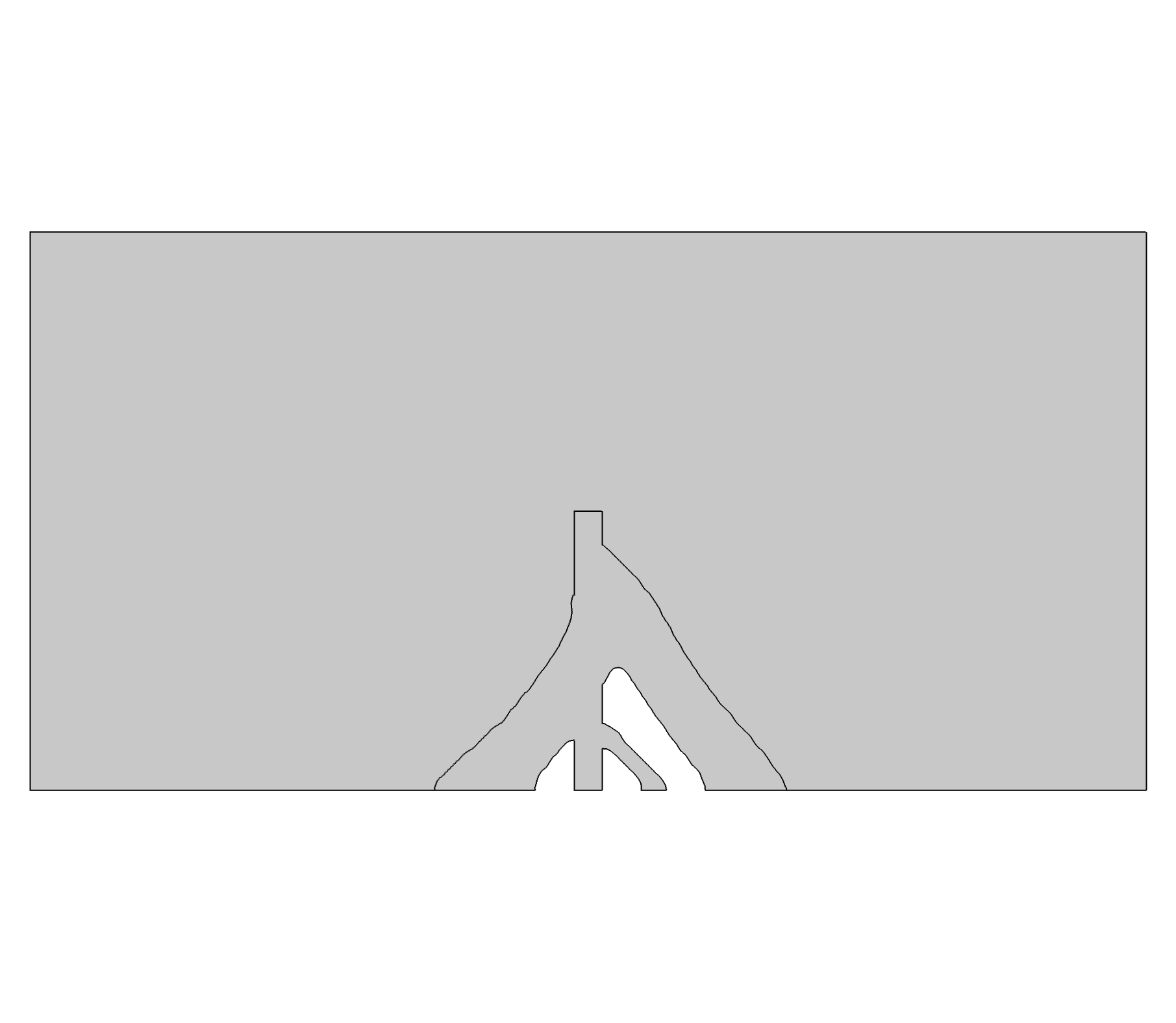} &
  \includegraphics[scale = 0.25, trim={0cm 3.5cm 0cm 3.5cm}, clip]{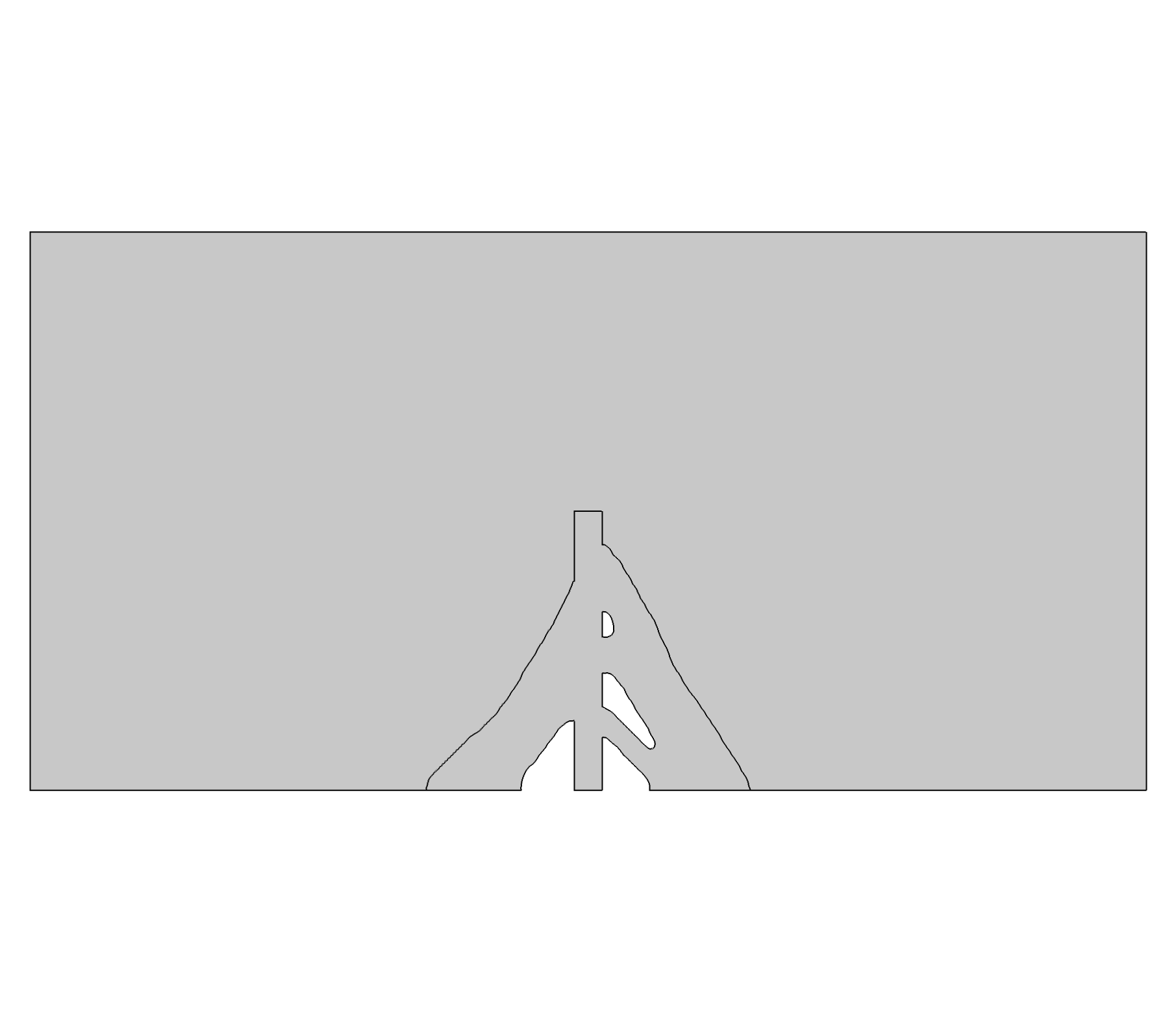} \\
  \textcolor{black}{(g)} Re = 5,000, $\gamma$ = 3 & \textcolor{black}{(h)} Re = 5,000, $\gamma$ = 5 & \textcolor{black}{(i)} Re = 5,000, $\gamma$ = 10 \\
  $C$ = 2.1089$\times10^{-6}$ Nm & $C$ = 1.8365$\times10^{-6}$ Nm & $C$ = 1.7755$\times10^{-6}$ Nm \\
\end{tabular}
 \caption{Topology solutions obtained by the \textcolor{black}{proposed} method for \textcolor{black}{(a-f)} low Reynolds and \textcolor{black}{(g-i)} turbulent flow ($k-\varepsilon$) cases using different $\gamma$'s.}
 \label{fig:the_wall_p}
\end{figure}

\textcolor{black}{To compare with the literature and further investigate the influence of the material models employed, the mean compliance minimization problem is solved subject to a volume fraction constraint of $\bar{V}$ = 10$\%$, as done by \cite{Lundgaard18}.} In this work, the \textcolor{black}{coupling condition (in practice, the FSI loading)} is linearly interpolated with the design variables $x_j$, similarly as in \cite{Yoon10}, and the \textcolor{black}{solid material property depends} on $x_j^\gamma$. Therefore, the optimization solutions depend on the interpolation parameter $\gamma$. \cite{Lundgaard18} considered a general interpolation for the \textcolor{black}{coupling condition} as well, but the authors ended up \textcolor{black}{using the same} linear interpolation. In our numerical experience, we also adopted linear interpolation on the \textcolor{black}{coupling condition} to obtain better convergence. Figure \ref{fig:the_wall_p} presents the \textcolor{black}{optimized designs} for the wall example using different $\gamma$'s for low Reynolds flows (Re = 10 and 100) and for turbulent flow (Re = 5,000).

\begin{figure}[ht]
\centering
\begin{tabular}{ccc}
  \includegraphics[scale = 0.25]{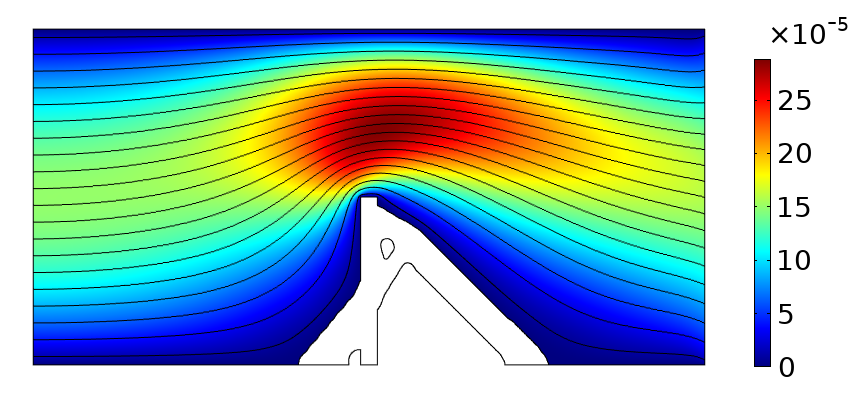} &
  \includegraphics[scale = 0.25]{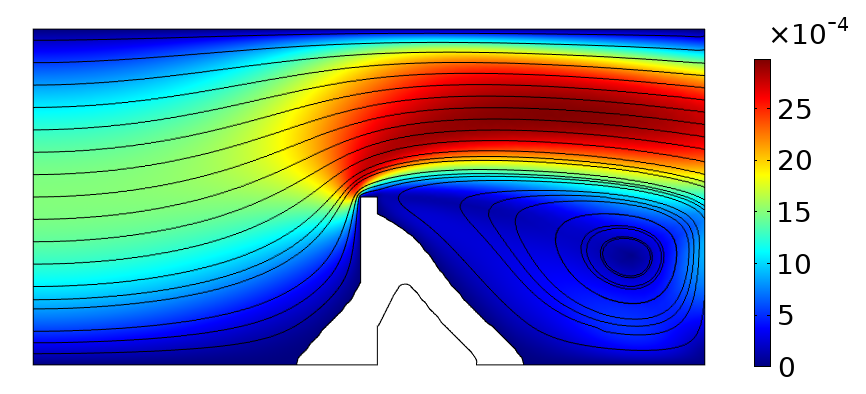} &
  \includegraphics[scale = 0.25]{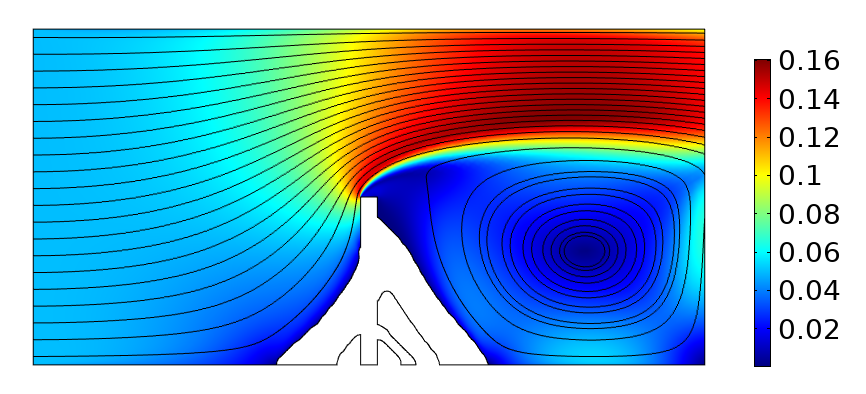} \\
  \textcolor{black}{(a)} Re = 10 & \textcolor{black}{(b)} Re = 100 & \textcolor{black}{(c)} Re = 5,000 \\
\end{tabular}
 \caption{Velocity fields (in m/s) of the wall example solutions by the TOBS-GT method for \textcolor{black}{(a-b)} low Reynolds and \textcolor{black}{(c)} turbulent flow ($k-\varepsilon$) cases with $\gamma$ = 5.}
 \label{fig:the_wall_turb_vel}
\end{figure}

In general, the \textcolor{black}{obtained} solutions for the wall example presented a front structural member positioned towards the bottom of the domain and a rear structure holding the wall \textcolor{black}{closer} to the top of the wall. For lower $\gamma$'s, the optimizer prioritizes the removal of the left boundaries, which have higher fluid pressure and shear loading. In this way, less solid material remains in the left region of the design domain. For higher $\gamma$'s, the distribution of solid material showed to be more balanced. None of these solutions resemble the ones obtained by \cite{Lundgaard18}, which present the tendency of including more material on the left region of the design domain than on the right. Herein, the final compliance values $C$ showed to be lower for higher $\gamma$'s. However, much larger $\gamma$'s will reduce the effects of the interpolation of the coupling condition, leading to a case similar \textcolor{black}{to the one present in Fig. \ref{fig:the_wall_sens_no_sens}(a-b)}, not suitable for high Reynolds flow. Figure \ref{fig:the_wall_turb_vel} presents the velocity fields (in m/s) of the wall example obtained by the \textcolor{black}{proposed} method for the low Reynolds and turbulent flow cases with \textcolor{black}{$\gamma$ = 5}. 

\begin{figure}[ht]
\centering
\begin{tabular}{ccc}
  \includegraphics[scale = 0.28, trim={0cm 0cm 0cm 0cm}, clip]{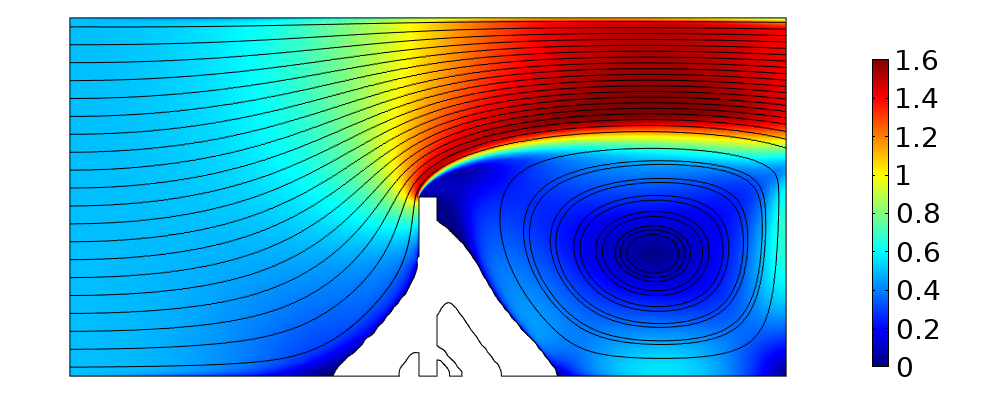} &
  \includegraphics[scale = 0.28, trim={0cm 0cm 0cm 0cm}, clip]{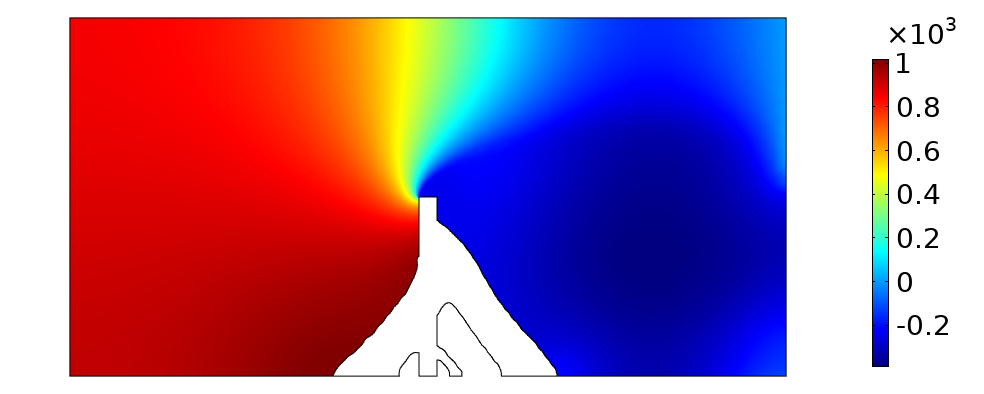} \\ (a) velocity (in m/s) & (b) pressure (in Pa)  \\
  \multicolumn{2}{c}{\includegraphics[scale = 0.28, trim={0cm 0cm 0cm 0cm}, clip]{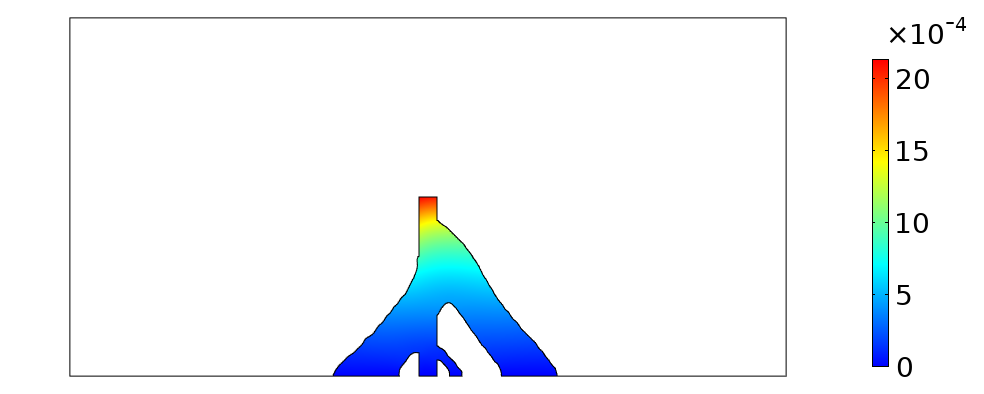}} &  \\
  \multicolumn{2}{c}{(c) \textcolor{black}{displacement (in m)}} &  \\
\end{tabular}
 \caption{Topology optimization solution for the wall example with the $k-\varepsilon$ turbulence model, Re = 50,000 and $\gamma$ = 5.}
 \label{fig:the_wall_turb_vel_Re5e4}
\end{figure}

\begin{figure}[ht]
\centering
\begin{tabular}{cc}
  \includegraphics[scale = 0.4, trim={3cm 8cm 3cm 8cm}, clip]{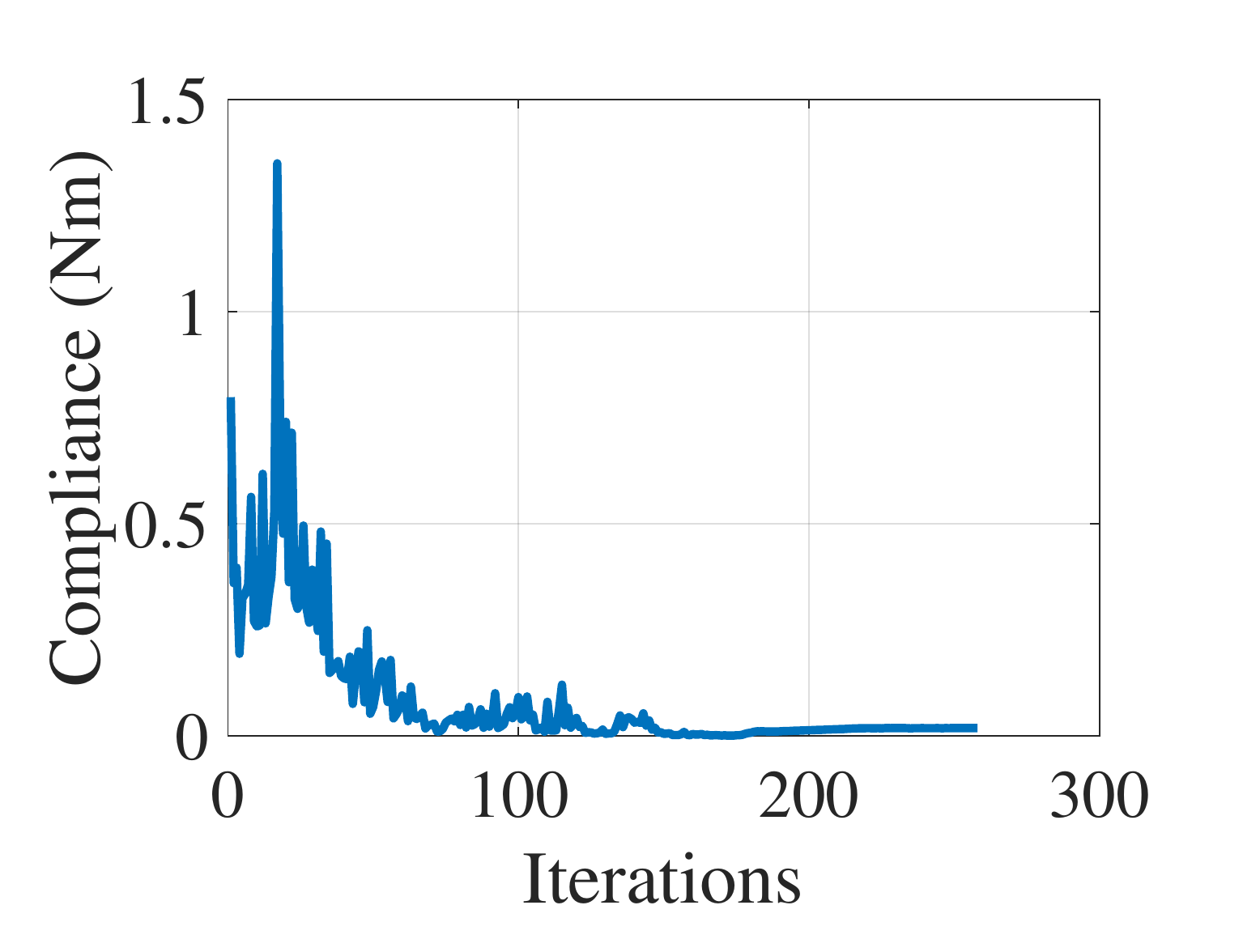} & \includegraphics[scale = 0.4, trim={3cm 8cm 3cm 8cm}, clip]{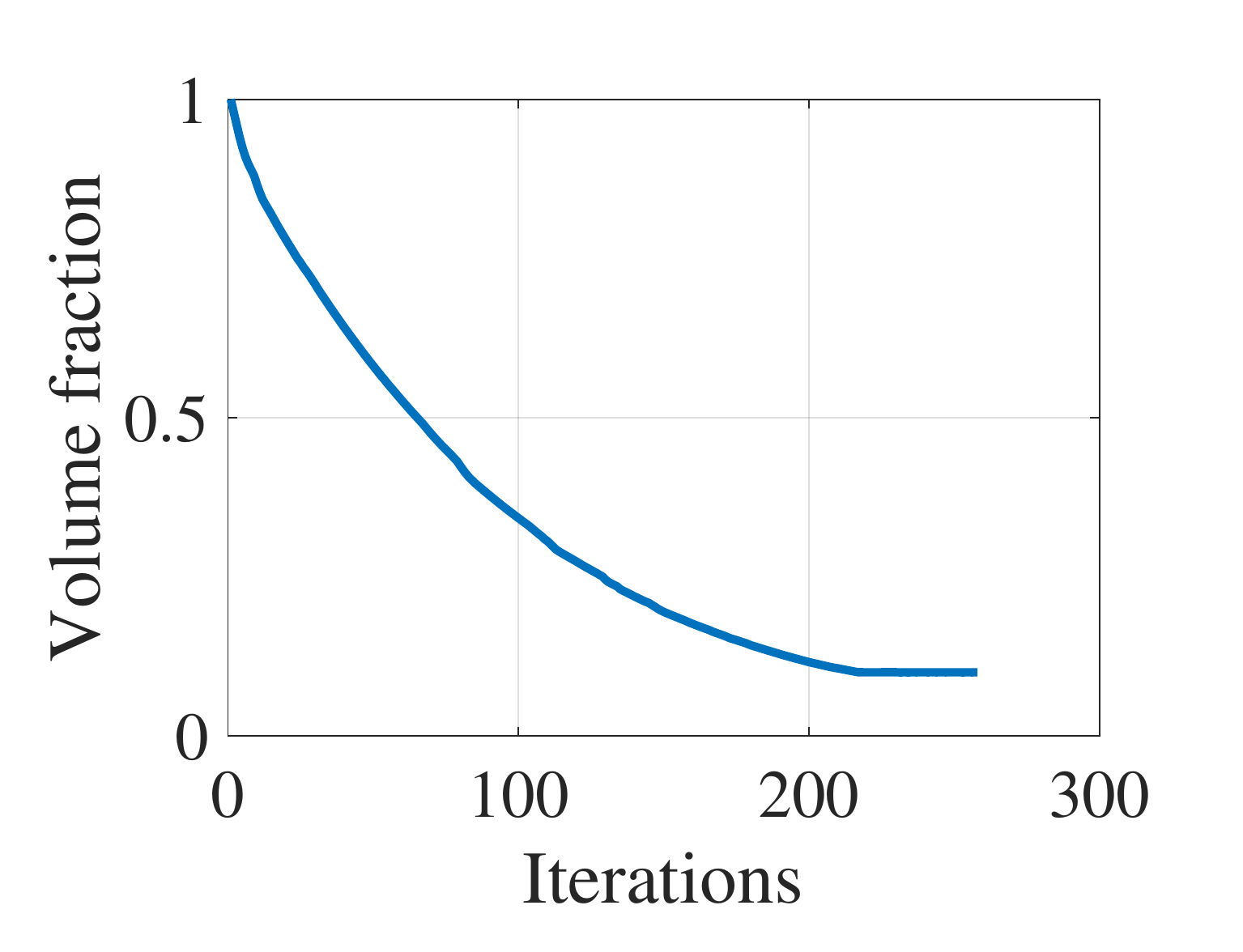} \\
  (a) & (b) \\
\end{tabular}
 \caption{History of the (a) compliance (objective) and (b) volume constraint functions for the case with Re = 50,000 and $\gamma$ = 5.}
 \label{fig:evo}
\end{figure}

\begin{figure}[ht]
\centering
  \includegraphics[scale = 0.75]{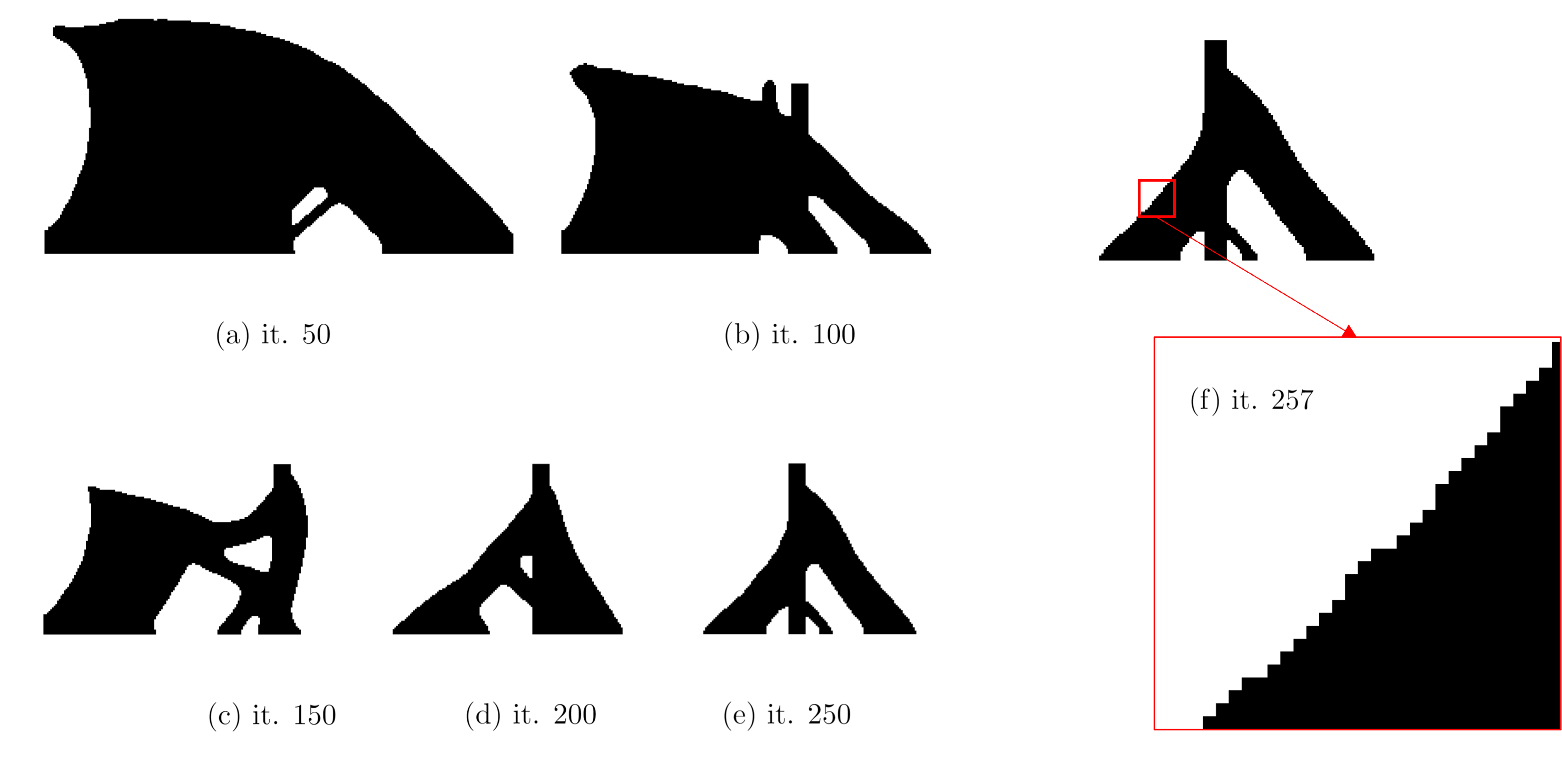} 
 \caption{Snapshots of the set of binary design variables during optimization with zoomed detail at the boundary of the final solution.}
 \label{fig:the_wall_turb_top_Re5e4}
\end{figure}

\begin{figure}[ht]
\centering
\begin{tabular}{cc}
  \includegraphics[scale = 0.5, trim={0cm 0cm 0cm 2cm}, clip]{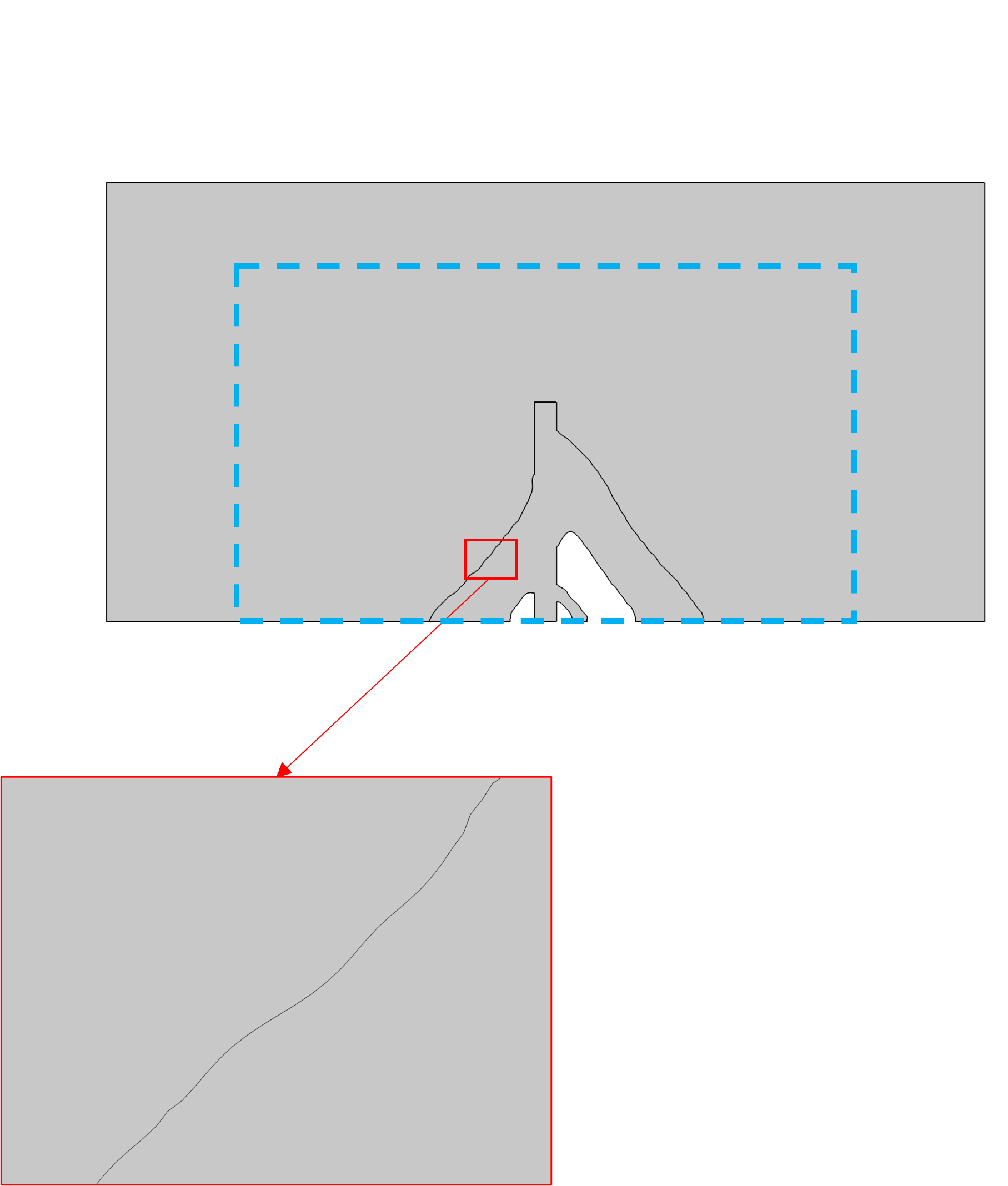} &
  \includegraphics[scale = 0.5, trim={0cm 0cm 0cm 2cm}, clip]{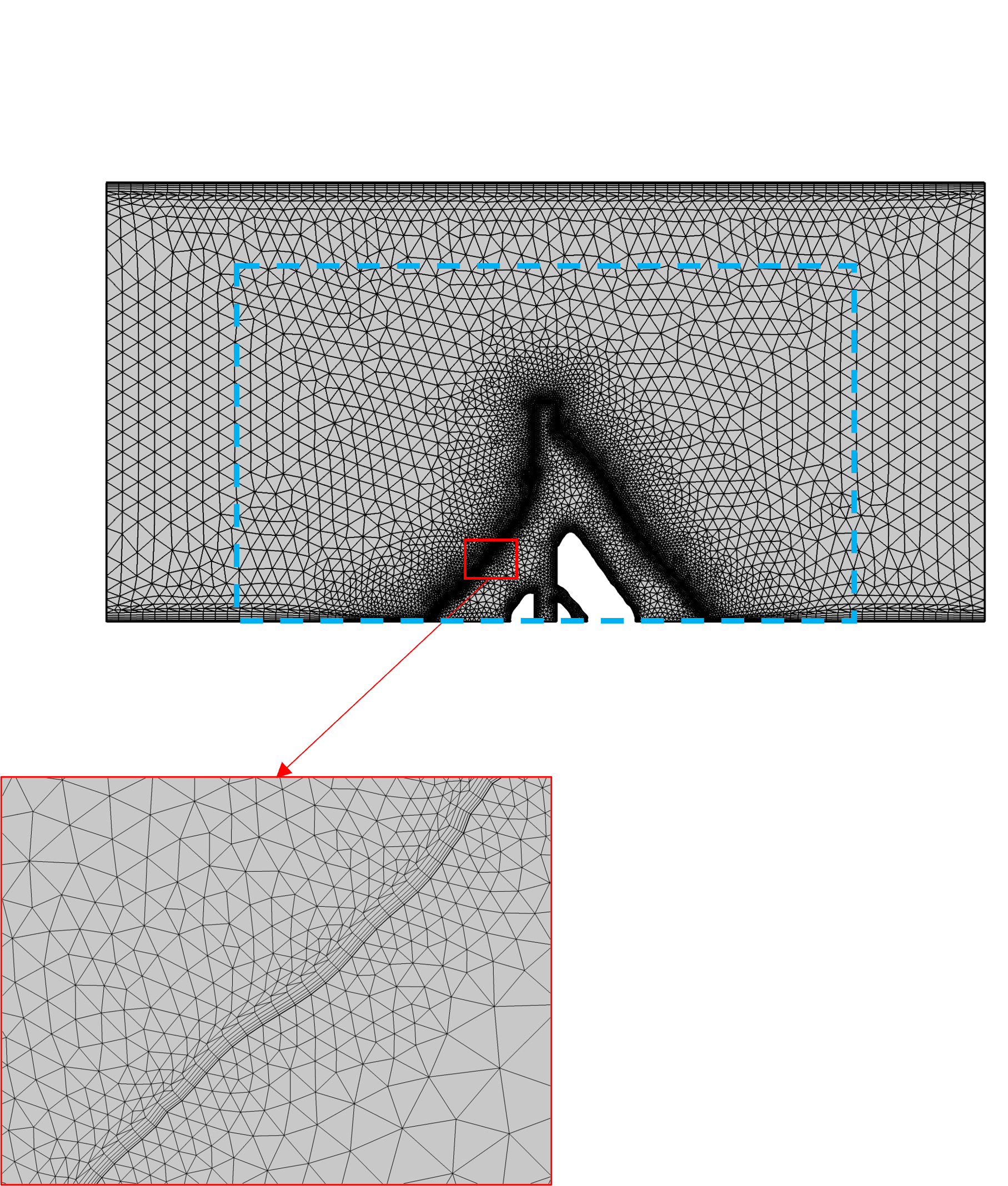}  \\
  (a) & (b) \\
\end{tabular}
 \caption{Zoomed details of the boundary of the optimized solution for Re = 50,000 and $\gamma$ = 5: (a) smooth CAD geometry and (b) finite element mesh. The dashed rectangles in blue indicate the prescribed design domain.}
 \label{fig:the_wall_turb_mesh_Re5e4}
\end{figure}

Theoretically, the \textcolor{black}{proposed} method is able to consider different fluid models or increase the fluid velocities without changing its algorithm. As shown in Fig. \ref{fig:the_wall_turb_vel}, the same method is capable of including low Reynolds and turbulent flows. The solution for more complex FSI design problems would become then a matter of improving the analysis or including more powerful wall smoothing or FEA solvers. In this context, smoothing the walls allowed the modelling of boundary layers and the increase in the velocity field. Figure \ref{fig:the_wall_turb_vel_Re5e4} presents the solution for the wall example under the turbulent regime with Re = 50,000. Plots of the velocity, pressure and \textcolor{black}{displacement} fields are given to illustrate the solution. The structure started from the initial full solid domain and converged in 257 iterations to a final topology with $C$ = 0.0187 Nm (see Fig. \ref{fig:evo} for convergence history). The snapshots of the set of design variables during optimization can be seen in Fig. \ref{fig:the_wall_turb_top_Re5e4}. A zoomed detail in Fig. \ref{fig:the_wall_turb_top_Re5e4}(f) shows the jagged boundaries produced by the binary design variables. The Savitzky–Golay filter \citep{Savitzky64} is used to smooth the jagged contour information. \textcolor{black}{There is not a general rule, but our numerical experience indicates that an optimization grid with a few hundred points per dimension (between 100 and 300) is enough to obtain a satisfactorily smooth boundary after the filter. Coarser grids would lead to rough edges and refined grids would require more data points to be interpolated.} Figure \ref{fig:the_wall_turb_mesh_Re5e4} presents the zoomed details of the smoothed walls obtained, including details of the finite element mesh generated by the FEA software considering the modelling of boundary layers. It can be highlighted that TOBS-GT is as an \textcolor{black}{explicit boundary} methodology that produces smooth enough boundaries but remains in the ``digital'' framework by solving a $\{0,1\}$ material distribution problem. In this case, the \textcolor{black}{method} allows further modelling of the boundaries (e.g., the inclusion of turbulence wall functions) that is not possible in density-based approaches and still remains relatively easier to be solved than \textcolor{black}{methods based on level-sets}.

\begin{figure}
    \centering
    \begin{tabular}{cc}
        \includegraphics[scale = 0.5, trim={3.5cm 8.5cm 3.5cm 8.5cm}, clip]{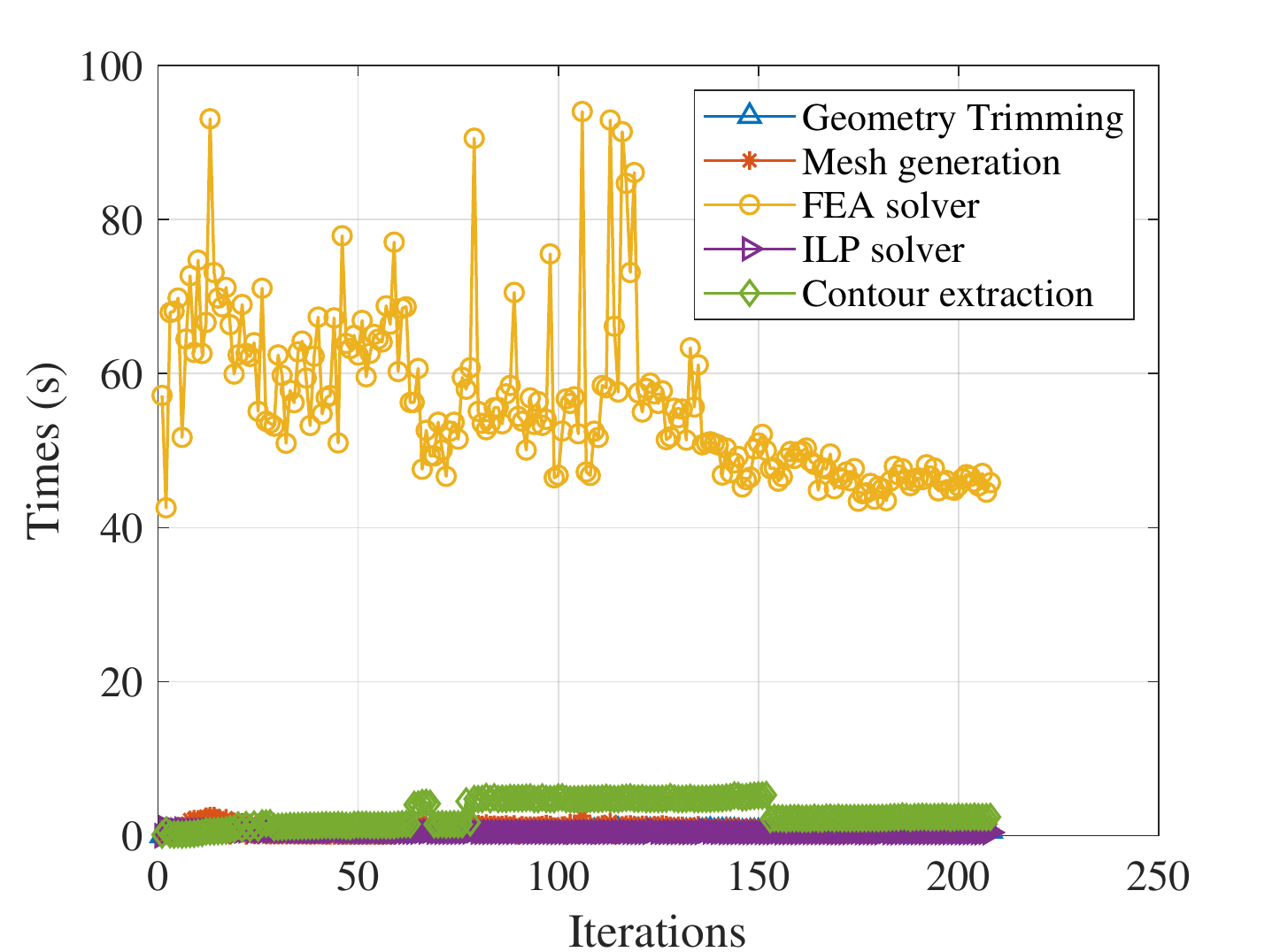} &
        \includegraphics[scale = 0.5, trim={3.5cm 8.5cm 3.5cm 8.5cm}, clip]{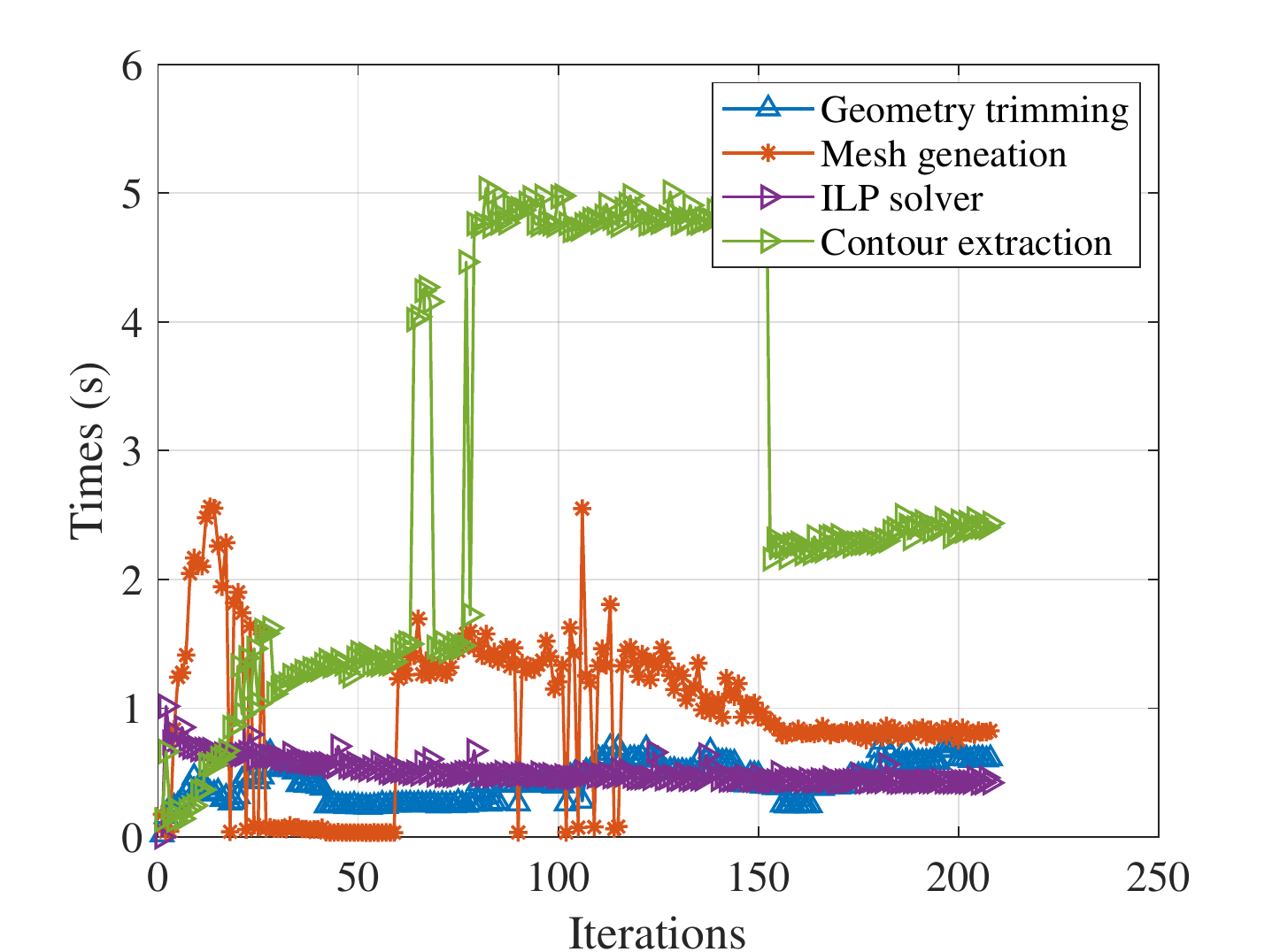} \\
        (a) & (b) \\
    \end{tabular}
    \caption{Breakdown computation times for the wall example optimized with the TOBS-GT method: (a) for all main steps and (b) omitting the FEA solver times. Case with Re = 5,000 and $\gamma$ = 10.}
    \label{fig:the_wall_times}
\end{figure}

Another \textcolor{black}{possible} advantage of the \textcolor{black}{proposed} method is that, by \textcolor{black}{separating} the optimization and analysis grids, some computational efficiency is gained. In this example, while a grid of 280$\times$160 optimization points (total of 44,800) is used, the final fluid-structure system could be modeled with 18,237 triangular and 2,469 quadrilateral elements (mesh shown in Fig. \ref{fig:the_wall_turb_mesh_Re5e4}b). This means that \textcolor{black}{less} finite elements \textcolor{black}{are required} to solve the forward and adjoint problems in this \textcolor{black}{explicit boundary} approach than in fixed grid methods. To further illustrate that, Fig. \ref{fig:the_wall_times} presents the breakdown computation times for the wall example optimized with the \textcolor{black}{proposed} method for Re = 5,000 and $\gamma$ = 5. It can be noticed that the FEA is \textcolor{black}{still} the computational bottleneck of the present algorithm, an expected behavior for topology optimization methods. The FEA could be run between 40 and 100 seconds for the final optimized wall Re = 5,000 and $\gamma$ = 5, time summed up for both forward and adjoint problems. The contour extraction (including smoothing) took up to 5 seconds to be carried out, depending on the complexity of the structure during optimization. Another important point is that the integer linear programming \textcolor{black}{solver} required less than a second of computation. This illustrates that the proposed TOBS-GT method can be a relatively cheap \textcolor{black}{design} method for highly complex \textcolor{black}{physics} problem.

\subsection{Multiple walls}
\label{sec:seal}

\begin{figure}[ht]
\centering
\includegraphics[scale=0.8]{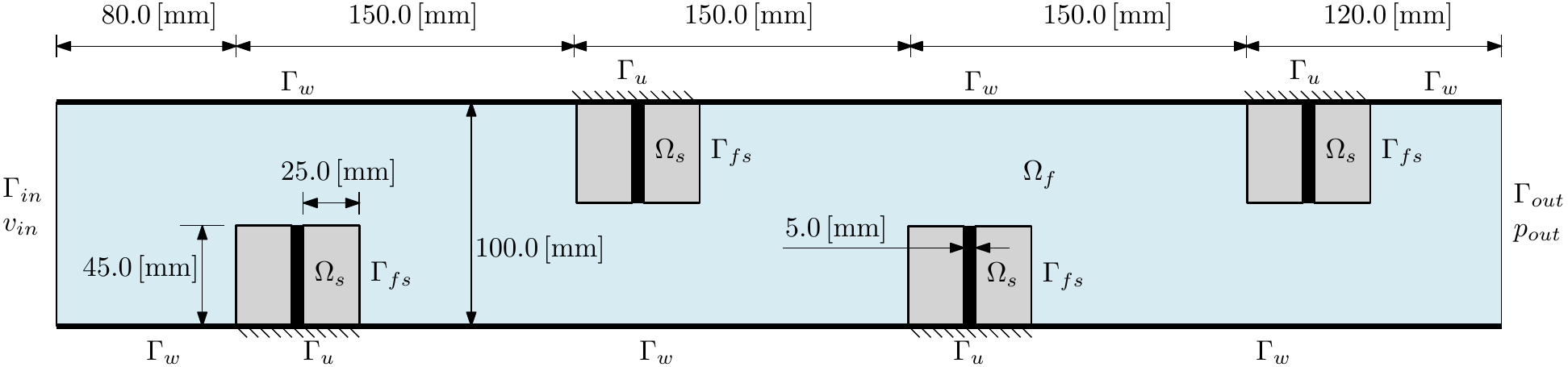}
\caption{The multiple walls example. The $5\times45$ mm solid walls (non-design domains in black region) must be supported by structures to be designed in the $25\times45$ mm areas in front and behind each wall.}
\label{fig:the_seal_model}
\end{figure}

This section presents the design of multiple walls inside a turbulent fluid flow channel. Figure \ref{fig:the_seal_model} illustrates the case of water flowing from the left \textcolor{black}{boundary} of a $650\times100$ mm channel to an outlet ($\hat{p}_0 \leq p_{out}$) at the right \textcolor{black}{boundary}. A normal constant inlet velocity $\mathbf{v} = v_{in}$ is imposed in a similar manner as the single wall example. The walls are considered to $5\times45$ mm in size. The regions of $25\times45$ mm in front and behind the walls are considered as design domain for the four structural supports. The flow path depends on the design of each support.

\begin{figure}[ht]
\centering
\begin{tabular}{c}
  \includegraphics[scale = 0.35, trim={0cm 6cm 0cm 7cm}, clip]{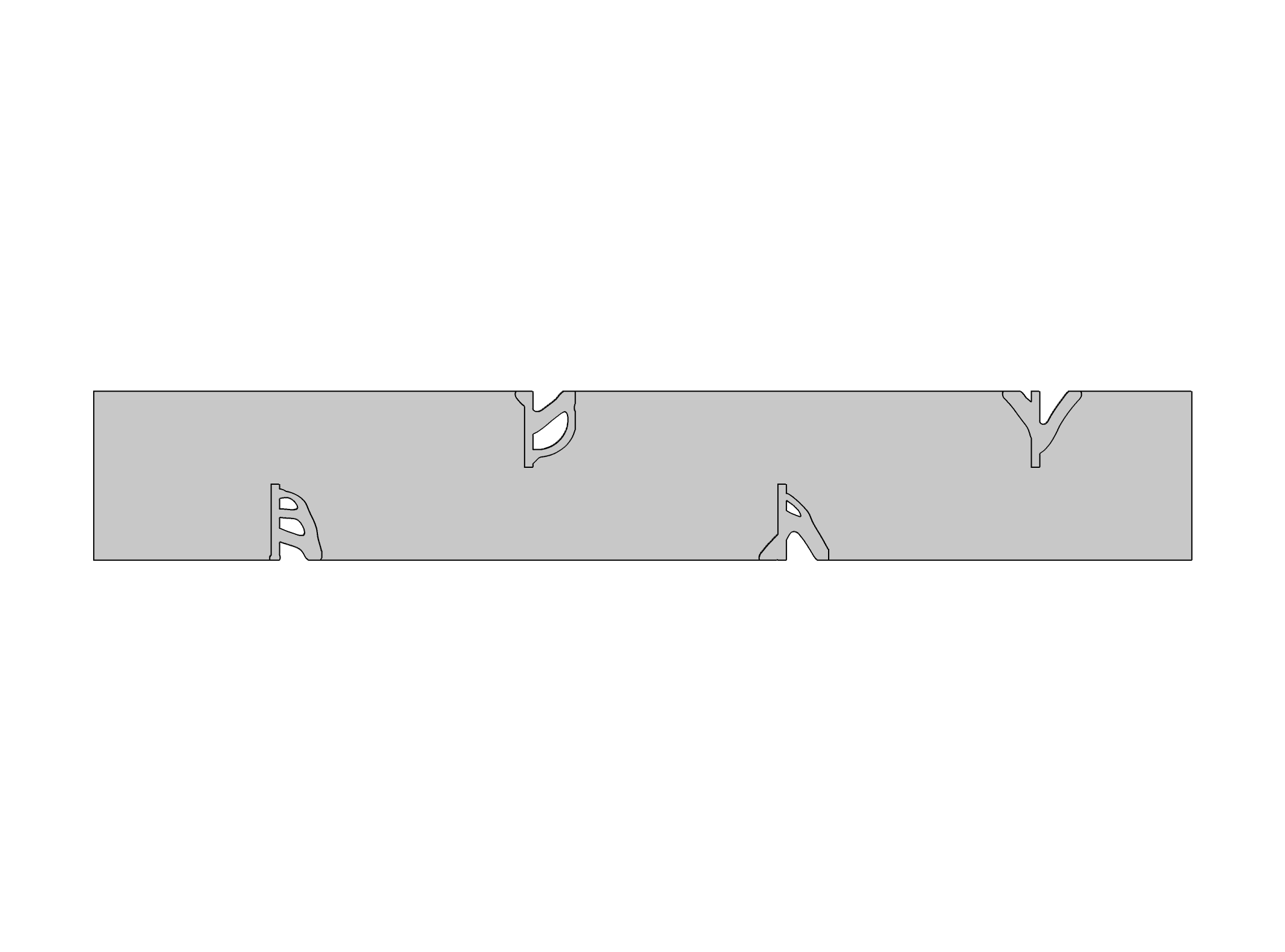} \\ (a) \\
  \includegraphics[scale = 0.35, trim={0cm 7cm 0cm 7cm}, clip]{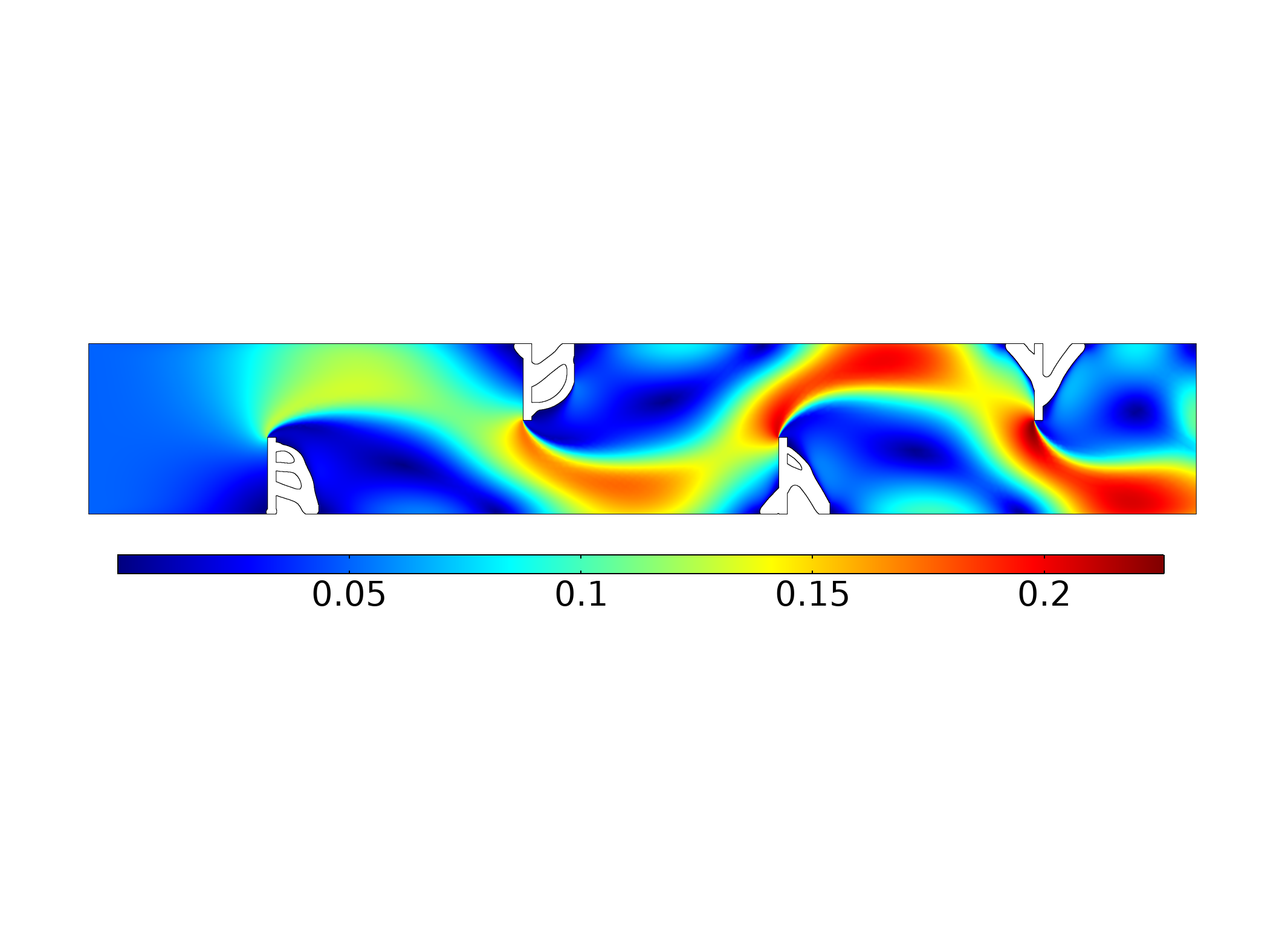} \\ (b) \\
  \includegraphics[scale = 0.35, trim={0cm 7cm 0cm 7cm}, clip]{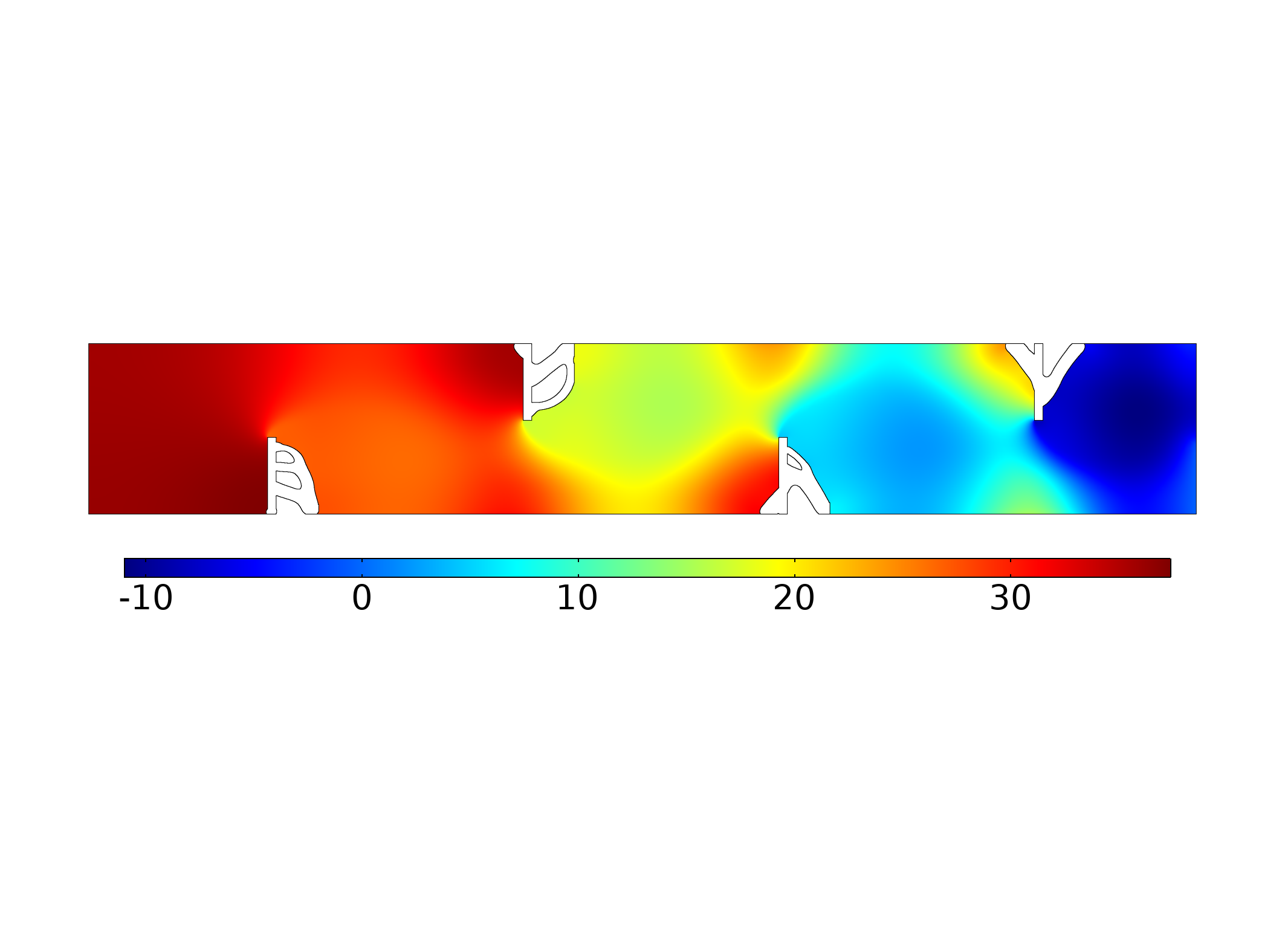} \\
  (c) \\
\end{tabular}
\caption{\textcolor{black}{The multiple walls example optimized for stiffness: (a) topology designs,(b) velocity field (in m/s) and (c) pressure field (in Pa), using the $k-\varepsilon$ turbulence model, Re = 5,000 and $p_{out} = 0$.}}
 \label{fig:the_seal_p0}
\end{figure}

The TOBS method is applied to the design the four structural supports. Compliance minimization is solved subject to four volume fraction constraints with $\bar{V}_i$ = 30$\%$. A set of 330$\times$270 optimization grid is used in each of the walls. A filter radius of 30 is chosen. The solid material penalization is $p = 5$. The constraint relaxation parameter is $\epsilon = 0.02$ and the truncation error constraint parameter is $\beta = 0.02$. The fluid flow is under the turbulent regime and the fluid-structure interaction problem is solved with Re = 5,000 and the $k-\varepsilon$ turbulence model. First, the problem is solved with outlet pressure $p_{out} = 0$. Figure \ref{fig:the_seal_p0} presents the optimized topologies and the velocity and pressure fields. It can be noticed that the first wall presents the least amount of solid material in the left part of the design domain. This makes sense as this is the region with higher pressure and shear so material is removed there to decrease the fluid flow loading, similarly to the previous single wall example. More material is chosen by the optimizer in the left part of the design domains as the pressure is lower. The first three support structures present only convex external shapes as they are being crushed by positive pressures on both sides. The fourth support structure presents a concave external shape in the region of negative pressure (dark blue in Fig. \ref{fig:the_seal_p0}). Not coincidentally, the fourth support structure looks alike the previous single wall designs. 

\begin{figure}[ht]
\centering
\begin{tabular}{c}
  \includegraphics[scale = 0.34, trim={0cm 5cm 0cm 6cm}, clip]{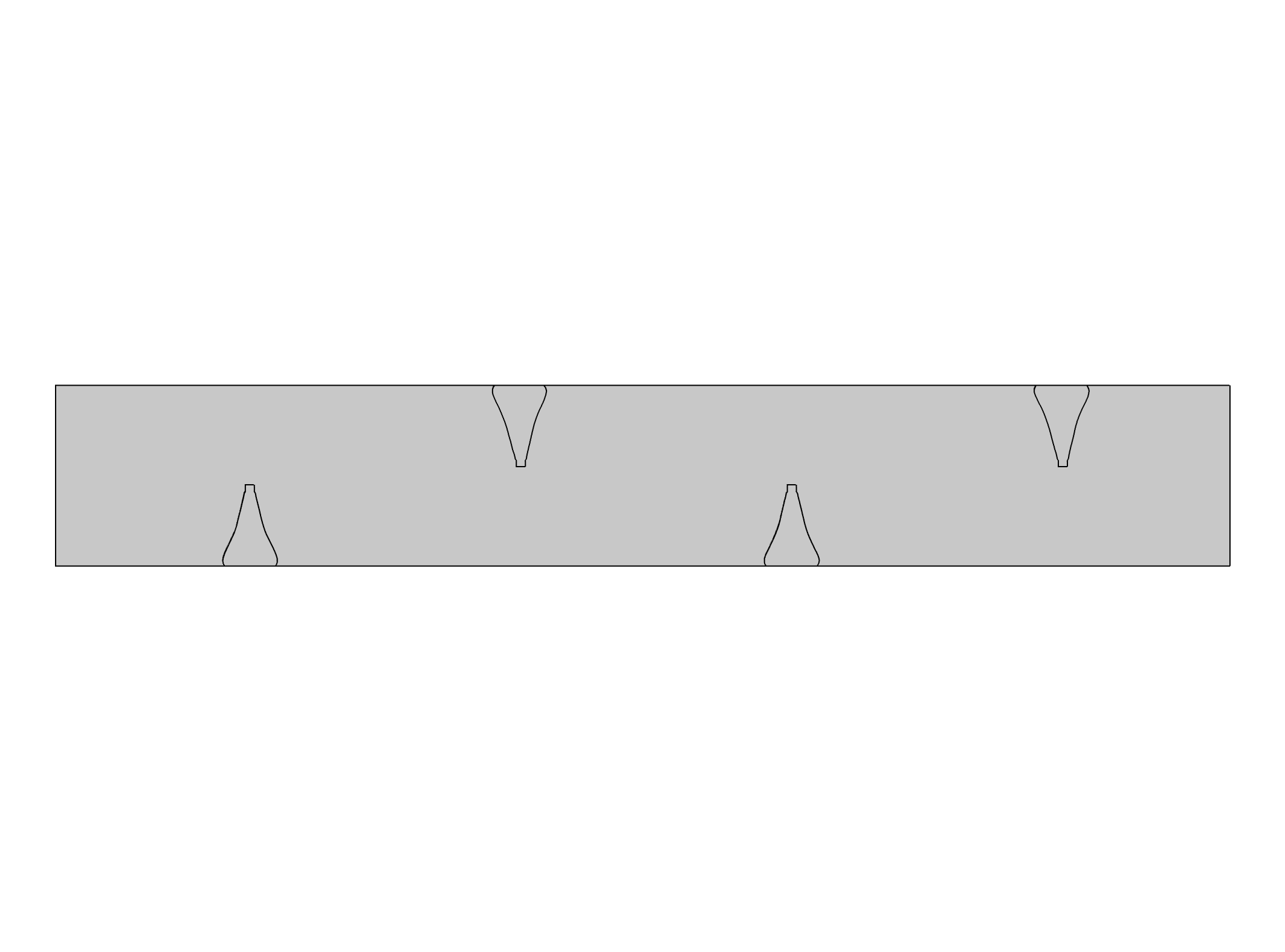} \\ (a) \\
  \includegraphics[scale = 0.34, trim={0cm 6cm 0cm 6cm}, clip]{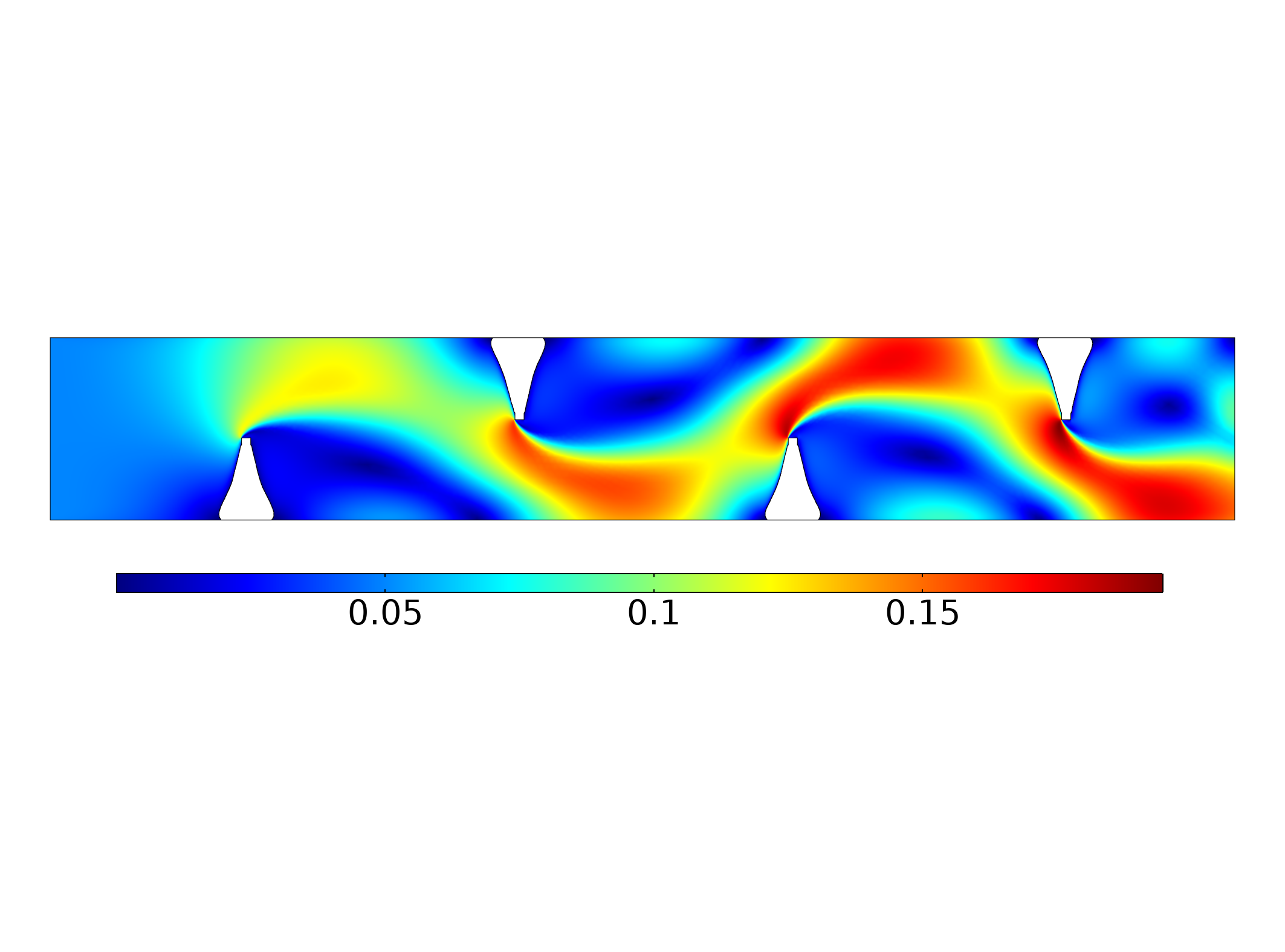} \\ (b) \\
  \includegraphics[scale = 0.34, trim={0cm 6cm 0cm 6cm}, clip]{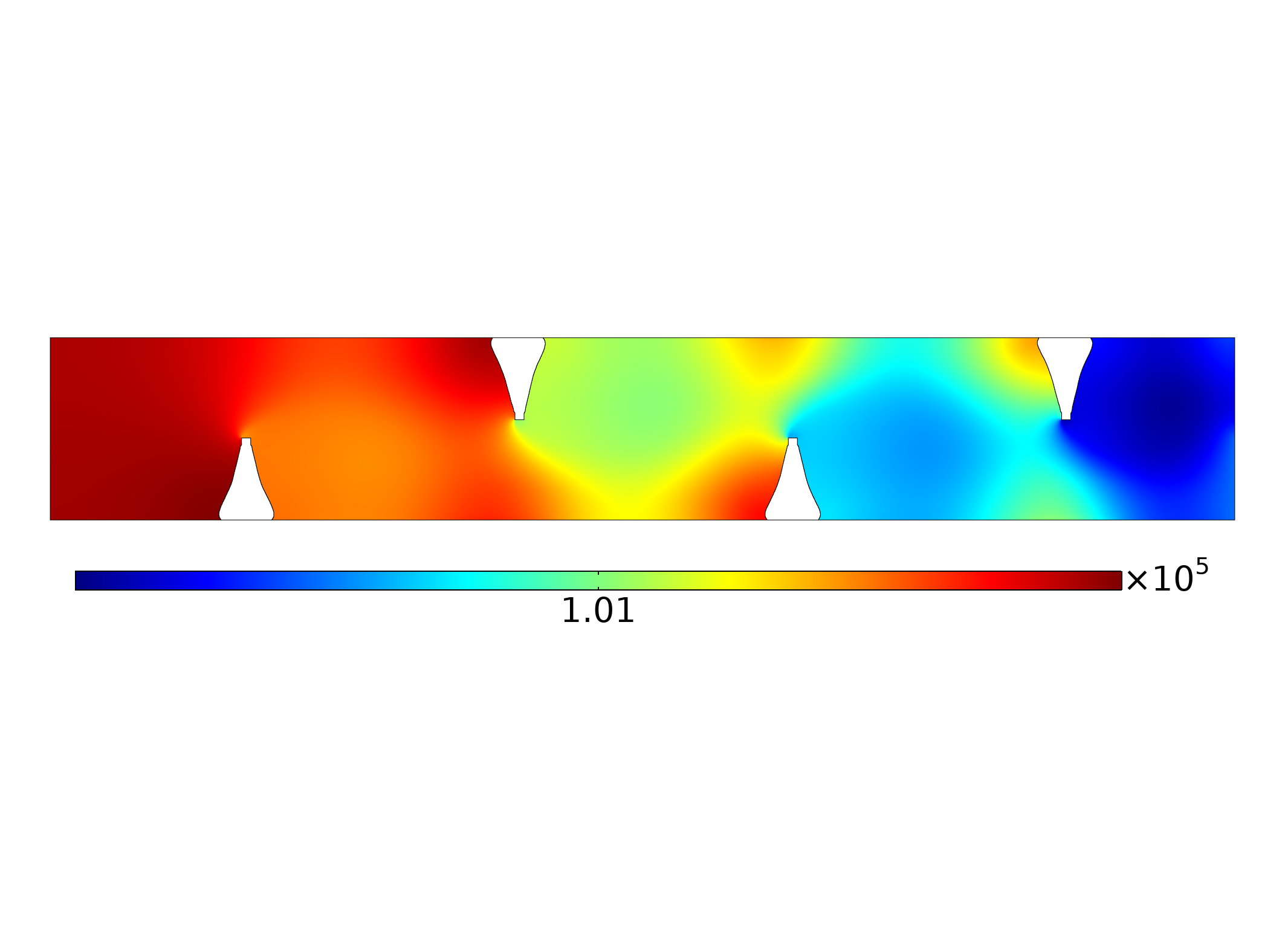} \\
   (c) \\
\end{tabular}
\caption{\textcolor{black}{The multiple walls example optimized for stiffness: (a) topology designs, (b) velocity field (in m/s) and (c) pressure field (in Pa), using the $k-\varepsilon$ turbulence model, Re = 5,000 and $p_{out} = 1$ atm.}}
 \label{fig:the_seal_p1}
\end{figure}

\begin{figure}[ht]
\centering
\begin{tabular}{c}
  \includegraphics[scale = 0.34, trim={0cm 0cm 0cm 0cm}, clip]{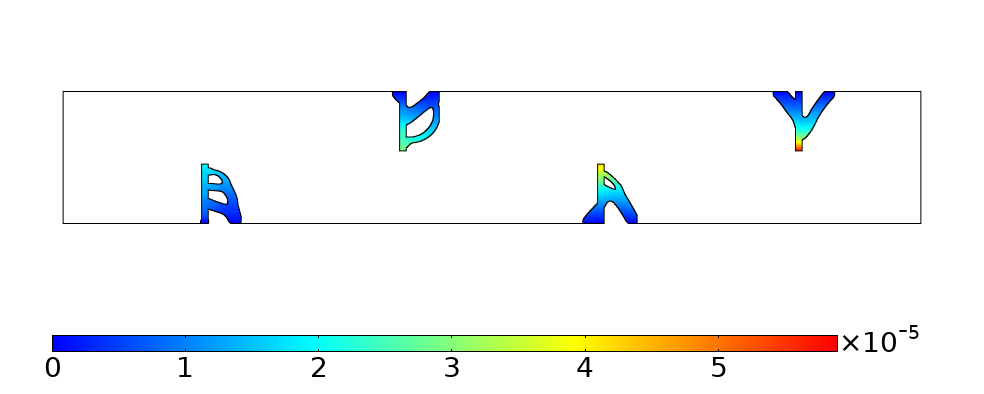} \\ (a) \\
  \includegraphics[scale = 0.34, trim={0cm 0cm 0cm 0cm}, clip]{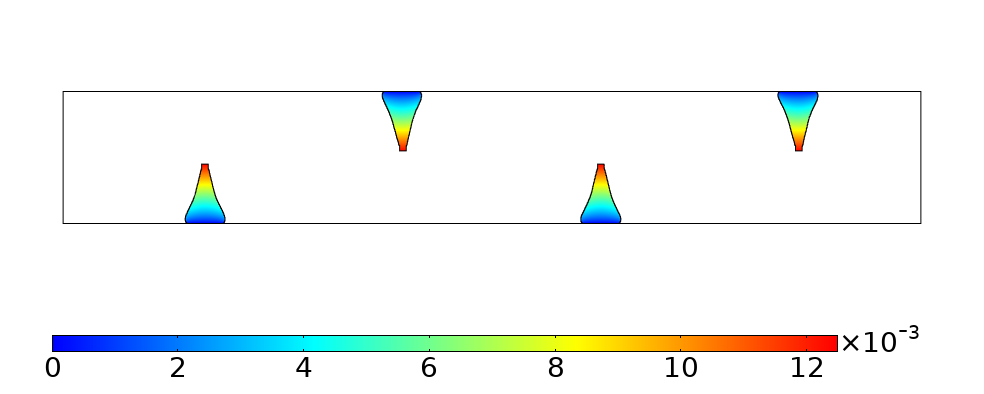} \\ (b) \\
\end{tabular}
\caption{\textcolor{black}{Displacement fields (in m) for the multiple walls final designs considering: (a) $p_{out} = 0$ atm and (b) $p_{out} = 1$ atm.}}
 \label{fig:the_seal_disps}
\end{figure}

The magnitude and signs of the pressure field in Fig. \ref{fig:the_seal_p0} are with respect to the reference pressure $p_{out} = 0$ from the outlet boundary condition and they are fundamental to understand obtained the designs. When solving the fluid flow using a different reference pressure, different designs are obtained. Figure \ref{fig:the_seal_p1} shows the optimized topologies when solving the problem using $p_{out} = 1$ atm. All the four walls present a similar design, thicker at the bottom and thinner at the top region to resist the crushing and practically constant loads. Although the pressure fields (Figs. \ref{fig:the_seal_p0}c and \ref{fig:the_seal_p1}c) are visually similar, the signs of the pressure dictate the structural deformation, \textcolor{black}{as seen in the displacement fields shown in Fig. \ref{fig:the_seal_disps}.}

\section{The traffic sign}

\begin{figure}[ht]
\centering
\includegraphics[scale=0.3]{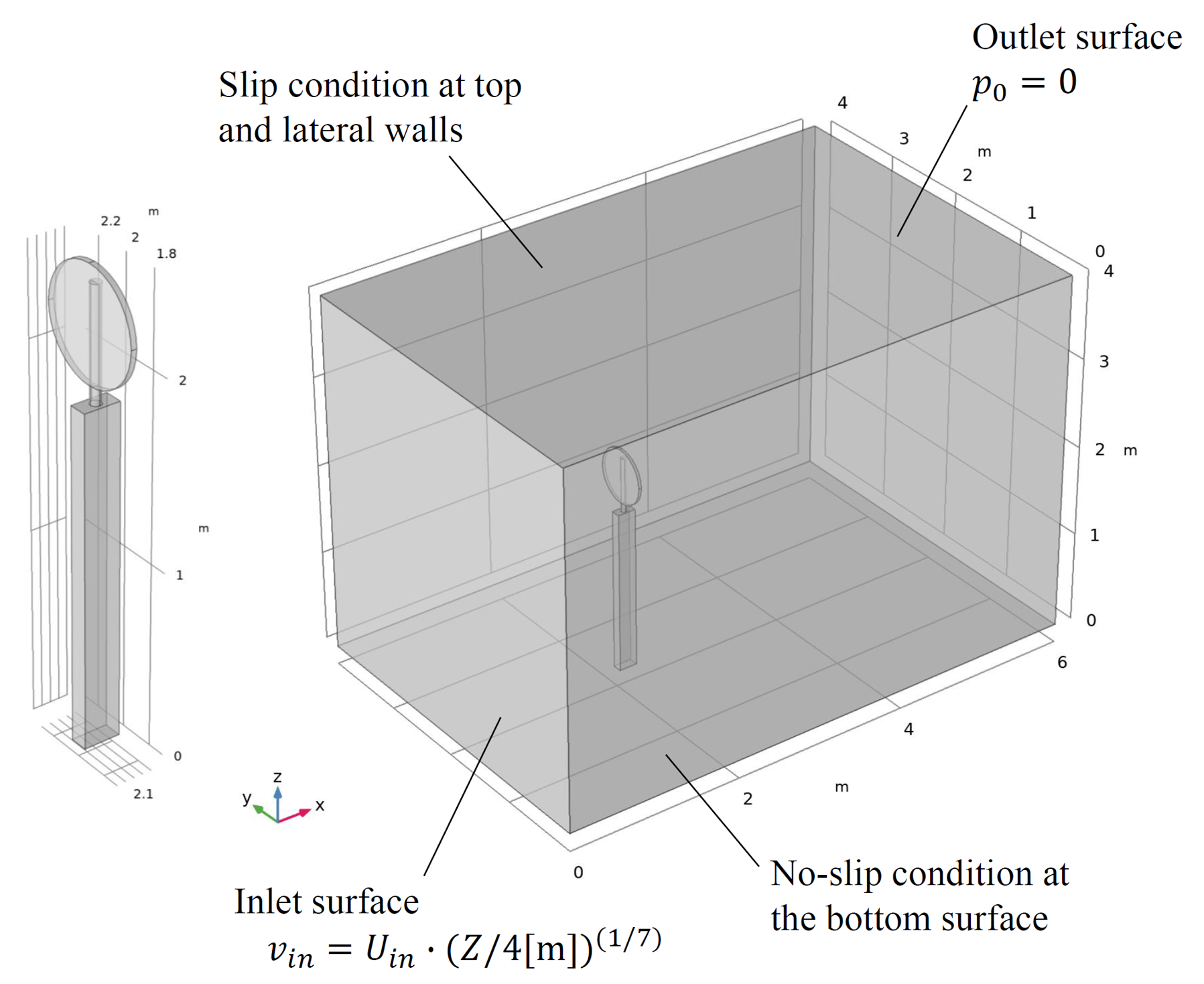}
\caption{The 3D traffic sign example. The structural support below the sign is considered as design domain and must comply with the turbulent ($k-\varepsilon$) wind load.}
\label{fig:the_3D_wind}
\end{figure}

This example investigates the 3D design using the \textcolor{black}{TOBS-GT} method of a traffic sign under wind loads. The fluid is considered to be air. The $k-\varepsilon$ turbulence model is employed. A traffic sign of height 2.5 m is placed inside a $6\times4\times4$ m fluid domain, as illustrated in Fig. \ref{fig:the_3D_wind}. The block of $0.2\times0.1\times1.8$ m below the sign is considered as design domain. The fluid flow enters the left \textcolor{black}{boundary} of the domain with inlet velocity $\mathbf{v} = v_{in}\cdot(z/H)^{\frac{1}{7}}$, where $H$ is the height of the fluid channel (4 m) and $z$ is the vertical coordinate at each point of the inlet. The velocity $v_{in}$ is set to be 15 m/s, the speed of a high wind, near gale, in the Beaufort scale. A reference pressure $p_{out} = 0$ is imposed at the outlet boundary of the fluid domain. The bottom boundary of the fluid is a no-slip wall and all the remaining boundaries are set to experience slip conditions. The bottom boundary of the structure has fixed displacements, $\mathbf{u} = 0$.

\begin{figure}[ht]
\centering
\begin{tabular}{ccc}
  \includegraphics[scale = 0.3, trim={4cm 0cm 4cm 0cm}, clip]{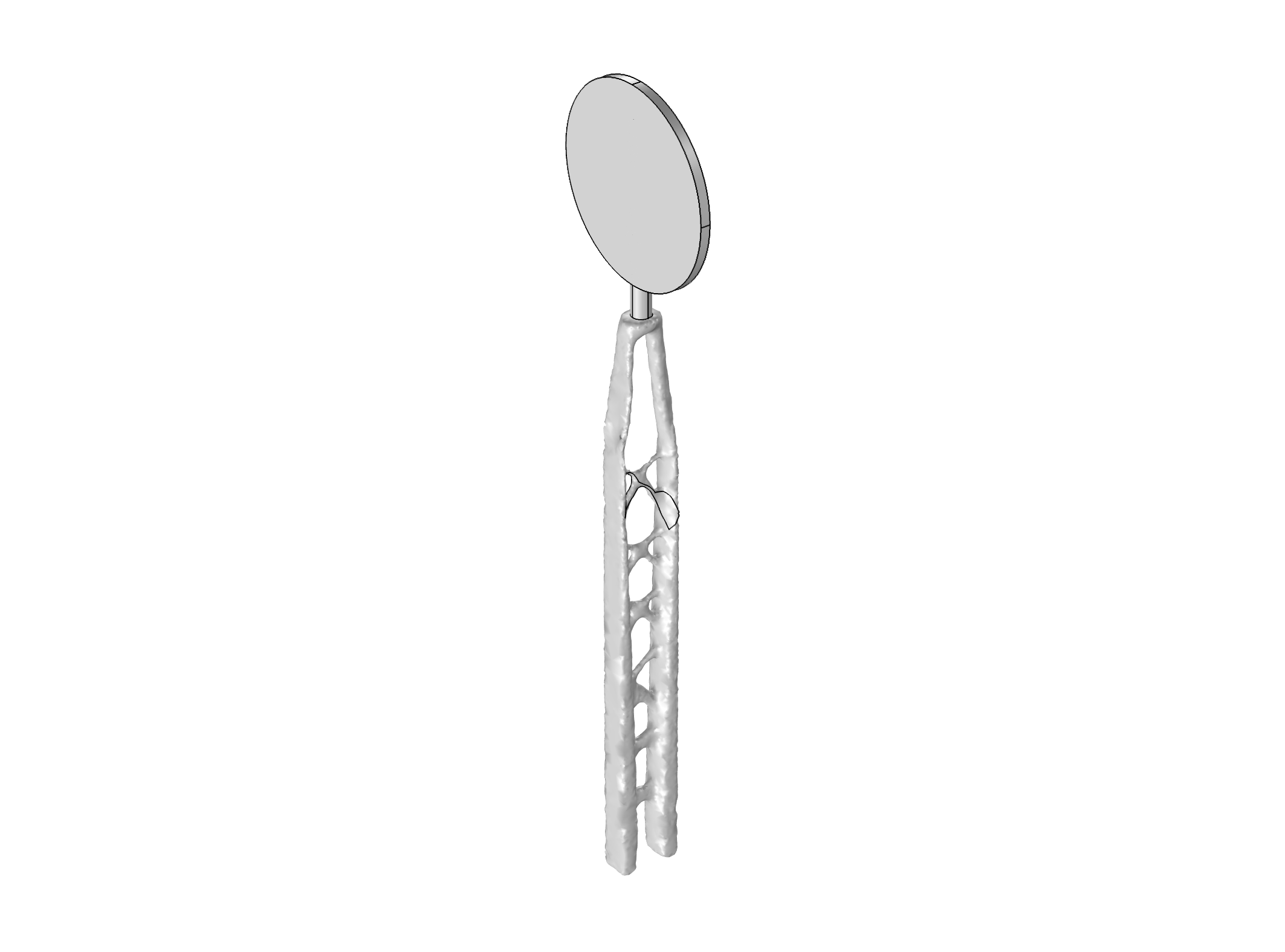} & \includegraphics[scale = 0.3, trim={4cm 0cm 4cm 0cm}, clip]{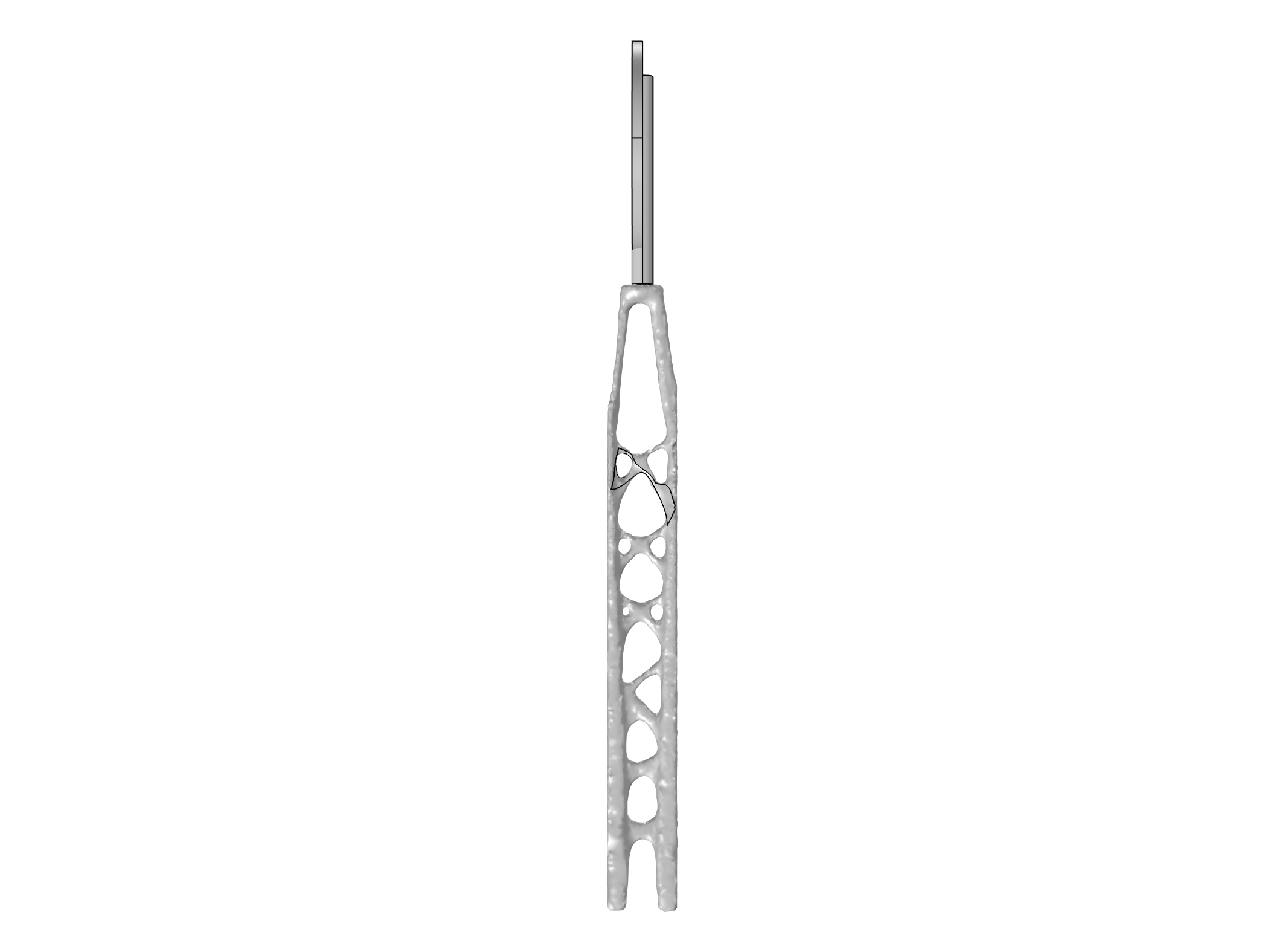} &
  \includegraphics[scale = 0.3, trim={4cm 0cm 4cm 0cm}, clip]{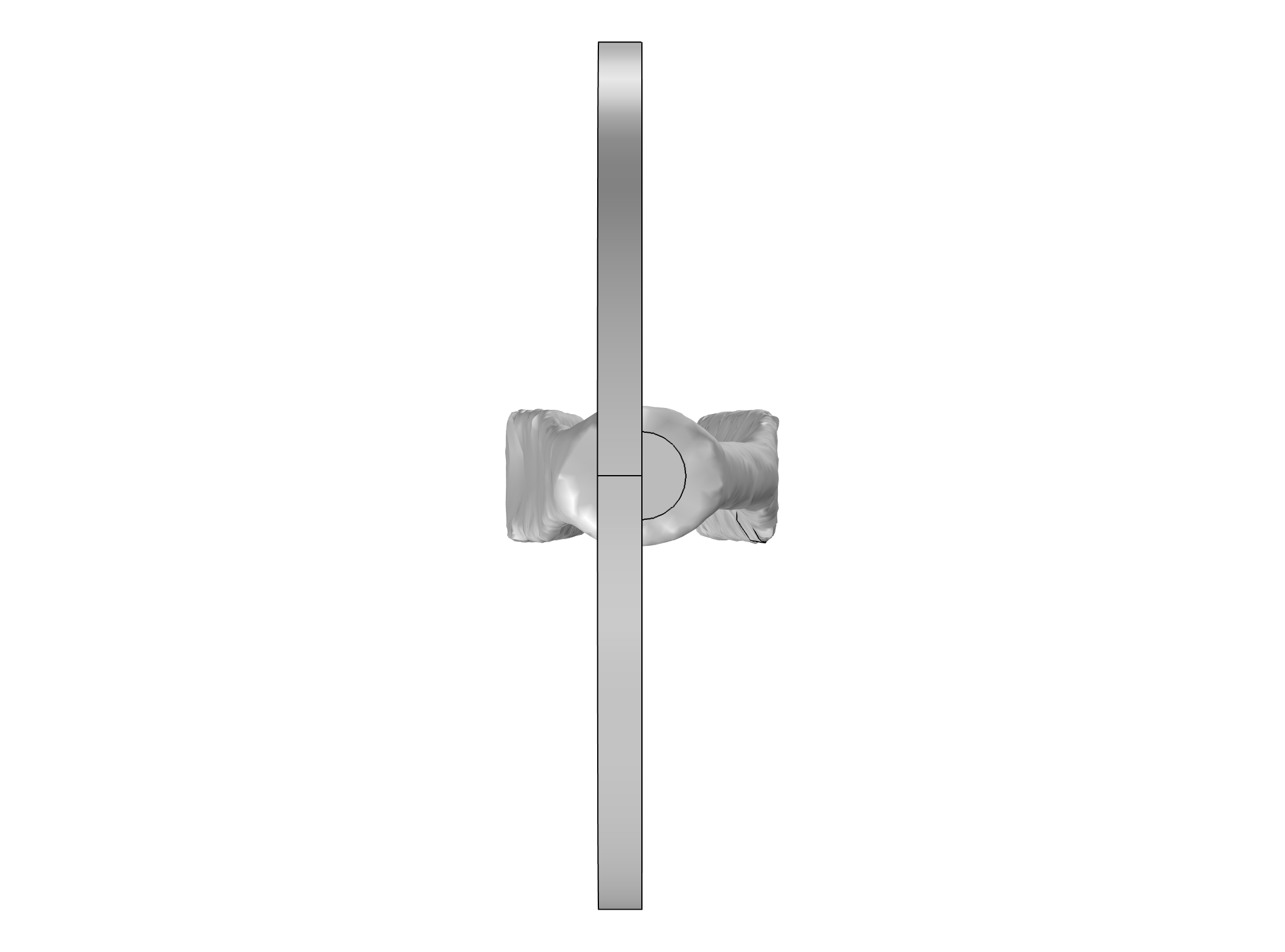} \\
\end{tabular}
 \caption{Different views of the optimized traffic sign.}
 \label{fig:the_traffic_sign_top}
\end{figure}

\begin{figure}[ht]
\centering
\begin{tabular}{cc}
  \includegraphics[scale = 0.3, trim={0cm 0cm 0cm 0cm}, clip]{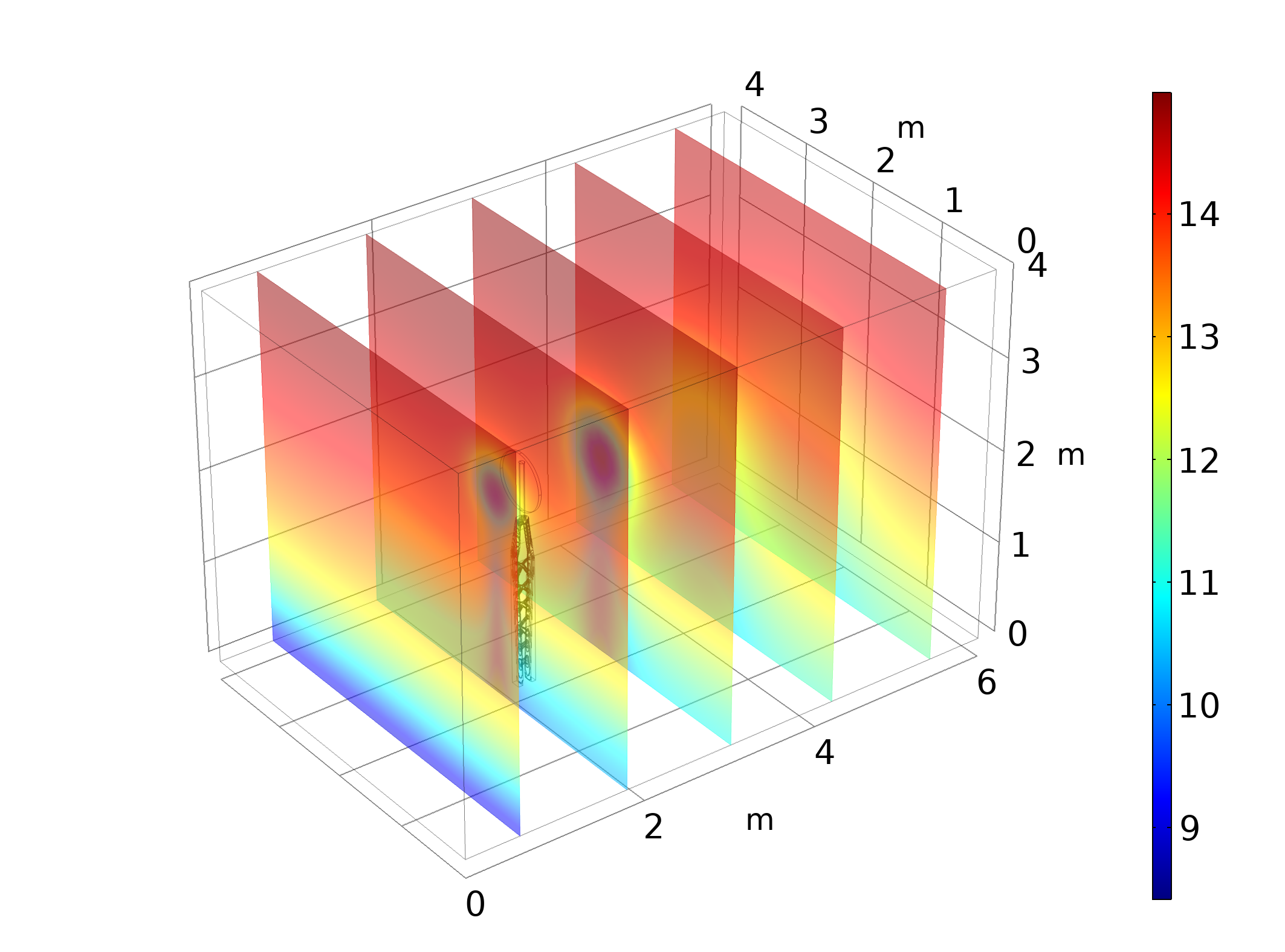} & \includegraphics[scale = 0.3, trim={0cm 0cm 0cm 0cm}, clip]{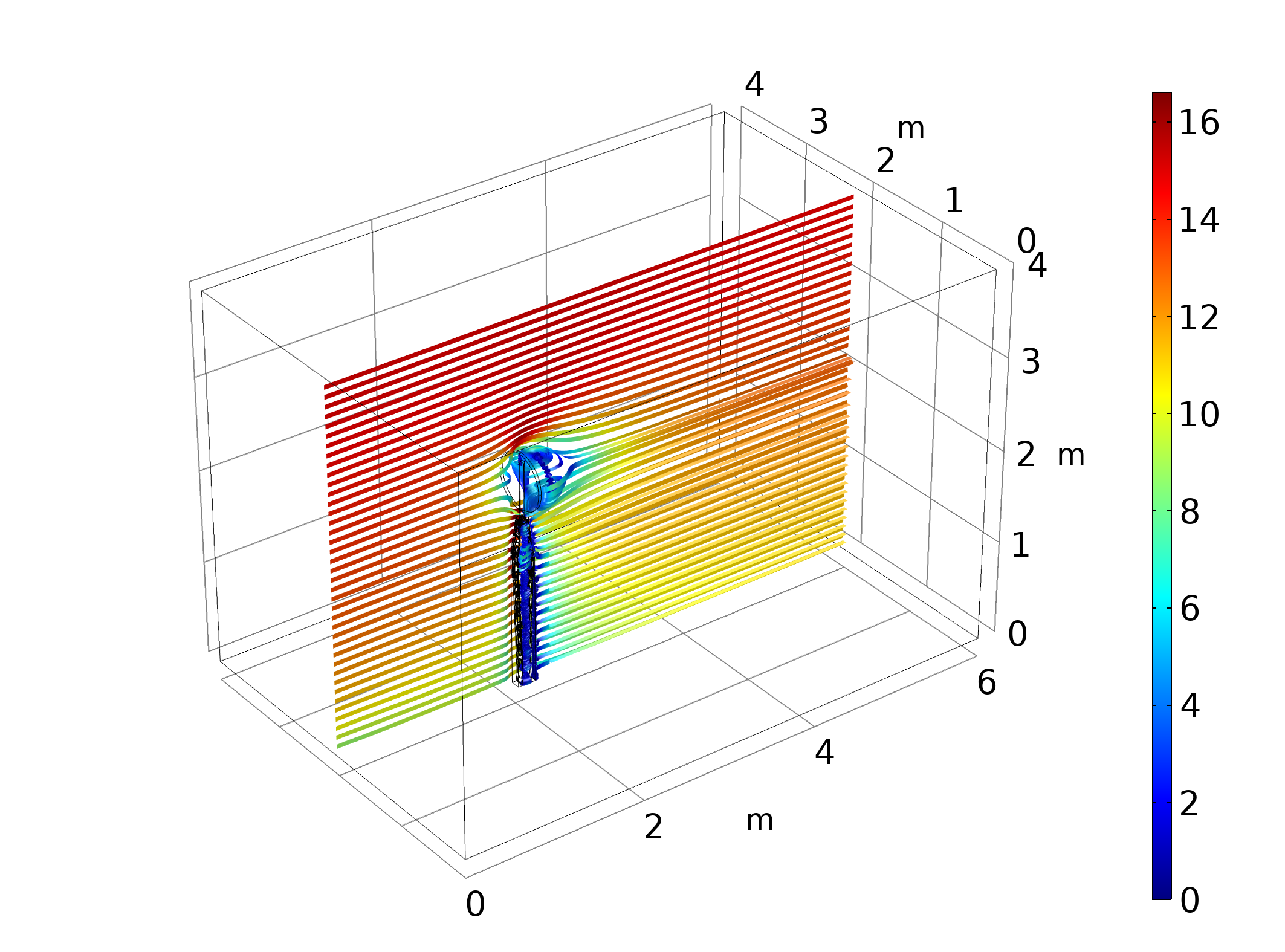} \\
  \includegraphics[scale = 0.3, trim={0cm 0cm 0cm 0cm}, clip]{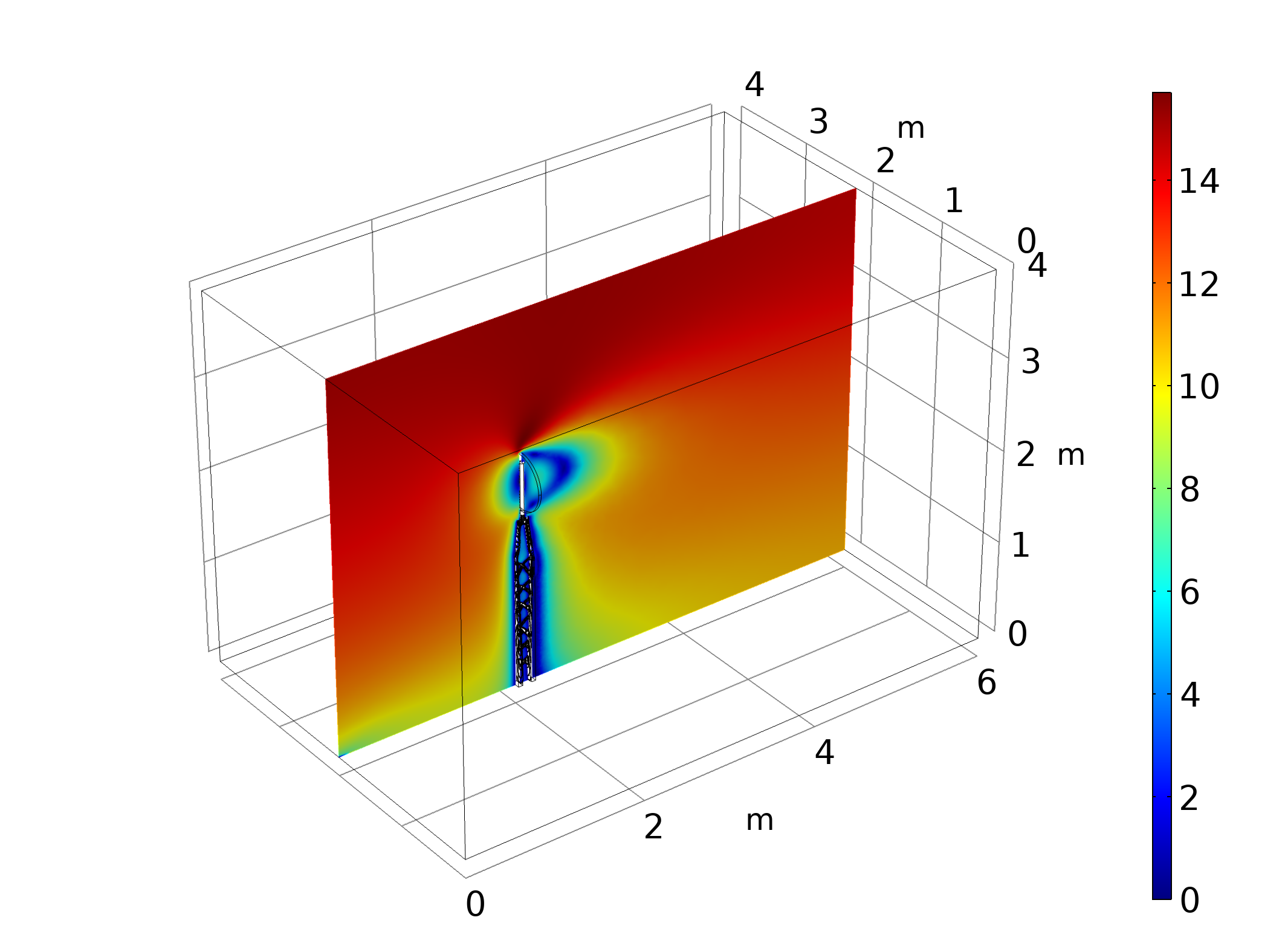} & \includegraphics[scale = 0.3, trim={0cm 0cm 0cm 0cm}, clip]{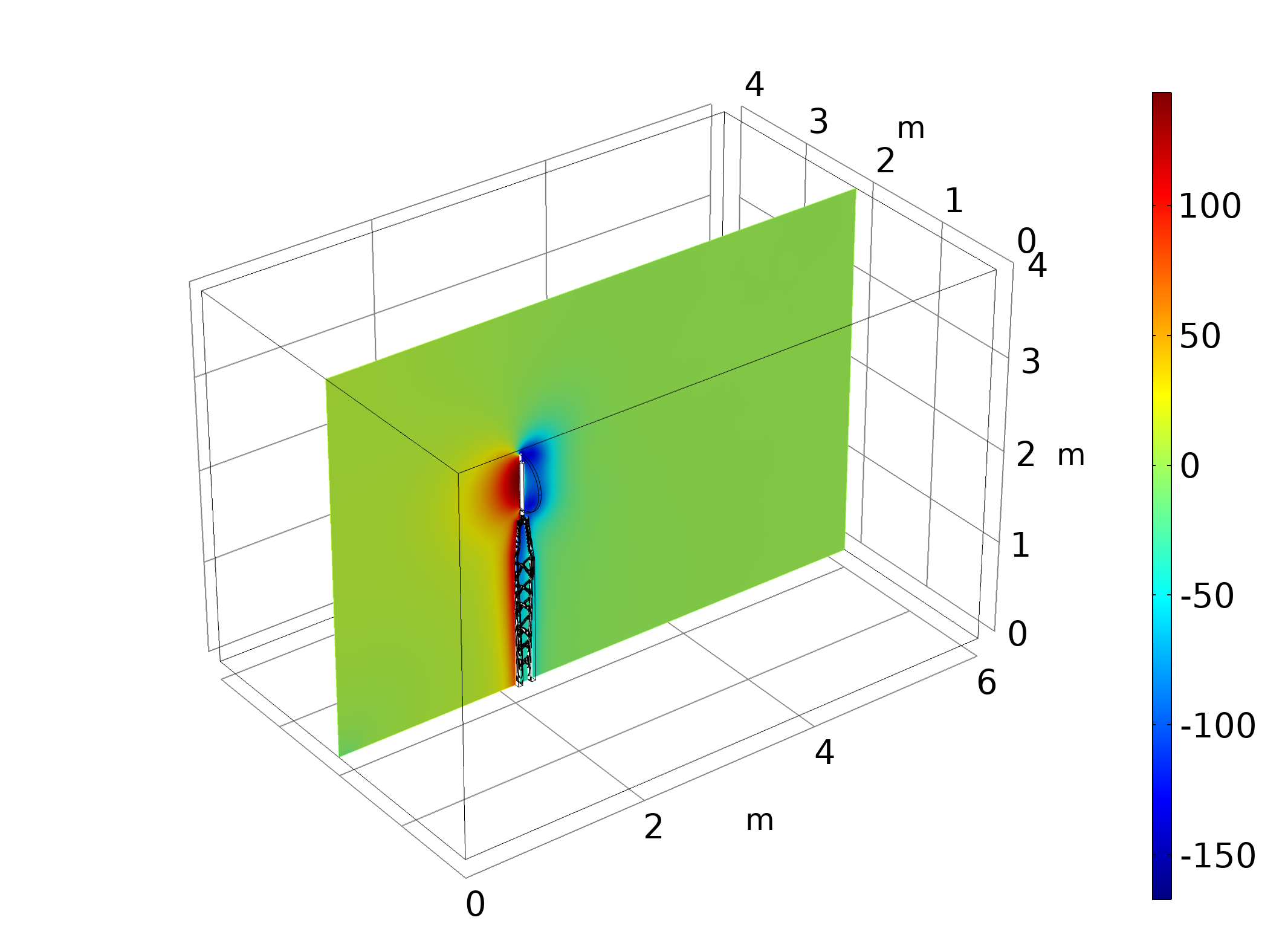} \\
\end{tabular}
 \caption{Velocity (in m/s) and pressure (in Pa) fields for a 15 m/s turbulent wind ($k-\varepsilon$) hitting a traffic sign designed with the TOBS method.}
 \label{fig:the_traffic_sign_results}
\end{figure}

The structural compliance is minimized using the TOBS-GT method subject to a volume fraction constraint of $\bar{V} = 40\%$. A $40\times360\times20$ optimization grid size is considered. A filter radius of 0.015 m is chosen. The constraint relaxation parameter is $\epsilon = 0.02$ and the truncation error constraint parameter is $\beta = 0.04$. The material penalization is set as $\gamma$ = 10. Figure \ref{fig:the_traffic_sign_top} presents three different views of the optimized traffic sign support. The structure is designed similarly as an I-beam, including holes that produce a truss-like design. The design of 3D structures via topology optimization considering FSI is challenging and they are not direct extensions of 2D designs. Besides, the fluid responses can be drastically different when solving 3D problems, such as presenting \textcolor{black}{vortices} flowing around the structure and presenting positive and negative pressure regions in all directions. Figure \ref{fig:the_traffic_sign_results} presents the velocity and pressure fields of the traffic sign example.

\section{Conclusions}
\label{sec:con}

This work developed a method for structural topology optimization that includes turbulent fluid flow loads. The \textcolor{black}{separation} of the optimization grid and the FEA mesh guarantees that the complex FSI analysis can be done with separate governing equations and domains whilst keeping the material distribution feature of the classic topology optimization idea. This method is called TOBS-GT. The optimization problem is formulated and solved with sequential integer linear programming by the standard TOBS method. Its association with a geometry trimming process enables the CAD modeling of the fluid-structure designs. These numerical ingredients lead to results that show topology optimization solving fluid-structure interaction problems with turbulence models for the first time. The $k-\varepsilon$ turbulence model with standard wall function at fluid and fluid-structure walls \textcolor{black}{was chosen to illustrate the method as this choice is independent of the proposed algorithm}. Structures could be designed with the TOBS-GT method under loads from high Reynolds flow (Re up to 50,000) for 2D and 3D cases. \textcolor{black}{Recommendations include to give attention to the penalized stiffness, as it changes the proportion between stiffness and fluid loading sensitivities. The optimization grid size can be set up around a few hundred units to obtain a smooth enough boundary. The magnitude of the outlet pressure field is relevant. The optimization parameters are usually standard when considering only volume constraints. It was shown that the computational bottleneck is still the FEA equations. Future works can address the present problem including \textcolor{black}{nonlinear constraints and} more complex physics such as compressible flow, rotational flow and conjugate heat transfer.}


\section{Replication of results}

The results presented in this work can be reproduced by following the algorithms and formulations presented in detail herein. The standard TOBS implementation is presented in \url{www.github.com/renatopicelli/tobs} and in \cite{Picelli20edu}.


{\setlength{\parindent}{0pt}\par\addvspace{17pt}\small\rmfamily\setstretch{-1.0} \textbf{Funding information} This research was partly supported by CNPq (Brazilian Research Council) and FAPESP (São Paulo Research Foundation). The authors thank the supporting institutions. The first author thanks FAPESP under the Young Investigators Awards program, grants 2018/05797-8 and 2019/01685-3. The fourth author thanks FAPESP under grant 2017/27049-0. The last author thanks the financial support of CNPq (National Council for Research and Development) under grant 302658/2018-1 and FAPESP under grant 2013/24434-0. The first, second, fourth and fifth authors also acknowledge the support of the RCGI (Research Centre for Greenhouse Gas Innovation), hosted by the University of São Paulo (USP) and sponsored by FAPESP (2020/15230-5) and Shell Brazil.}


\section*{Compliance with ethical standards}

{\setlength{\parindent}{0pt}\par\addvspace{17pt}\small\rmfamily\setstretch{-1.0} \textbf{Conflict of interest} The authors declare that they have no conflict of interest.}


\bibliographystyle{spbasic}      
\bibliography{mybibfile}   	 


\section*{\textcolor{black}{Appendix A}}

\begin{figure}[!h]
\centering
  \includegraphics[scale = 0.8]{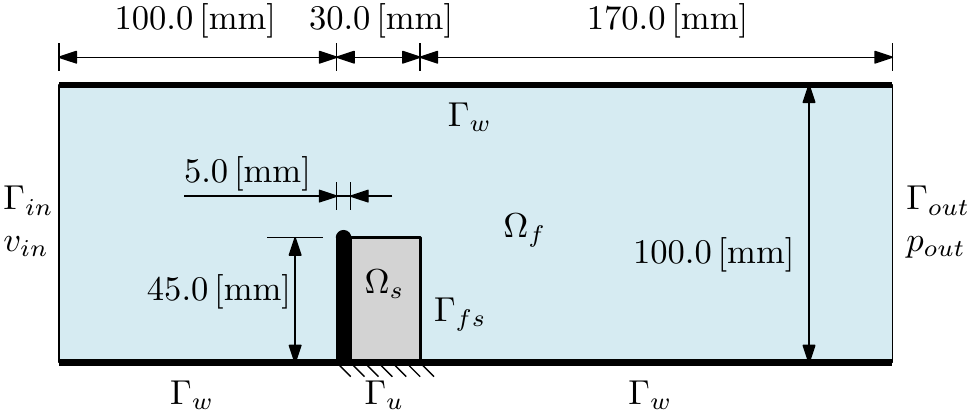}  \\
 \caption{\textcolor{black}{The wall example used to carry out the finite differences check.}}
 \label{fig:fd_model}
\end{figure}

\begin{figure}[!h]
\centering
\begin{tabular}{cc}
  \includegraphics[scale = 0.4]{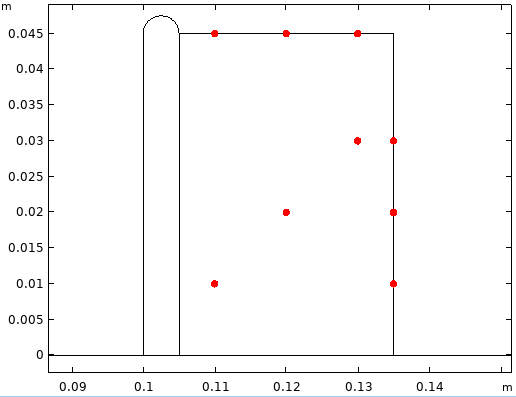} & \includegraphics[scale = 0.45, trim={0cm 0cm 0cm 0cm}, clip]{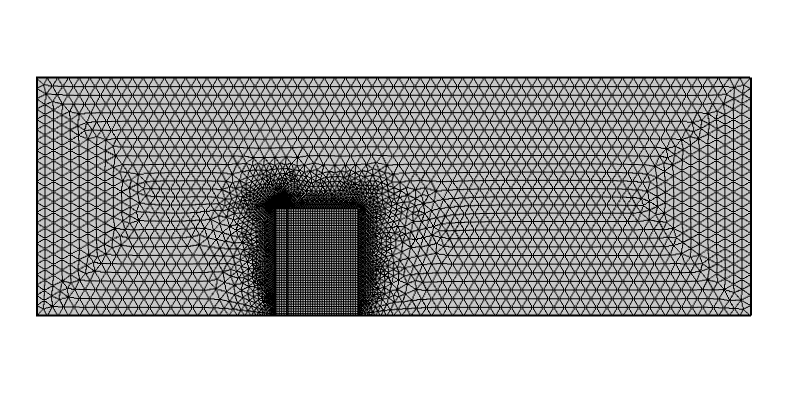} \\
  \textcolor{black}{(a) points for finite differences check} & \textcolor{black}{(b) finite element mesh} \\
\end{tabular}
 \caption{\textcolor{black}{Model used to carry out the finite differences check: (a) points in the design domain and (b) finite element mesh.}}
 \label{fig:fd}
\end{figure}

\begin{figure}[!h]
\centering
\begin{tabular}{cc}
  \includegraphics[scale = 0.3, trim={0cm 0cm 0cm 0cm}, clip]{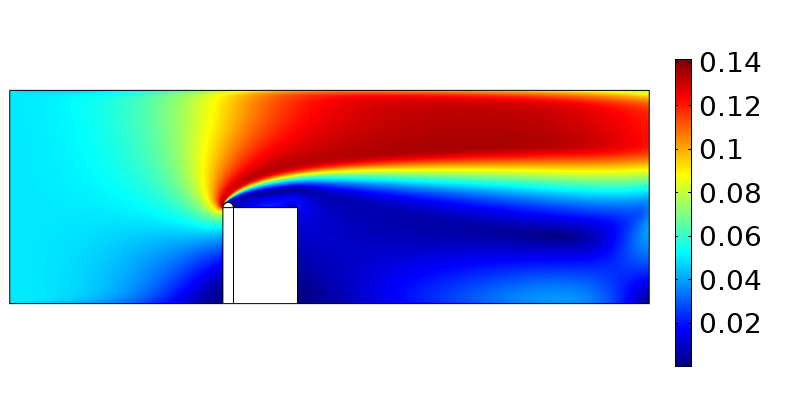} & \includegraphics[scale = 0.3, trim={0cm 0cm 0cm 0cm}, clip]{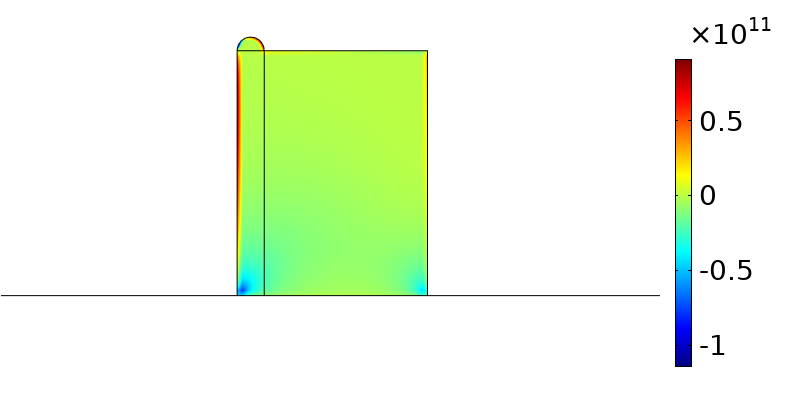} \\
  \textcolor{black}{(a) velocity (in m/s)} & \textcolor{black}{(b) sensitivities (in 1/m$^2$)} \\
  \includegraphics[scale = 0.3, trim={0cm 0cm 0cm 0cm}, clip]{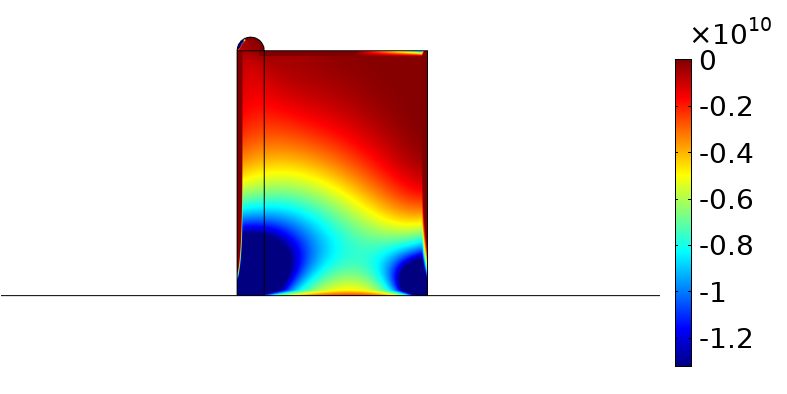} & \includegraphics[scale = 0.3, trim={0cm 0cm 0cm 0cm}, clip]{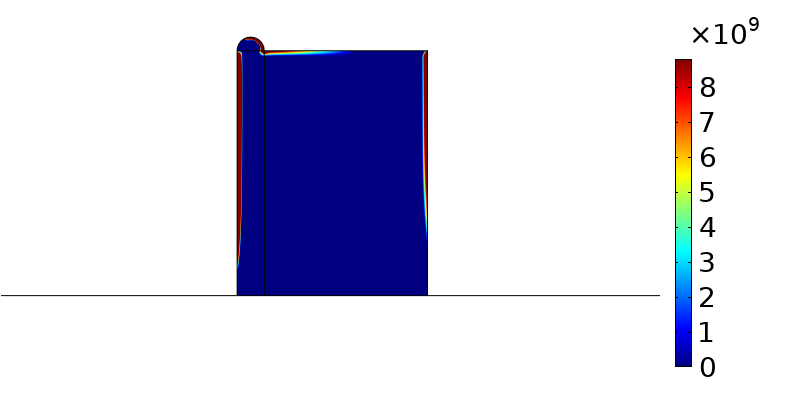} \\
  \textcolor{black}{(c) negative portion of the sensitivities (in 1/m$^2$)} & \textcolor{black}{(d) positive portion of the sensitivities (in 1/m$^2$)} \\
\end{tabular}
 \caption{\textcolor{black}{Initial full solid design: (a) fluid velocity field (in m/s) of the model considered in the finite differences check using the $k-\varepsilon$ turbulence model, (b) sensitivity field obtained with semi-automatic differentiation, (c) negative portion of the  sensitivities and (d) positive portion of the sensitivities.}}
 \label{fig:fd_vel}
\end{figure}

\begin{figure}[!h]
\centering
\begin{tabular}{cc}
  \includegraphics[scale = 0.3, trim={0cm 0cm 0cm 0cm}, clip]{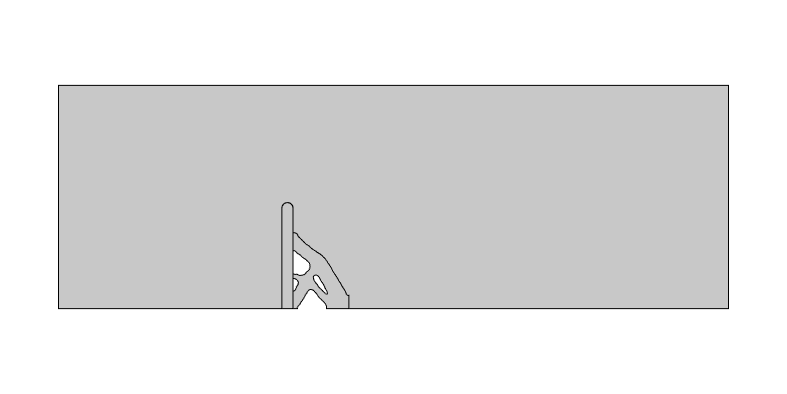} & \includegraphics[scale = 0.3, trim={0cm 0cm 0cm 0cm}, clip]{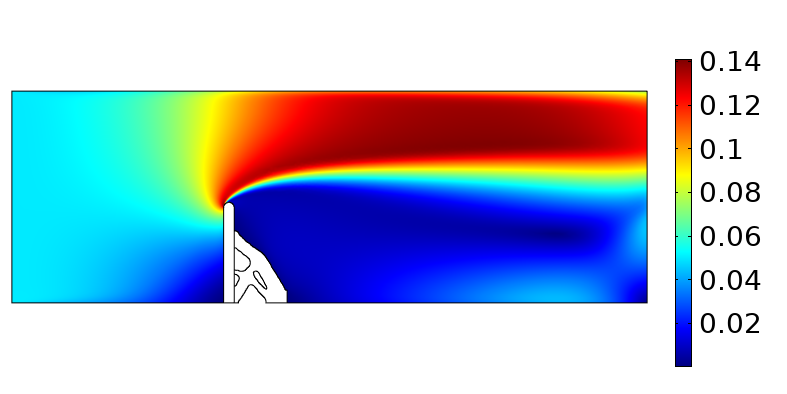} \\
  \textcolor{black}{(a) optimized topology} & \textcolor{black}{(b) velocity (in m/s)} \\
\end{tabular}
\caption{\textcolor{black}{Optimized design with the TOBS-GT method: (a) topology solution and (b) fluid velocity field (in m/s) using the $k-\varepsilon$ turbulence model with Re = 5,000.}}
 \label{fig:fd_top_vel}
\end{figure}

\textcolor{black}{This appendix presents the analysis by finite differences used to verify the sensitivities from Eq. \ref{eq:adjoint2} obtained via semi-automatic differentiation. The analysis is carried out for the wall example illustrated in Fig. \ref{fig:fd_model}. The turbulent fluid (using the $k-\varepsilon$ model with standard wall functions) flows through the inlet with Re = 5,000 and exits at the outlet with $p_{out} = 0$. The fluid is considered to be water and the solid material is chosen to have Young's modulus $E$ = 1$\times$10$^{-8}$ Pa and Poisson's ratio $\nu$ = 0.3. The penalty on stiffness is chosen to be $p = 5$. The semi-automatic differentiated sensitivities in a distribution of nine points around the design domain, being six of them at the fluid-structure boundaries, are verified against finite differences. \textcolor{black}{As the design variables in the FEA software are defined as a scalar field, we have chosen the field of the full solid design (all variables equal to 1) and have added a step of $1\times10^{-5}$ on a backward finite differences scheme.} Figure \ref{fig:fd}(a) and (b) present, respectively, the considered points in the finite differences analysis and the finite element mesh used. Figure \ref{fig:fd_vel}(a) presents the velocity field for this example using the and Fig. \ref{fig:fd_vel}(b) shows the sensitivity field computed with semi-automatic differentiation. Figures \ref{fig:fd_vel}(c-d) present the clipped sensitivity field to show its negative and positive portions. Table \ref{tab:fd} shows the sensitivity values obtained via semi-automatic differentiation and via finite differences at the considered points. It can be noticed that some of the points at the fluid-structure interfaces present positive values, which indicate the regions where material can be removed in order to decrease the overall loading on the structure. It is important then to verify the accuracy of these sensitivities as well. The maximum relative difference between the semi-automatic differentiation and the finite differences methods was 0.0217$\%$, which is small and validates the usability of the semi-automatic differentiation tool available in the software. Figure \ref{fig:fd_top_vel} presents the optimized topology for this example using the TOBS-GT method using $\epsilon = 0.02$ and $\beta = 0.02$.}

\begin{table}[h!]
  \begin{center}
    \caption{Finite differences analysis results; SAD = semi-automatic differentiated sensitivities; FD = sensitivities obtained by \textcolor{black}{backward} finite differences.}
    \label{tab:fd}
    \begin{tabular}{c|c|c|c|c}
      \hline
      Point  & $(x, y)$ [m] & AD [kNm] & FD [kNm] & difference [$\%$] \\
      \hline
      \hline
      1 & (0.110, 0.010) & -6.8460 & -6.8460 & 0.0004  \\
      2 & (0.120, 0.020) & -2.5138 & -2.5138 & 0.0004  \\
      3 & (0.130, 0.030) & -0.2004 & -0.2004 & \textcolor{black}{0.0001}  \\
      4 & (0.135, 0.010) &  3.7527 &  3.7535 & \textcolor{black}{0.0211}  \\
      5 & (0.135, 0.020) & 11.8272 & 11.8284 & \textcolor{black}{0.0100}  \\
      6 & (0.135, 0.030) & 18.6793 & 18.6811 & 0.0095  \\
      7 & (0.110, 0.045) &  9.3537 &  9.3549 & 0.0132  \\
      8 & (0.120, 0.045) &  1.5983 &  1.5985 & 0.0126  \\
      9 & (0.130, 0.045) & -5.9411 & -5.9417 & \textcolor{black}{0.0110}  \\
      \hline
    \end{tabular}
  \end{center}
\end{table}

\end{document}